\documentclass[twocolumn, astrosymb, tighten, trackchanges, twocolappendix, floatfix]{aastex631}

\usepackage{graphics,graphicx}
\usepackage{soul}
\usepackage{bm}
\usepackage{wrapfig}
\usepackage[normalem]{ulem}
\usepackage[most]{tcolorbox}
\usepackage{romannum}
\usepackage{tablefootnote}
\usepackage{color}

\newcommand{\nev}{\hbox{[Ne\sc v]}}
\newcommand{\nevlong}{\hbox{[Ne\sc v]\,3426\AA}}
\newcommand{\oii}{\hbox{[O\sc ii]}}
\newcommand{\oiilong}{\hbox{[O\sc ii]\,3726,3729\AA}}
\newcommand{\oiialong}{\hbox{[O\sc ii]\,3726\AA}}
\newcommand{\oiiblong}{\hbox{[O\sc ii]\,3729\AA}}
\newcommand{\neiii}{\hbox{[Ne\sc iii]}}

\newcommand{\neiiialong}{\hbox{[Ne\sc iii]\,3869\AA}}
\newcommand{\oiii}{\hbox{[O\sc iii]}}
\newcommand{\oiiilong}{\hbox{[O\sc iii]\,4959,5007\AA}}

\newcommand{\oiiiblong}{\hbox{[O\sc iii]\,5007\AA}}
\newcommand{\oi}{\hbox{[O\sc i]}}

\newcommand{\oialong}{\hbox{[O\sc i]\,6300\AA}}
\newcommand{\nii}{\hbox{[N\sc ii]}}
\newcommand{\niilong}{\hbox{[N\sc ii]\,6548,6583\AA}}
\newcommand{\niialong}{\hbox{[N\sc ii]\,6548\AA}}
\newcommand{\niiblong}{\hbox{[N\sc ii]\,6583\AA}}
\newcommand{\sii}{\hbox{[S\sc ii]}}
\newcommand{\siilong}{\hbox{[S\sc ii]\,6716,6731\AA}}
\newcommand{\siialong}{\hbox{[S\sc ii]\,6716\AA}}
\newcommand{\siiblong}{\hbox{[S\sc ii]\,6731\AA}}

\newcommand{\ha}{H$\alpha$}
\newcommand{\hb}{H$\beta$}

\newcommand{\lumcgs}{$\mathrm{erg}\,\mathrm{s}^{-1}$}

\newcommand{\msun}{$M_{\odot}$}

\newcommand{\lsun}{$L_{\odot}$}
\newcommand{\kms}{km\,s$^{-1}$}
\newcommand{\voff}{$v_\mathrm{s}$}
\newcommand{\voffabs}{$|v_\mathrm{s}|$}

\newcommand{\sfrunit}{\hbox{$M_{\odot}$\,yr$^{-1}$}}
\newcommand{\ccm}{cm$^{-3}$}

\received{August 19, 2024}
\revised{October 28, 2024}
\accepted{November 14, 2024}
\submitjournal{ApJ}

\begin{document}

\title{Implication of a galaxy-scale negative feedback by one of the most powerful multi-phase outflows \\
in a hyper-luminous infrared galaxy at the intermediate redshift}

\correspondingauthor{Xiaoyang Chen}
\email{xiaoyang.chen@nao.ac.jp}

\author[0000-0003-2682-473X]{Xiaoyang Chen}
\affiliation{National Astronomical Observatory of Japan, 2-21-1 Osawa, Mitaka, Tokyo 181-8588, Japan}

\author[0000-0002-2651-1701]{Masayuki Akiyama}
\affiliation{Astronomical Institute, Tohoku University, 6-3 Aramaki, Aoba-ku, Sendai, Miyagi 980-8578, Japan}

\author[0000-0002-4377-903X]{Kohei Ichikawa}
\affiliation{Faculty of Science and Engineering, Waseda University, 1-6-1 Nishi-Waseda, Shinjuku-ku Tokyo 169-8050 Japan}

\author[0000-0002-3531-7863]{Yoshiki Toba}
\affiliation{National Astronomical Observatory of Japan, 2-21-1 Osawa, Mitaka, Tokyo 181-8588, Japan}
\affiliation{Academia Sinica Institute of Astronomy and Astrophysics, 11F of Astronomy-Mathematics Building, AS/NTU, No.1, Section 4, Roosevelt Road, Taipei 10617, Taiwan}
\affiliation{Research Center for Space and Cosmic Evolution, Ehime University, 2-5 Bunkyo-cho, Matsuyama, Ehime 790-8577, Japan}

\author[0000-0002-3866-9645]{Toshihiro Kawaguchi}
\affiliation{Department of Economics, Management and Information Science, Onomichi City University, Hisayamada 1600-2, Onomichi, Hiroshima 722-8506, Japan}

\author[0000-0001-9452-0813]{Takuma Izumi}
\affiliation{National Astronomical Observatory of Japan, 2-21-1, Osawa, Mitaka, Tokyo 181-8588, Japan}

\author[0000-0002-2501-9328]{Toshiki Saito}
\affiliation{Department of Global Interdisciplinary Science and Innovation, Shizuoka University, 836 Ohya, Suruga-ku, Shizuoka-Shi, 422-8529, Japan}

\author[0000-0002-2364-0823]{Daisuke Iono}
\affiliation{National Astronomical Observatory of Japan, 2-21-1, Osawa, Mitaka, Tokyo 181-8588, Japan}

\author[0000-0001-6186-8792]{Masatoshi Imanishi}
\affiliation{National Astronomical Observatory of Japan, 2-21-1, Osawa, Mitaka, Tokyo 181-8588, Japan}

\author[0000-0003-4814-0101]{Kianhong Lee}
\affiliation{Astronomical Institute, Tohoku University, 6-3 Aramaki, Aoba-ku, Sendai, Miyagi 980-8578, Japan}
\affiliation{National Astronomical Observatory of Japan, 2-21-1, Osawa, Mitaka, Tokyo 181-8588, Japan}

\author[0000-0003-0292-3645]{Hiroshi Nagai}
\affiliation{National Astronomical Observatory of Japan, 2-21-1, Osawa, Mitaka, Tokyo 181-8588, Japan}

\author[0000-0001-6020-517X]{Hirofumi Noda}
\affiliation{Astronomical Institute, Tohoku University, 6-3 Aramaki, Aoba-ku, Sendai, Miyagi 980-8578, Japan}

\author[0000-0002-5258-8761]{Abdurro'uf}
\affiliation{Department of Physics and Astronomy, Johns Hopkins University, 3400 N. Charles Street, Baltimore, MD 21218, US}

\author[0000-0001-6402-1415]{Mitsuru Kokubo}
\affiliation{National Astronomical Observatory of Japan, 2-21-1, Osawa, Mitaka, Tokyo 181-8588, Japan}

\author[0000-0002-8299-0006]{Naoki Matsumoto}
\affiliation{Astronomical Institute, Tohoku University, 6-3 Aramaki, Aoba-ku, Sendai, Miyagi 980-8578, Japan}


\begin{abstract}
Powerful, galactic outflows driven by Active Galactic Nuclei (AGNs) are commonly considered as a main mechanism to regulate star formation in massive galaxies. Ultra- and hyper-luminous IR galaxies (U/HyLIRGs) are thought to represent a transition phase of galaxies from a rapidly growing period to a quiescent status as gas swept out by outflows, providing a laboratory to investigate outflows and their feedback effects on the hosts. In this paper we report recent Gemini and ALMA observations of a HyLIRG, J1126 at $z=0.46842$, which has been identified with a puzzling co-existence of a fast ionized outflow ($>2000$ km s$^{-1}$) and an intense starburst (star formation rate of 800 $M_{\odot}$ yr$^{-1}$). The Gemini observation shows the fast ionized outflow is extended to several kpc with a mass-loss rate of 180 $M_{\odot}$ yr$^{-1}$. A massive molecular outflow with a high mass-loss rate (2500 $M_{\odot}$ yr$^{-1}$) is revealed by ALMA. The multi-phase outflows show large factors of momentum boost and loading of kinetic power, indicating a driving by thermal pressure of a nuclear hot wind and/or radiation pressure of a highly obscured AGN. In addition to ejection of kinetic energy, it is also found that the powerful outflow can induce an ionizing shock in the galaxy disk and enhance the excitation and dissociation of molecular gas. The powerful outflow probably results in an instantaneous negative feedback and shows potential to regulate the host growth in a long term.
\end{abstract}



\section{Introduction} 
\label{sec:Intro}

It is widely accepted that the supermassive black holes (SMBHs) 
in the center of galaxies can affect the evolution of their host galaxies. 
Such association is supported by observational evidences, e.g., 
the relation between SMBH masses and properties of host stellar spheroids (masses and velocity dispersions;
e.g., \citealt{Ferrarese2000,Gebhardt2000,King2005,Kormendy2013}),
and the low gas cooling rate in massive galaxies in the center of clusters
\citep[e.g.,][]{Ciotti1997,Peterson2003}.
The feedback by SMBHs is also required in simulations to explain 
the reduced star formation efficiency in massive galaxies
as well as the 
bi-modal distribution of local galaxies in the stellar mass and star formation rate (SFR) diagram 
\citep[e.g.,][]{Benson2003,Bower2006,Schaye2015,Khandai2015,Nelson2018}. 
Powerful, galactic scale outflows are considered as
a main feedback mechanism of SMBHs
\citep[e.g.,][]{Fabian2012,King2015,Tombesi2015}, 
which can either quench the star formation in the hosts
via expelling out or heating the cold fueling gas
(i.e., a negative feedback; e.g., \citealt{Ellison2021,Saito2022CI,Lammers2023}), 
or promote the star formation by compressing the ambient gas
(i.e., a positive feedback; e.g., \citealt{Cresci2015,Maiolino2017,Shin2019}).
Those outflows are observed in multiple gas phases, i.e.,
from hot/warm ionized gas \citep[e.g.,][]{Liu2013,Harrison2014,Zakamska2016,Toba2017a,Chen2019}
to cold neutral atomic \citep[e.g.,][]{Perna2020,Avery2022}
and molecular gas \citep[e.g.,][]{Sturm2011,Veilleux2013,Cicone2014}.
Although the contribution of an intense starburst
to launch a powerful outflow is argued by several recent works
\citep[e.g.,][]{Geach2014,Rupke2019},
it is usually considered that the outflows, especially those with a high velocity, are driven by SMBHs in a fast accretion mode, 
which are observed as Active Galactic Nuclei 
(AGNs; e.g., \citealt{Cicone2014,Zakamska2016,Fiore2017}). 

Ultra- and hyper-luminous IR galaxies (ULIRGs, $L_{\mathrm{IR}}>10^{12}$\lsun; HyLIRGs, $L_{\mathrm{IR}}>10^{13}$\lsun) are thought to 
represent a transiting population of galaxies 
from a merger-induced, rapidly growing phase of SMBHs and stellar spheroids, 
to a quiescent status as the fueling gas swept out by powerful outflows
\citep[e.g.,][]{Hopkins2008,Sturm2011,Veilleux2013,Toba2022}. 
The ubiquitousness of powerful outflows in U/HyLIRGs
makes them laboratories to investigate 
the properties of outflows (e.g., velocity, extent, mass-loss rate)
and their feedback effects on the hosts. 
\cite{Chen2020} reported a sample of $\sim200$ U/HyLIRGs at the intermediate redshifts ($0.1<z<1.0$)
selected from the AKARI far-IR all-sky survey, 
which are cross-matched with SDSS (optical) and WISE (near- and mid-IR) photometric catalogs
and spectroscopically identified with SDSS and Subaru/FOCAS
observations. 
It is found that fast ionized outflows 
(e.g., $v_{10}>1500$ \kms, where $v_{10}$ is the 10th percentile velocity of \oiii\ line)
and intense starbursts with IR-based SFR over 300 \sfrunit\ co-exist in a few of U/HyLIRGs in the sample. 
The co-existence of fast outflows and intense starbursts
suggest a possibility of an apparently ``failed'' feedback, 
i.e., the star formation activities have not been profoundly suppressed by the outflow. 
The puzzling results can be likely explained by 
a compact distribution of the outflow, i.e., 
only a limited volume of the star-forming regions
can be affected by the outflow; 
or the different timescales 
of the outflow (several Myr; e.g., \citealt{Harrison2024})
and the IR-based SFR estimator (several tens to 100 Myr; e.g., \citealt{Murphy2011}), i.e., 
a possible instantaneous feedback cannot be reflected 
by the SFR in a long-term. 
It is also possible that the fast outflow only occurs in the ionized gas phase while the molecular gas clouds 
in which stars form
are not effectively accelerated/dispersed \citep[e.g.,][]{Toba2017b}. 
A ``delayed'' feedback scenario may also provide an alternative explanation for the apparent non-quenching of star formation 
\citep[e.g.,][]{Woo2017}. 

\begin{figure*}[!ht]
	\begin{center}
		\hspace{3mm}
		\includegraphics[trim=0 -81 0 0, clip, width=0.64\columnwidth]{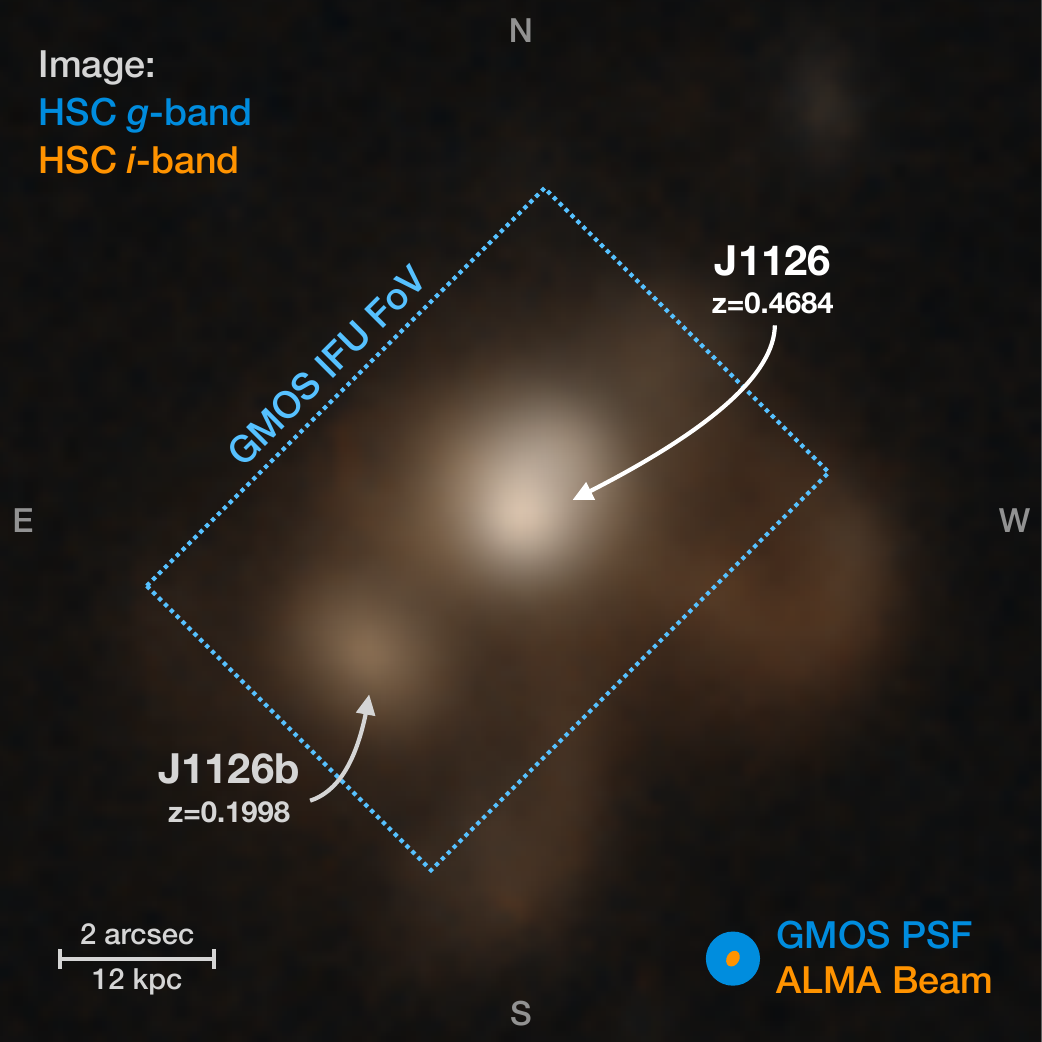}
		\hspace{2mm}
		\includegraphics[trim=0 0 0 0, clip, width=1.28\columnwidth]{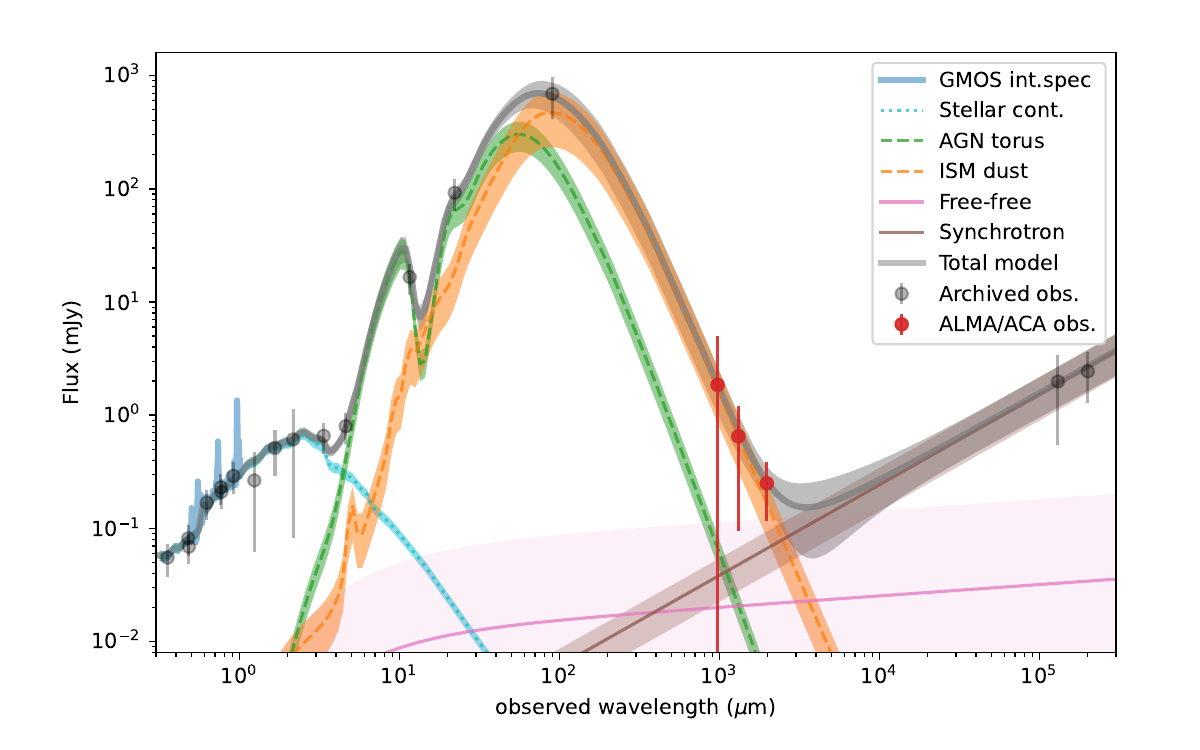}
	\end{center}
	\vspace{-9mm}
	\caption{
		\textbf{Left:}
		HSC $gi$-composite image of J1126 and a nearby faint galaxy, J1126b. 
		J1126 was observed by HSC only in the two bands. 
		The PSF and FoV of the GMOS observation are shown in blue circle and dotted rectangle, respectively.
		The orange ellipse denotes the beam of the Band 4 observation of ALMA. 
		\textbf{Right:}
		Archived data (grey dots) and the new ALMA observations of continua (red dots) of J1126 with the best-fit SED fitting results.  
		The error bars show $\pm3\sigma$ errors,
		which are the original measurement errors plus 10\% of the fluxes
		to account for the systematic errors among instruments. 
		The GMOS spectrum is shown in blue. 
		The total best-fit model is shown in the grey curve with $\pm3\sigma$ uncertainty range in the grey shadow region. 
		All of the fitting components with their $\pm3\sigma$ uncertainties are also shown with the markers denoted in the legend.
		See details of the archived observations in Section \ref{subsec:Reduction_archived} and the SED fitting in 
		Section \ref{subsec:ALMA_dust}. 
	}
	\label{fig:J1126_image_SED}
\end{figure*}

In order to shed light on the mechanisms behind the co-existence
of fast ionized outflow and intense starbursts, 
we conducted follow-up observations for the U/HyLIRG sample 
with multi-wavelength telescopes. 
In this paper we report the recent observations with Gemini-N telescope 
and the Atacama Large Millimeter/submillimeter Array (ALMA)
for one HyLIRG in the sample, 
AKARI-FIS-V2 J1126579+163917 (hereafter J1126; Figure \ref{fig:J1126_image_SED}) at
$z=0.4684$.
J1126 has a total IR (1--1000 \micron) luminosity of $1.2\times10^{13}$ \lsun\ 
and harbors a Type-2 AGN identified with optical emission lines
and mid-IR radiation \citep{Chen2020}. 
J1126 possesses one of the fastest ionized outflow 
($v_{10}>2000$ \kms)
and the most vigorous starburst (SFR $\sim800$ \sfrunit)
among U/HyLIRGs and AGNs at $z<1$ 
\citep[e.g.,][]{Arribas2014,Liu2013,Harrison2014,Fiore2017}, 
which make it an excellent target to investigate
the complex interplay between outflow and star formation 
in the intensest case. 
As an analog of the high-$z$ luminous HyLIRGs/quasars
\citep[e.g.,][]{Harrison2012,Perrotta2019}, 
J1126 at a lower redshift provides the opportunity
to understand the evolution of those high-$z$ galaxies
with a higher S/N and spatial resolution
(e.g., 6 kpc per arcsec at $z=0.5$ vs. 8.5 kpc per arcsec at $z=2$).  

The paper is organized as follows.
The reduction of the Gemini and ALMA observations are described in Section \ref{sec:Reduction}.
The analyses of the Gemini data
including the spatially-resolved kinematics of ionized gas
as well as the association between outflow and gas ionization, 
are reported in Section \ref{sec:GMOS}.
The analyses of the ALMA observations, 
e.g., the identification of molecular outflow, 
are reported in Section \ref{sec:ALMA}.
In Section \ref{sec:Discussion} we discuss the dynamical properties, 
the driving mechanisms, and the feedback effects of the multi-phase outflows. 
The results of the analyses are summarized in Section \ref{sec:Summary}. 
We adopt the cosmological parameters $H_0=$ 69.32 \kms\,Mpc$^{-1}$, $\Omega_\text{M}=0.29$ and $\Omega_{\Lambda}=0.71$ \citep{COSMO} throughout the paper. 

\section{Data reduction} 
\label{sec:Reduction}

\subsection{Gemini-N/GMOS IFU observations}
\label{subsec:Reduction_GMOS}

J1126 was observed by Gemini-N/GMOS in 2021 March
(ID: GN-2020B-Q-117, PI: Chen).
The observation was conducted with a total of 2 hours on-source exposure in two nights 
in the 2-slit IFU mode of GMOS. 
The 2-slit IFU mode has a field of view (FoV) of $5\arcsec\times7\arcsec$, 
which is set to cover J1126 and a faint companion galaxy (Figure \ref{fig:J1126_image_SED}, left panel). 
The Point Spread Function (PSF) of the IFU data
is estimated with 
the size of stars, $\sim0.7\arcsec$, in acquisition images taken in $i$-band ($\sim7500$\AA)
and the wavelength-dependent relation, $\propto \lambda^{-1/5}$, 
of the Kolmogorov turbulence model \citep[e.g.,][]{Fried1965,Roddier1981}.
The grating R150 was utilized to achieve a wide wavelength coverage, 
i.e., from \oiiblong\ line to \siiblong\ line, 
which provides a spectral resolution of R $\sim$ 1000 ($\sim$ 300 \kms) around 7000\AA.
The wavelength accuracy estimated with night sky emission lines is $\sim$ 50 \kms\ . 

The GMOS data is reduced with the Gemini IRAF package \citep{GeminiIRAF}, e.g., to calibrate the wavelength, mask bad pixels, 
remove the scattered light, correct for the atmospheric dispersion, 
and subtract the night sky emissions. 
The cosmic rays are rejected with the L.A.Cosmic algorithm \citep{vanDokkum2001}. 
The atmospheric extinction is corrected with the extinction curve at Mauna Kea \citep{Buton2013}. 
The Galactic extinction is corrected with the extinction curve of \cite{Cardelli1989}
and the Galactic dust reddening map of \cite{Schlafly2011}. 
The standard star Feige 34 was observed with the same configurations as those of the observation of J1126
to derive the spectral response curve of the instruments; 
we also utilize the observation of a red star KF08T3 from the program (ID: GN-2022A-Q-221, PI: Chen)
to reduce the contamination of the 2nd-order spectrum in the derivation of the response curve. 
The spectra of J1126 is corrected with the response curve
and then calibrated with the SDSS photometry 
in the $r^\prime$, $i^\prime$, and $z^\prime$ bands, 
which are fully covered by the GMOS observations. 
Images are created by convolving the GMOS cube with the SDSS transmission curves of these bands 
and used to calculate the magnitudes with a de Vaucouleurs profile; 
the derived magnitudes are then compared to SDSS model magnitudes (based on the same profile)
to obtain the calibration factor. 
These image are also used to derive the absolute astrometry of GMOS data
by matching the intensity centers of them
to those in the corresponding SDSS images of J1126. 

\begin{figure*}[!ht]
	\begin{center}
		\includegraphics[trim=0 8 0 20, clip, width=.9\textwidth]{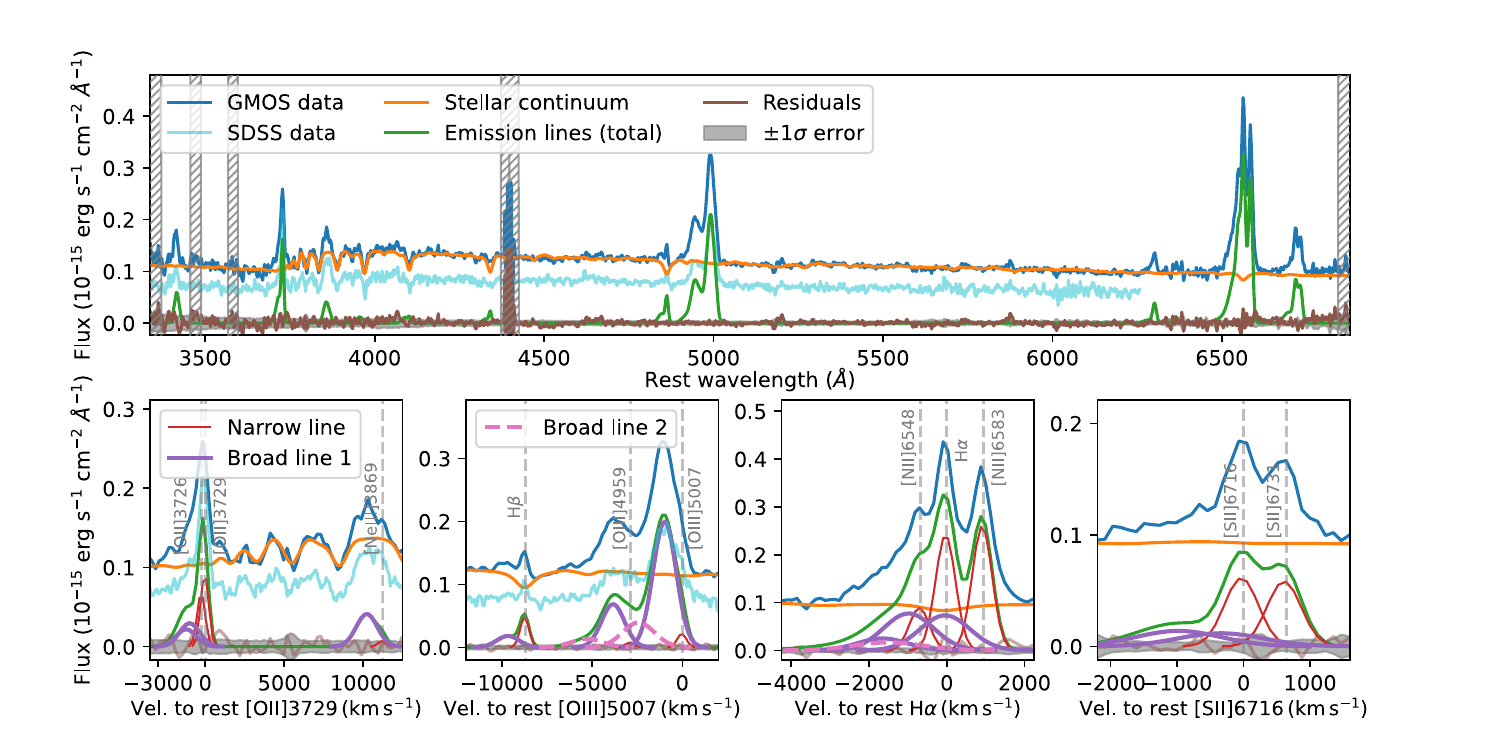}
	\end{center}
	\vspace{-7mm}
	\caption{
		\textbf{Top:} Integrated GMOS spectrum of J1126 (blue) with the best-fit 
		stellar continuum (orange) and emission lines (green). 
		The fitting residuals are shown in brown and the $\pm1\sigma$ error range in grey shadow region. 
		The grey hatched regions denote the wavelength ranges with large instrumental errors, 
		which are masked out in the fitting. 
		The SDSS spectrum of J1126 is shown in cyan for a reference, 
		which is fainter than the GMOS spectrum due to a small aperture. 
		\textbf{Bottom:} 
		Zoom-in windows of \oii\ doublets and \neiii\ (left),
		\hb\ and \oiii\ doublets (middle-left),
		\ha\ and \nii\ doublets (middle-right), 
		as well as \sii\ doublets (right). 
		The grey vertical lines represent the rest wavelengths of these lines.
		The red thin lines show the best-fit narrow line profiles; 
		the violet lines show the broad line profiles for each line, 
		while the purple dashed lines show the additional broad profiles for \oiii, \ha, and \nii. 
		The $x$-axis is converted to velocity in relative to the rest wavelengths
		of \oiiblong, \oiiiblong, \ha, and \siialong, respectively, in the four panels. 
		Other legends are the same as those in the top panel. 
	}
	\label{fig:J1126_GMOS_spec_int}
\end{figure*}

With the adopted configuration using R150, 
the (1st-order) spectra of the GMOS blue-slit 
can be contaminated by the 0th-order spectra of the GMOS red-slit
in the range of rest 4370\AA--4420\AA\ (as shown in Figure \ref{fig:J1126_GMOS_spec_int}), 
which is masked out in the later analyses. 

\subsection{ALMA observations}
\label{subsec:Reduction_ALMA}

J1126 has been observed by ALMA in Band 4 and 6 for 
the detections of CO(2-1) and CO(3-2) emission lines
with rest frequencies of 230.538 and 345.796 GHz
, respectively. 
The Band 4 observation was conducted in Cycle 8 (project ID: 2021.1.01496.S, PI: Chen)
in an extended array configuration with baselines from 41.4 to 3396.4 meters that yields a mean angular resolution (AR) of 0.17\arcsec and 
and a maximum recoverable scale (MRS) of 2.4\arcsec,
plus a compact array configuration with baselines from 14.9 to 876.6 meters that yields a mean AR of 0.69\arcsec and 
and a MRS of 7.6\arcsec. 
The achieved sensitivity is 0.06 and 0.13 mJy beam $^{-1}$ over 100 \kms, 
for observations with the extended and compact arrays, respectively. 
The Band 6 observation was conducted in Cycle 9 (project ID: 2022.1.01376.S, PI: Chen) 
in an extended array configuration with baselines from 79.4 to 5323.0 meters that yields a mean AR of 0.09\arcsec and and a MRS of 1.0\arcsec. 
The sensitivity of the Band 6 observation is 0.18 mJy beam $^{-1}$ over 100 \kms.

The Band 4 and 6 data sets are calibrated using CASA\footnote{
	Common Astronomy Software Applications \citep{McMullin2007}. 
} and the pipeline scripts delivered by ALMA. 
In order to maximize the uv coverage, 
we combine the Band 4 observations with the extended and compact arrays with the CASA task \texttt{concat} with the default weight
in the calibrated measurement sets. 
The continuum in the Band 4 and 6 data sets is subtracted 
in the uv plane using the CASA task \texttt{uvcontsub\_old}\footnote{
	We use the old version of \texttt{uvcontsub} since the current version does not support fitting across multi spectral windows. 
} in line-free frequency ranges. 
The continuum-subtracted visibilities are then deconvolved utilizing the CASA task \texttt{tclean} with the weighting option of Briggsbwtaper and a robust number of 0.5. 
The reduced Band 4 and 6 data sets have synthesized beams of
$0.20\arcsec\times0.16\arcsec$ with a position angle (PA) of $331^\circ$, and
$0.12\arcsec\times0.09\arcsec$ with a PA of $323^\circ$, respectively. 

In order to perform a direct comparison between observations of the two CO lines, e.g., to derive the line ratio, 
we create modified CO(2-1) and CO(3-2) cubes with matched beams and MRS.
The modified CO(2-1) cube is obtained via a cut of the visibilities with baselines shorter than 119.1 meters, i.e., to reduce the MRS.
The modified CO(3-2) cube is obtained via a Gaussian taper, 
$0.14\arcsec\times0.09\arcsec$ with a PA of $350^\circ$,
in the \texttt{tclean} process. 
The consequent modified cubes have 
a synthesized beam of $0.20\arcsec\times0.16\arcsec$ ($1.2\times1.0$ kpc) with a PA of $331^\circ$, and a MRS of $\sim1.0\arcsec$ (6.0 kpc). 

In addition to the CO line image cubes, 
we also create the continuum images in Band 4 and 6
(1340 and 890 \micron\ in the rest frame)
with the data sets in line-free frequency ranges
using \texttt{tclean} in a MFS mode with a weighting option of Briggs and a robust number of 0.5.

J1126 was also observed by Atacama compact array (ACA) in Band 7
in Cycle 7 (project ID: 2019.2.00085.S, PI: Chen)
with a sensitivity of 2.43 mJy beam $^{-1}$ over 100 \kms.
The Band 7 continuum and CO(4-3) line 
(461.041 GHz in rest frame)
are reduced following 
similar processes as done for Band 4 and 6 data sets.
The mean AR of the ACA observation is 4.9\arcsec (29 kpc), 
i.e., J1126 is unresolved in the observation, 
and we only use the spatially integrated continuum and CO(4-3) line
in the analyses. 

\subsection{Definitions of the systemic redshift and the position of the galaxy center}
\label{subsec:v0_redshift_center}

In order to perform a self-consistent analysis for GMOS and ALMA observations, 
we adopt uniform definitions of the systemic redshift and the position of the galaxy center
throughout the paper.
The systemic redshift of the galaxy can be estimated from 
the absorption features in stellar continuum (GMOS) and 
the narrow components of CO emission lines (ALMA). 
Throughout the paper we adopt the systemic redshift, $0.468417\pm0.000002$, 
measured from the CO(2-1) emission in the central region ($r$$<$0.5 kpc; Section \ref{subsec:ALMA_CO_main}), 
which shows a higher accuracy than those from the stellar continuum fitting, $0.4685\pm0.0001$ (Section \ref{subsec:GMOS_stellar}). 
The difference between the two estimates is within the uncertainty of GMOS wavelength calibration ($\sim50$ \kms). 

The position of the galaxy center can be estimated with 
intensity maps of stellar continuum (GMOS), 
dust continuum, and CO emission lines (ALMA). 
Throughout the paper we adopt the galaxy center as the center 
of the CO(2-1) intensity map (Section \ref{subsec:ALMA_CO_main}), 
11$^\mathrm{h}$26$^\mathrm{m}$57$^\mathrm{s}$.771 +16$^\mathrm{d}$39$^\mathrm{m}$11$^\mathrm{s}$.79, 
which is estimated with a two-dimensional Gaussian fitting
and has an angular accuracy of $\sim0.01\arcsec$ under the array configuration.
The Band 4 dust continuum has the same center as CO(2-1). 
The center of the stellar continuum in $V$-band (Section \ref{subsec:GMOS_stellar})
has a relative distance of 0.1\arcsec\ from the center of CO(2-1), 
which could be due to the positional uncertainty of GMOS ($\sim0.1\arcsec$)
and SDSS ($\sim0.1\arcsec$, for astrometric calibration). 

\subsection{Archived observations}
\label{subsec:Reduction_archived}

We collect the multi-wavelength archived observations of J1126 of 
the following telescopes:
SDSS ($u^\prime$ to $z^\prime$ bands), Subaru/HSC ($g$ and $i$ bands), 
2MASS ($J$, $H$, and $K_s$ bands), WISE (3.4, 4.6, 12, and 22 \micron), 
AKARI (90 \micron), and VLA (13 and 20 cm). 
The integrated fluxes in each band are shown in Figure \ref{fig:J1126_image_SED} (right panel).
An updated multi-band spectral energy distribution (SED) 
fitting using the archived data
and the new ALMA observations in (sub-)millimeter bands
are discussed in Section \ref{subsec:ALMA_dust}. 

There is a companion galaxy (J1126b), in the southeast of J1126 with a distance of $3\arcsec$ 
(Figure \ref{fig:J1126_image_SED}, left panel).
J1126b is faint with an HSC $i$-band magnitude of 19.8.
GMOS observation of J1126b results in $z=0.1998$ from \ha\ line, suggesting that 
it locates in the foreground of J1126. 
The spectral fitting for J1126b (see the method in Appendix \ref{appendix:GMOS_fitting}) reveals 
a stellar mass of $2.6\times10^9$ \msun\
with a SFR of 0.6 \sfrunit, i.e., it is not an active star-forming galaxy.
Non-detection of \oiii\ line suggests no AGN in J1126b. 
In addition, J1126b is not detected in the ALMA observations in Band 4 and 6.
Therefore, the contribution of J1126b in the unresolved
IR/radio observations (e.g., WISE and VLA) can be considered to be 
negligible in the later analyses (e.g., SED fitting of J1126). 


\section{Analysis method and results of the GMOS IFU observations}
\label{sec:GMOS}

J1126 is identified as a Type-2 AGN since there is no features of Type-1 AGNs in its optical spectra (e.g., extremely broad permitted lines, Balmer continuum, and iron pseudo continua; \citealt{Chen2020}). 
We fit the GMOS spectra using the PopStar stellar templates \citep{Millan-Irigoyen2021} plus Gaussian profiles for emission lines. 
Two Gaussian profiles are used to fit each line, 
in which a narrow profile with a full-width-half-maximum (FWHM)
$<$ 700 \kms\ reflects the line emitting from gas on the galaxy disk, 
and a broad line profile (FWHM $>$ 700 \kms) indicates the line emitting from the outflowing gas. 
An additional broad component  
with a velocity shift ($v_\mathrm{s}$) $<-2000$ \kms\ is required 
for the bright blueshifted wings of \oiii\ doublets and \ha-\nii\ complex. 
The spectral fitting is performed for each pixel of the original GMOS data cube. 
Details on the fitting method are described in Appendix \ref{appendix:GMOS_fitting}. 

The GMOS data shows a moderate resolution, $\sim0.7\arcsec$ (4 kpc).
If J1126 has a compact outflow \citep[e.g., $<$\,1 kpc;][]{Genzel2014,Carniani2015},
it can be blurred by the PSF and observed to be extended. 
On the other hand, 
the highly-blueshifted \oiii\ line profile (Figure \ref{fig:J1126_GMOS_spec_int})
suggest a face-on view of the ionized outflow. 
Even if the outflow is intrinsically extended, 
it could show a bright core component in the galaxy center where 
the line of sight is close to the central axis of the outflow cone. 
In addition, it is reported that ULIRGs usually have compact starburst nuclei, 
which have a typical size of $\sim100$ pc and contribute the bulk of the star formation
in the host galaxy \citep[e.g.,][]{Privon2017,Falstad2021,GarciaBernete2022}.
Such bright core components of outflows and star formation could dominate the observation, 
while the emission in the outskirt could be hidden.
In order to determine whether the outflow in J1126 is indeed extended, 
and investigate the properties (e.g., kinematics and ionization of gas)
in the galaxy outskirt to understand the feedback effect of outflow on the entire galaxy, 
we also perform the spectral fitting for the cube after removing an unresolved core component.
The core component is obtained using 
the wavelength-dependent PSF with its peak
scaled by the mean spectrum of the central pixels within $\sim0.2\arcsec$. 
The core component contributes to 60\%, 42\%, and 36\% 
of the total fluxes of the \oiii-traced outflow, the narrow \ha\ line, 
and the $V$-band stellar continuum of the entire galaxy.

The best-fit fitting results of the original and the core-removed cubes
are discussed in the following subsections. 
The estimation of stellar mass is shown in Section \ref{subsec:GMOS_stellar}.
The kinematics and ionizing mechanisms of ionized gas are discussed in Section \ref{subsec:GMOS_kinematics} 
and \ref{subsec:GMOS_ionization}, respectively. 
The comparison between the extinction estimated from the stellar
and nebular emission is discussed in Section \ref{subsec:GMOS_extinction}. 

\subsection{Properties from the stellar continuum}
\label{subsec:GMOS_stellar}

An integrated spectrum is extracted within a radius of 1.5\arcsec\ (9.0 kpc)
and used to estimate properties of the entire galaxy.
The best-fit result for the integrated spectrum
is shown in Figure\,\ref{fig:J1126_GMOS_spec_int}, in which 
the stellar continuum is marked in orange. 
The continuum fitting is performed with  
an non-parametric star formation history 
(SFH; e.g., \citealt{Hernandez2000})
and the PopStar templates. 
We assume the initial mass function of \cite{Kroupa2001} and a fixed solar metallicity. 
The best-fit SFH from the integrated spectrum  
consists of an old stellar population with $\sim1$ Gyr
and a starburst in the recent several tens Myr\footnote{
	Details of the SFH estimated with the stellar continuum
	and other estimators are discussed in Appendix \ref{appendix:SFH}.
}. 
The total stellar mass ($M_\star$) is estimated to be  
$(2.0\pm0.5)\times10^{11}$ \msun\  
after corrected for the remaining mass fraction 
provided by the PopStar model.
The best-fit stellar continuum has
a velocity dispersion ($\Delta v_{\sigma}$) of $343\pm21$ \kms, which is
estimated with the absorption features (e.g., H- and K-lines of calcium around rest 4000\AA)
and corrected for the instrumental broadening (FWHM $\sim$ 300 \kms). 
The estimated $M_\star$ and $\Delta v_{\sigma}$ are consistent with the  
Faber–Jackson $M_\star$-$\Delta v_{\sigma}$ relation 
observed in massive galaxies \citep[e.g.,][]{Barat2019}.

\begin{figure}[!h]
	\vspace{-2mm}
	\begin{center}
		\hspace{-9mm}
		\includegraphics[trim=0 0 230 40, clip, width=0.90\columnwidth]{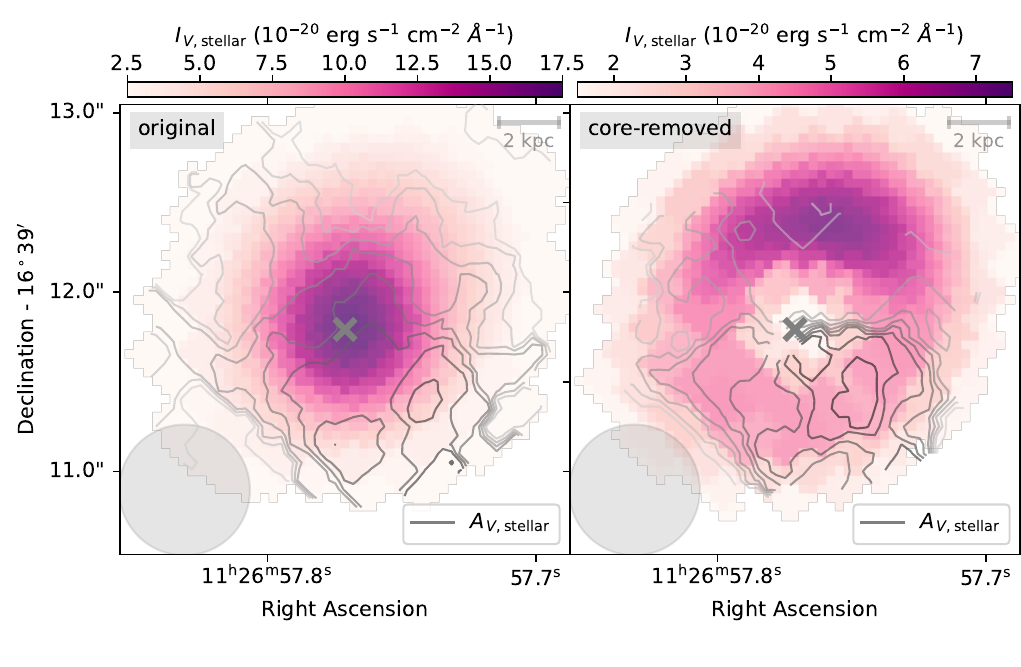}
	\end{center}
	\vspace{-9mm}
	\caption{
		Intensity maps of the rest $V$-band stellar continuum 
		from the original (left) and core-removed (right) GMOS data. 
		The grey contours show the extinction estimated
		from the fitting of stellar continuum. 
		Only pixels with $V$-band S/N $>$ 5 are shown in the panels.
		PSF is shown in grey circles and the galaxy center in grey crosses.
	}
	\label{fig:J1126_GMOS_IntenV_AV}
\end{figure}

In the stellar continuum fitting, 
it is assumed that old stars and young stars that migrate from the dense birth clouds, 
i.e., in the diffuse interstellar medium (ISM), 
suffer the same amount of extinction. 
The best-fit SFH under the assumption suggests that 
the recently formed stars, 
which is younger than 30 Myr
and older than the birth clouds timescale (e.g., 6 Myr; \citealt{Kennicutt2009}),
contribute to 98\% of the stellar ionizing flux in the diffuse ISM
and 85\% of the observed stellar continuum flux in $V$-band. 
The map of the $V$-band stellar continuum 
is obtained by integrating the best-fit stellar continuum (before extinction correction)
with the $V$-band transmission curve in the rest frame (Figure \ref{fig:J1126_GMOS_IntenV_AV}).
The intensity map has an effective radius (i.e., the half-light radius) of 3.0 kpc (0.5\arcsec) after correction for the PSF. 
The stellar light is more extended towards the north as shown in the core-removed intensity map,  
which suggests an intense recent star formation in the north. 
The dust extinction ($A_{V,\mathrm{\,stellar}}$) from the stellar continuum fitting
is shown as contours in Figure \ref{fig:J1126_GMOS_IntenV_AV}. 
The extinction is higher in the south in both of 
the original and core-removed maps, which is probably 
associated with the dust carried out from the nuclear dusty region
by the outflow.
We discuss the relative contribution of young stars and the AGN on gas ionization in Section \ref{subsec:GMOS_ionization};
and compare the extinction estimated from stellar continuum and nebular emission in Section \ref{subsec:GMOS_extinction}.

\subsection{Kinematics of ionized gas}
\label{subsec:GMOS_kinematics}

\begin{figure*}[!ht]
	\begin{center}
		\includegraphics[trim=0 54 0 50, clip, width=0.62\textwidth]{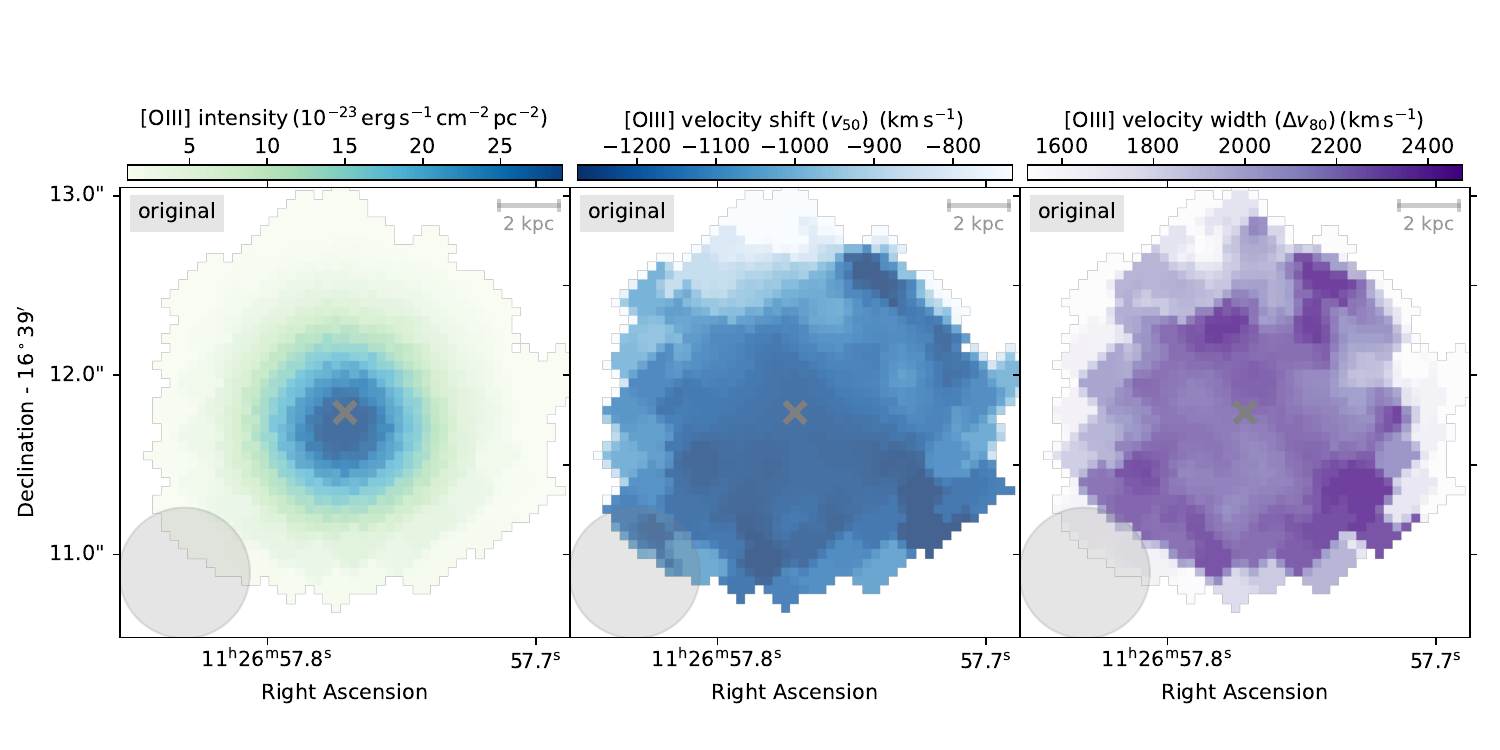}
		\includegraphics[trim=0  0 0 64, clip, width=0.62\textwidth]{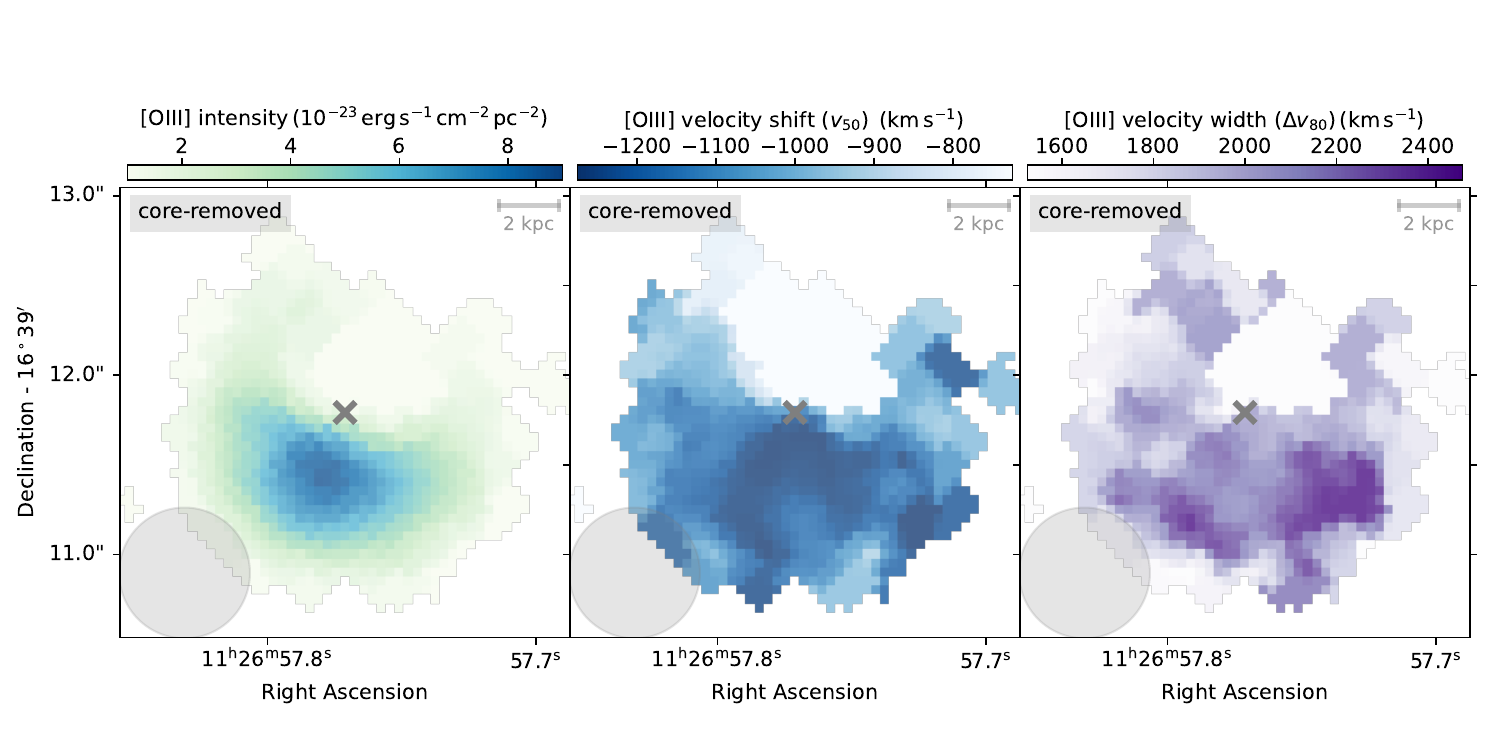}
	\end{center}
	\vspace{-9mm}
	\caption{
		\textbf{Top:} Intensity (left), velocity shift ($v_{50}$, middle) 
		and width ($\Delta v_{80}$, right) maps of \oiii. 
		Only pixels with S/N\,$>$\,3 for \oiii\ flux are shown. 
		PSF is shown in grey circles and the galaxy center in grey crosses.
		\textbf{Bottom:} The same as the top panels for the core-removed GMOS data. 
	}
	\label{fig:J1126_GMOS_OIII}
\end{figure*} 

Thanks to a wide wavelength coverage of the GMOS observations, 
a series of emission lines are detected 
including transitions of ions with low ionization potential (IP, e.g., $<15$ eV), 
i.e., the hydrogen Balmer lines, \oiiblong, \niiblong, and \siilong\ doublets;
high-IP ionized lines (e.g., IP $>30$ eV), i.e., \oiiiblong, \neiiialong, and \nevlong;
and an emission line of the neutral atomic oxygen, \oialong. 
The kinematic properties of the emission lines are reported 
in this subsection. 

The best-fit narrow and broad line components for the GMOS integrated
spectrum are shown in Figure \ref{fig:J1126_GMOS_spec_int} (bottom panels). 
Each of the kinematic components are fixed to the same 
shift velocity (\voff) and FWHM in the fitting
for each emission line.
The narrow component has a \voff\ of $-33\pm3$ \kms\ and a FWHM of $530\pm10$ \kms, 
while the (primary) broad component has a \voff\ of $-980\pm20$ \kms\ and a FWHM of $1460\pm40$ \kms. 
Those two components are used for each emission line. 
An additional broad component is required to fit the broad wings
of \oiii\ doublets and \ha-\nii\ complex
(Figure \ref{fig:J1126_GMOS_spec_int}, bottom-middle panels), 
which has a more blueshifted profile with 
a \voff\ of $-2410\pm160$ \kms\ and a FWHM of $2080\pm220$ \kms. 
The line widths have been corrected for the instrumental broadening of GMOS observation ($\sim300$ \kms). 

The flux fraction of the broad components over the entire line profile ($f_\mathrm{broad}$)
has been calculated for each emission line. 
We also calculate the 50th percentile velocity ($v_{50}$)
and the line width ($\Delta v_{80}=v_{90}-v_{10}$)
with the best-fit line models
, as the kinematic parameters of the entire profile of each line. 
For instance, \oiiiblong\ has a predominant broad component 
that contributes to 97\% of its total flux 
with $v_{50}=-1210\pm10$ \kms\ and $\Delta v_{80}=2370\pm40$ \kms,  
which provides an excellent tracer of the ionized gas outflow.
The other high-IP lines, i.e., \neiii\ and \nev, also have a $f_\mathrm{broad}$ over 90\%, 
indicating that they are mainly emitted from the outflowing gas.
On the contrary, the low-IP lines, e.g., \oii, and \ha, have a relatively low $f_\mathrm{broad}$ of $\sim$40\%, 
suggesting a mixed contribution of emission by gas in the outflow
and in the galaxy disk. 
The detailed values of the kinematic parameters for each emission line
are listed in Table \ref{tab:lines}. 

We then investigate the spatial distribution of intensity and kinematics of these emission lines. 
Figure\,\ref{fig:J1126_GMOS_OIII} shows maps of intensity (before correction for intrinsic extinction), velocity shift ($v_{50}$) and velocity width ($\Delta v_{80}$)
of the tracer of the ionized outflow, \oiiiblong, 
for both of the original and core-removed data cubes. 
The original intensity map 
has a PSF-corrected effective radius of 1.5 kpc. 
The intensity maps, especially the core-removed one, show a spatial distribution tilted to the south. 
The velocity maps indicate the fast outflow (i.e., high $v_{50}$ and $\Delta v_{80}$) is extended to the southern outskirt region ($\sim$ 3 kpc). 
The other high-IP lines, \neiii\ and \nev\, have a similar south-tilted morphology (Figure \ref{fig:J1126_GMOS_mHiIP_intensity}). 
We discuss the dynamical properties of the outflow (e.g., the mass-loss rate) and its feedback effect in Section \ref{sec:Discussion}. 

\begin{figure*}[!ht]
	\begin{center}
		\includegraphics[trim=0 54 0 50, clip, width=0.62\textwidth]{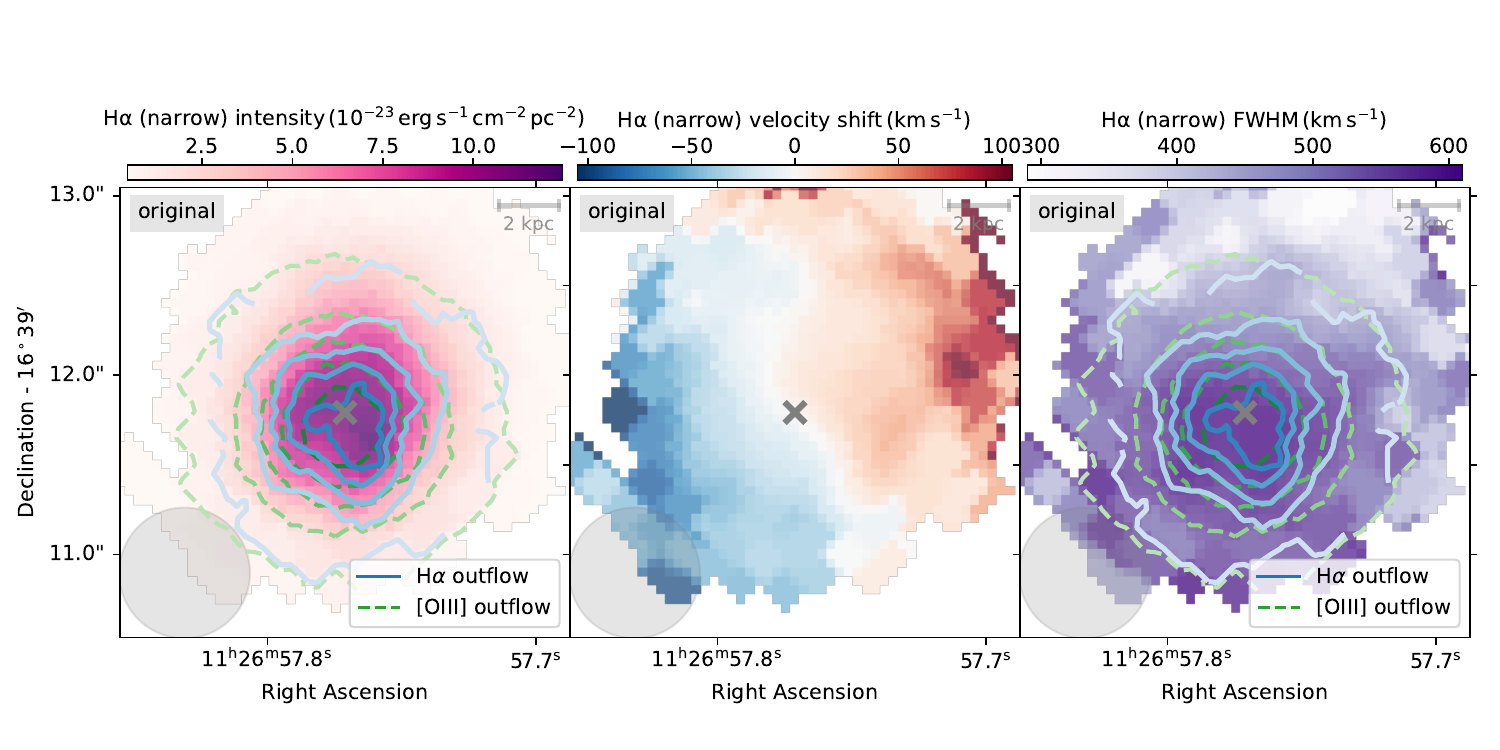}
		\includegraphics[trim=0  0 0 64, clip, width=0.62\textwidth]{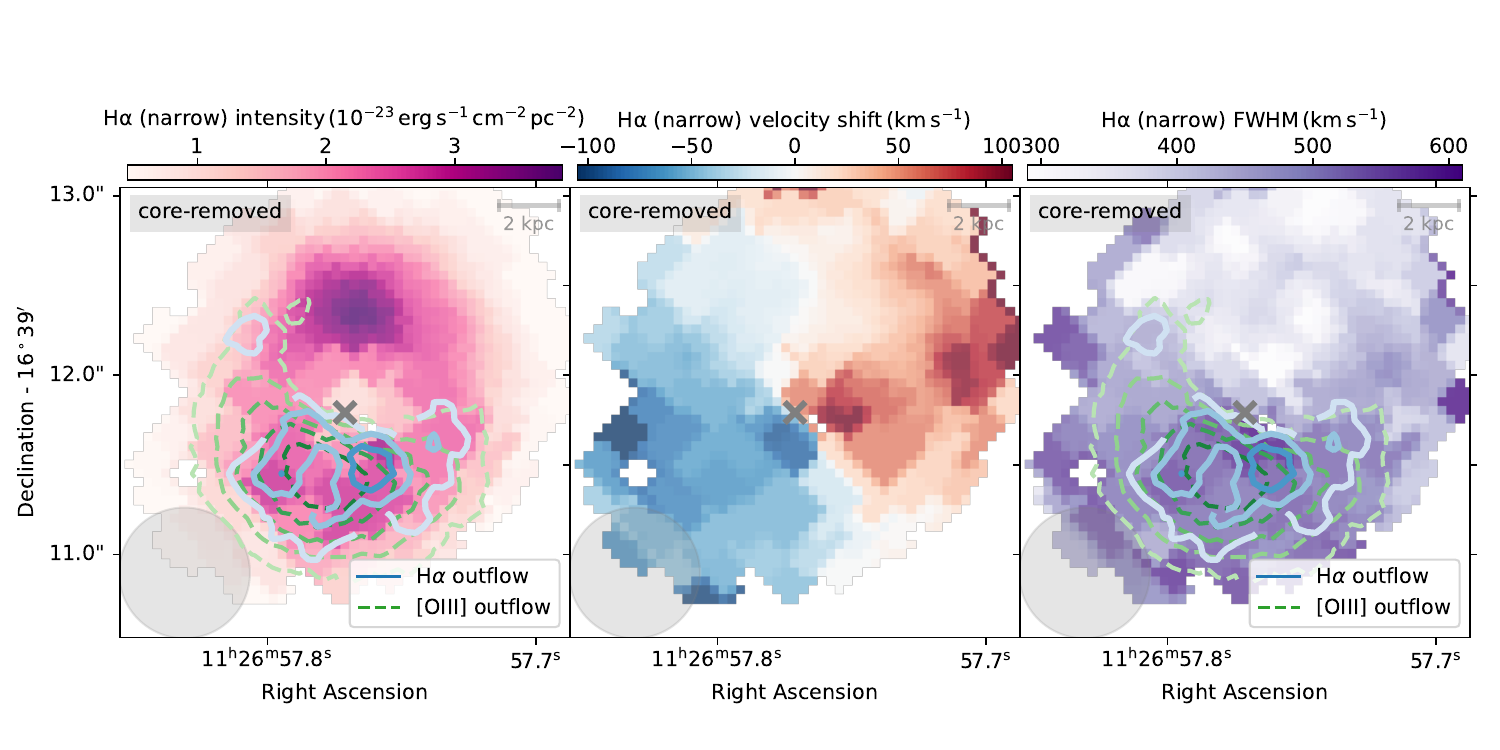}
	\end{center}
	\vspace{-9mm}
	\caption{
		\textbf{Top:} Intensity (left), velocity shift (\voff, middle) 
		and width (FWHM, right) maps of narrow component of \ha\ line. 
		The \oiii\ outflow intensity is also shown in green dashed contours 
		to indicate the alignment between the outflows traced by \ha\ and \oiii. 
		The shown velocity shift in the middle panel is in relative to the best-fit \voff\ of narrow line of the integrated spectrum,
		which has a shift of $-33$ \kms\ for \voff\ in relative to systemic velocity using CO(2-1) (Table \ref{tab:kinematics}). 
		An S/N\,$>$\,3 cut is used for the narrow \ha\ line maps and the contours of \ha\ and \oiii\ outflows. 
		PSF is shown in grey circles and the galaxy center in grey crosses. 
		\textbf{Bottom:} The same as the top panels for the core-removed GMOS data. 
	}
	\label{fig:J1126_GMOS_Ha}
\end{figure*}

Unlike \oiii\ that is dominated by the outflow, 
the low-IP lines, e.g., \ha, have a mixed contribution of 
both of the fast outflow and the narrow line component. 
The intensity maps of the narrow and broad components
of \ha\ are shown in Figure \ref{fig:J1126_GMOS_Ha} (left panels).
The effective radii of the narrow and broad components are 2.4 and 1.2 kpc, respectively, after correction for the PSF blurring.
The broad component
shows a south-tilted distribution, 
which is similar to \oiii\ and suggests the outflowing hydrogen and oxygen gas locate in the same region\footnote{
	The broad component of \ha\ could have a large uncertainty in the faint outskirt region
	due to the blurring of the adjacent broad \nii\ lines. See discussion in Appendix \ref{appendix:GMOS_broad_Ha_NII}.
}. 
The narrow \ha\ line has a different distribution, which is more extended to the north, i.e., opposite to the outflow, in the core-removed map. 
The velocity shift maps of the narrow \ha\ line show a pattern of a rotating disk. 
Interestingly, although the velocity FWHM of the narrow \ha\ line (300--600 \kms) is
only 1/5--1/3 of those of the fast outflow components, 
it shows an increasing trend towards the direction of the fast outflow 
(e.g., the bottom-right panel of Figure \ref{fig:J1126_GMOS_Ha}). 
This enhanced dispersion along the outflow direction 
suggests a strong effect of the fast outflow on the disk gas, 
e.g., with a large opening angle or an orientation close to the disk plane. 
The intensity maps of the other low-IP lines and the atomic oxygen line
are shown in Figure \ref{fig:J1126_GMOS_Hb_intensity} and \ref{fig:J1126_GMOS_mLoIP_intensity}.
All of these lines have a broad component towards the south, 
i.e., following the direction shown by the fast \oiii\ outflow.
However, their narrow component show different morphologies: 
\hb\ extended to the north, i.e., similar to \ha;
while the intensity of \nii, \oi, and \sii\ is enhanced 
in the southern region. 
The different morphologies suggest different mechanisms of gas ionization with their ratios, which will be discussed in Section \ref{subsec:GMOS_ionization}. 


\subsection{Mechanisms of gas ionization}
\label{subsec:GMOS_ionization}

\begin{figure} 
	\begin{center}
		\includegraphics[trim=0 0 0 20, clip, width=1\columnwidth]{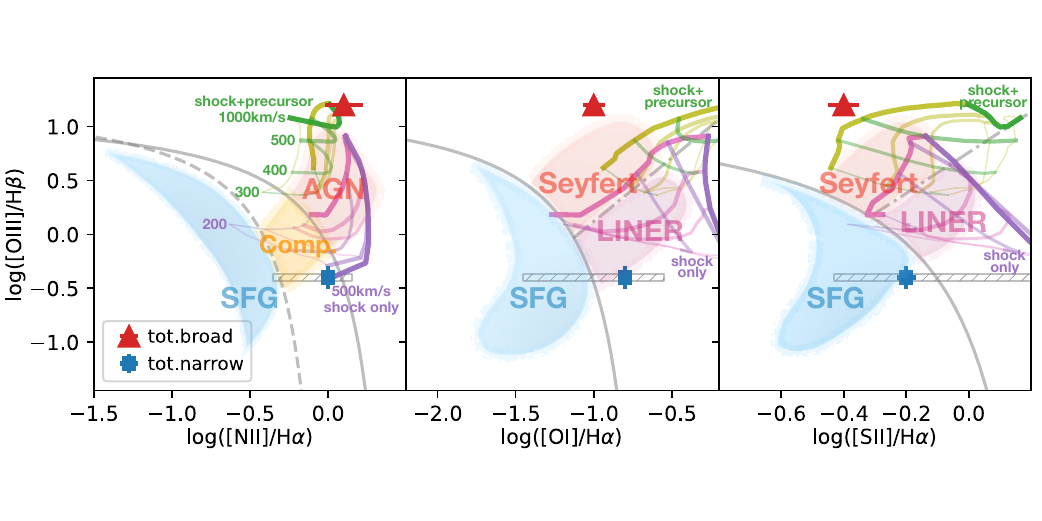}
	\end{center}
	\vspace{-11.5mm}
	\caption{
		\oiii/\hb\ vs. 
		\nii/\ha\ (left), \oi/\ha\ (middle), and \sii/\ha\ (right) ionization diagrams with the 
		emission line ratios of narrow (blue) and broad (red) components of integrated spectrum of J1126.
		The solid and dashed line represents the classification lines 
		of \cite{Kewley2001} and \cite{Kauffmann2003}, respectively.
		The dot-dashed lines in the middle and right panels represent the empirical
		classification line of Seyferts and LINERs by \cite{Kewley2006}. 
		The range of SDSS galaxies with types of SFG (blue), composite (yellow), 
		AGN/Seyfert (red), and LINER (violet) are denoted in the filled regions \citep{Kewley2006}. 
		The shock+precursor models 
		with velocity of 300--1000 \kms\ (green) and magnetic field intensity of $10^{-4}$--10 $\mu$G (yellowgreen), 
		as well as the shock-only models
		with 200--500 \kms\ (violet) and $10^{-4}$--10 $\mu$G (purple),
		are shown as grids (thicker lines for higher values)
		adopting the solar abundance and a preshock density of 1 \ccm\ \citep{Allen2008}.
		The horizontal hatches show the ranges of color bars in Figure \ref{fig:J1126_GMOS_BPT_map}. 
	}
	\label{fig:J1126_GMOS_BPT_int}
	\vspace{-1mm}
\end{figure}

\begin{figure*}
	\vspace{-4mm}
	\begin{center}
		\includegraphics[trim=0 54 0 50, clip, width=0.62\textwidth]{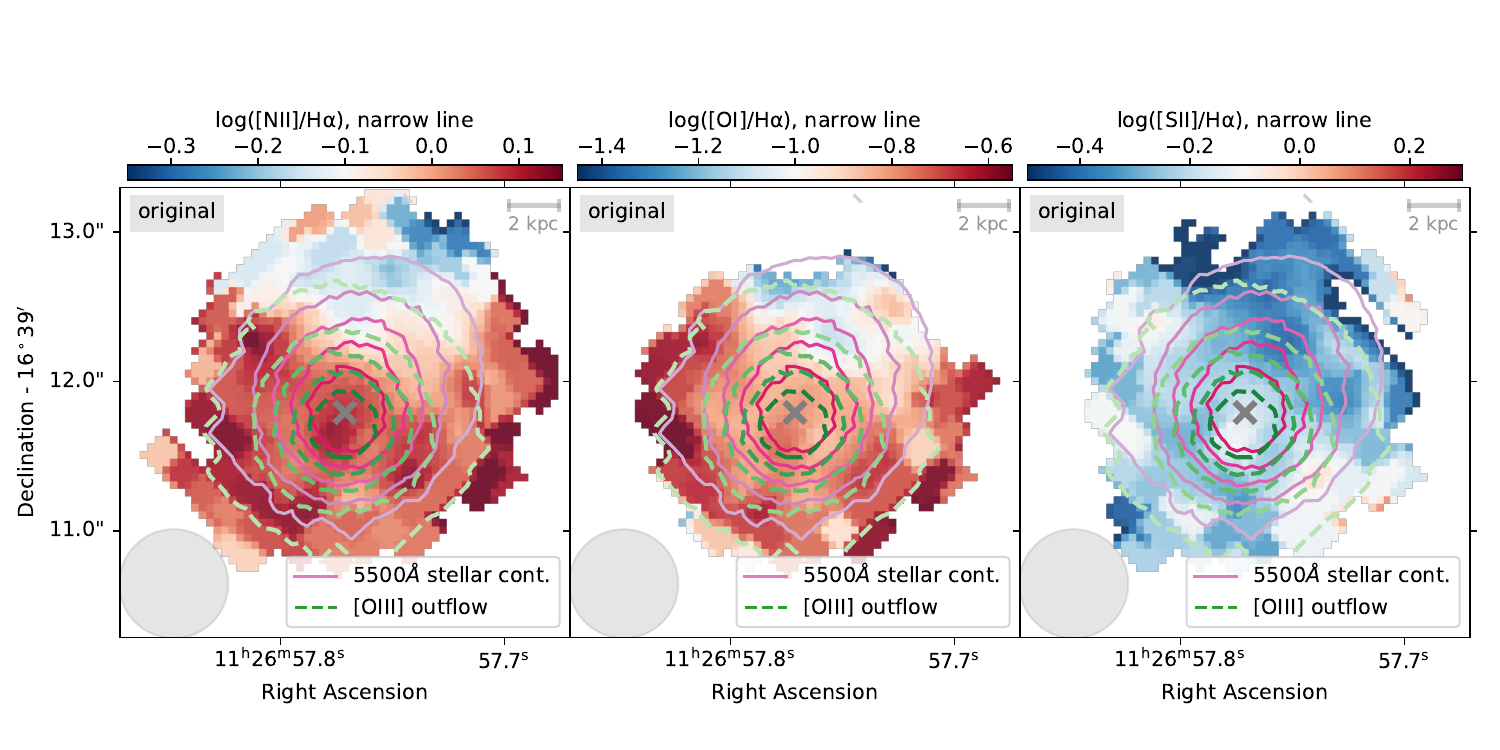}
		\includegraphics[trim=0  0 0 90, clip, width=0.62\textwidth]{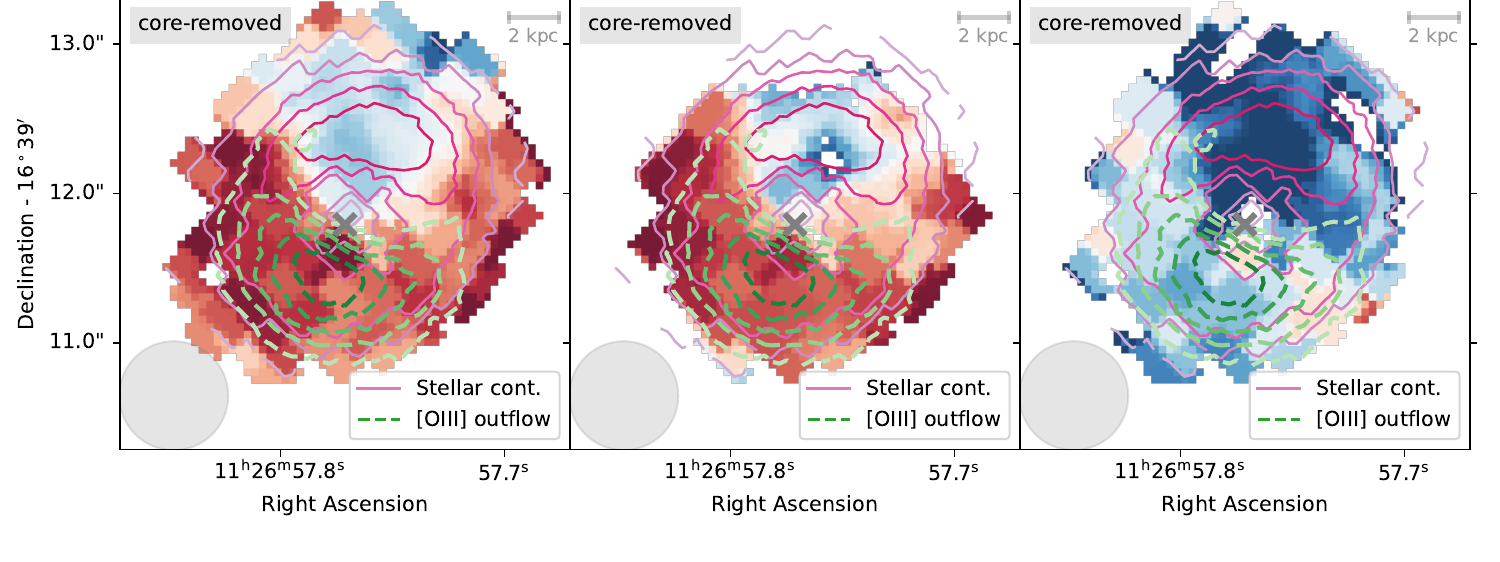}
	\end{center}
	\vspace{-9mm}
	\caption{
		Maps of \nii/\ha\ (left), \oi/\ha\ (middle), and \sii/\ha\ (right) of narrow lines from the original (top)
		and core-removed (bottom) GMOS data. 
		The bounds of color bar of \nii/\ha\ maps are set as the boundaries of the composite ionization (Figure \ref{fig:J1126_GMOS_BPT_int}, left)
		with $\log{(f_\mathrm{[OIII]}/f_\mathrm{H\beta})}=-0.35$ from the integrated spectrum. 
		The ranges of color bars of \oi/\ha\ and \sii/\ha\ maps are also shown in Figure \ref{fig:J1126_GMOS_BPT_int}. 
		In all of the maps, a bluer color denotes a larger contribution of star formation 
		and a redder color shows a larger contribution of shock.
		Only pixels with S/N $>$ 3 for both of the two lines to derive the ratios are shown in the panels.
		The distributions of $V$-band stellar continuum and \oiii\ outflow are shown in purple solid and green dashed contours, respectively,
		to indicate the associations of them to different ionizing mechanisms.
		PSF is shown in grey circles and the galaxy center in grey crosses.
	}
	\label{fig:J1126_GMOS_BPT_map}
	\vspace{-1mm}
\end{figure*}

We use the line ratio diagnostics to understand the energy source of gas ionization in the galaxy, which is the so-called BPT diagrams
\citep{Baldwin1981,Kewley2001,Kauffmann2003}.
The predominant broad component of the bright \oiii\ line
results in an extremely high \oiii/\hb\ ratio, i.e., $\log{(f_\mathrm{[OIII]}/f_\mathrm{H\beta})}=1.16\pm0.02$
in the broad component of the GMOS integrated spectrum, 
which corresponds to the maximum value of the local SDSS galaxies \citep{Kewley2006}. 
The broad line ratios of J1126 locate in the typical AGN/Seyfert regions in the BPT diagrams (Figure \ref{fig:J1126_GMOS_BPT_int}),
suggesting a significant of contribution of AGN radiation
on the ionization of the outflowing gas. 
The high \oiii/\hb\ ratio can also be explained by a fast shock 
as the fast outflow (i.e., 1000 \kms) penetrates into the ambient ISM in the outer regions of the galaxy. 
As the shock velocity increases, the cooling of hot gas behind the shock front 
generates a strong radiation field of extreme UV and soft X-ray photons, 
and produces a supersonic photoionization front that moves ahead of the shock front and pre-ionizes the gas,
which is the so-called photo-ionizing precursor \citep[e.g.,][]{Allen2008,Kewley2013}. 
The shock+precursor models of \cite{Allen2008} are shown as green grids in Figure \ref{fig:J1126_GMOS_BPT_int}.
The precursor can make a significant contribution to the ionized gas emission lines such as \oiiiblong\ \citep[e.g.,][]{Santoro2018}. 
Since J1126 possesses a luminous AGN (e.g., bright AGN radiation in mid-IR) and a fast, extended outflow, 
there could be a mixed contribution of the direct AGN radiation and the shock precursor
on the ionization of the outflowing gas. 
Such a mixture of ionization mechanisms in outflows of several nearby quasars is reported in \cite{Hinkle}. 

On the other hand, the narrow lines of the integrated spectrum
show a moderate \oiii/\hb\ ratio, i.e., $\log{(f_\mathrm{[OIII]}/f_\mathrm{H\beta})}=-0.35\pm0.11$, 
due to the faintness of narrow \oiii\ line.
The narrow line ratios locate in 
the composite region in the \nii/\ha\ diagram, 
the low-ionization narrow emission line regions (LINER) in the \oi/\ha\ diagram, 
and the star formation region in the \sii/\ha\ diagram 
(Figure \ref{fig:J1126_GMOS_BPT_int}), 
suggesting a mixed contribution of LINER and star formation ionization. 
The power sources of LINER include
inefficiently accreting low-luminosity AGN (LLAGN), shock, 
and aged stars in post-starburst galaxies \citep[][and references therein]{Kewley2013}. 
Since J1126 is a starburst galaxy with a luminous AGN, 
a low-speed shock (i.e., without the supersonic precursor), 
which are shown as violet grids in Figure \ref{fig:J1126_GMOS_BPT_int},
can be the most likely ionization mechanism \citep[e.g.,][]{Allen2008,Rich2010}.
Therefore, the narrow lines of J1126 (i.e., gas on the galaxy disk)
can be ionized by a combination of starburst and shock mechanisms. 

We use the maps of line ratios
to determine the ionizing mechanisms in the galaxy disk. 
The map of line ratios, \nii/\ha, \oi/\ha, and \sii/\ha\ 
are shown in Figure \ref{fig:J1126_GMOS_BPT_map} for both of the original and core-removed IFU data. 
Since the narrow \oiii\ line is faint, we assume a common \oiii/\hb\ ratio from the integrated spectrum, 
i.e., $\log{(f_\mathrm{[OIII]}/f_\mathrm{H\beta})}=-0.35$, for the entire galaxy. 
The color bars of the maps are set with the adopted \oiii/\hb\ ratio
and the boundary equations of different ionizing mechanisms of \cite{Kewley2001} and \cite{Kauffmann2003}.
In all of the maps, a bluer color indicates an increasing contribution from star formation activity, 
while a redder color indicates an increasing contribution from LINER (mainly due to shock, as discussed above). 
The \nii/\ha\ and \oi/\ha\ maps indicate that 
the ionization in the south is mainly contributed by shock
whereas star formation dominates ionization in the northern region.
The star formation ionization in the north is more apparently exhibited in the core-removed maps.
Similar to \nii/\ha\ and \oi/\ha,
the \sii/\ha\ map also shows a star formation dominated ionization in the north
and an increasing trend towards the south. 
Although the line ratio in the south, $\log{f_\mathrm{[SII]}/f_\mathrm{H\alpha}}\sim-0.2$, 
is still in the regime of star formation ionization with the classification of \cite{Kewley2001}, 
recent studies report that such ratio can be produced by shock excitation in supernova remnants or outflow \citep[e.g.,][]{Kopsacheili2020}.
To summarize, the maps of \nii/\ha, \oi/\ha, and \sii/\ha\ 
suggest a stronger star formation ionization in the northern outskirt of J1126, 
and a larger contribution by shock ionization towards the south. 

The maps of narrow line ratios are also compared to the spatial distribution of the stellar continuum
and the fast \oiii\ outflow (contours in Figure \ref{fig:J1126_GMOS_BPT_map}). 
A tight association is exhibited in the core-removed maps:
the star formation ionization is related to the stellar light from young stars in the north;
and the shock ionization is aligned with the distribution of fast outflow in the south.
The association between shock ionization of the disk gas and the fast outflow 
implies that the ionizing shock could be related to the large gas dispersion (e.g., Arp 220, \citealt{Perna2020}) in the galaxy disk, 
which is promoted by the fast outflow 
(Section \ref{subsec:GMOS_kinematics}; Figure \ref{fig:J1126_GMOS_Ha}, bottom-right panel).
It is also possible that the disk gas is ionized by diffuse ionizing photons from the shock precursor
of the fast outflow along its orientation. 
Either of the mechanisms indicates a significant effect of the fast outflow
on the gas ionization in the disk. 

The intensity map of the broad \ha\ line exhibits a cavity in the south outflow region (Figure \ref{fig:J1126_GMOS_Ha}), 
where the broad \oi, \nii, and \sii\ lines show enhanced intensity (Figure \ref{fig:J1126_GMOS_mLoIP_intensity} and \ref{fig:J1126_GMOS_NIIHaB_intensity}).
These results could suggest strong shock in the outflowing gas as implied by 
the integrated broad line ratios (Figure \ref{fig:J1126_GMOS_BPT_int}). 
However, a spatially-resolved line ratio diagnostics is not applicable
due to the large uncertainty of the decomposition of broad \ha\ and \nii\ lines in faint outskirt region.
A detailed discussion on the fitting uncertainty is shown in Appendix \ref{appendix:GMOS_broad_Ha_NII}. 


\subsection{Dust attenuation estimated with stellar and nebular observations}
\label{subsec:GMOS_extinction}

With the integrated spectrum, 
the dust attenuation of the stellar radiation is estimated to be $A_{V,\mathrm{\,stellar}}=1.5\pm0.1$ via the spectral fitting of the continuum. 
The extinction of the nebular region is $A_{V,\mathrm{\,narrow}}=3.0\pm0.2$ for the galaxy disk and $A_{V,\mathrm{\,broad}}=1.7\pm0.5$ for the gas in outflow, 
which are estimated using the Balmer decrement of the narrow and broad components, respectively. 
All of the estimations follow the extinction curve of \cite{Calzetti2000} with $R_\mathrm{V}=4.05$. 
An intrinsic \ha/\hb\ ratio of 2.86 from 
Case B recombination with an electron temperature of $10^4$ K
and an electron density of 100 \ccm\ \citep{Storey1995} 
is used to estimate the nebular extinction.
The ratio is usually adopted as the intrinsic value in 
star-forming regions \citep[e.g.,][]{Battisti2017}
and is also used in outflow studies \citep[e.g.,][]{Fiore2017}.
A higher \ha/\hb\ ratio, 3.1, is employed for AGNs in the literature
due to enhanced collisional excitation of \ha\ \citep[e.g.,][]{Ferland1983,Veilleux1995}.
Since it is difficult to quantitatively separate the contributions of star-forming region and the AGN
in gas ionization, we adopt a single ratio, 2.86, in the analysis. 
Lower extinction values, i.e., $A_{V,\mathrm{\,narrow}}=2.7$ and $A_{V,\mathrm{\,broad}}=1.4$, 
can be derived if 3.1 is adopted. 

The nebular region on the disk is more obscured than the stellar light
with a stellar-to-nebular (disk) attenuation ratio of $A_{V,\mathrm{\,stellar}}/A_{V,\mathrm{\,narrow}}=0.5\pm0.1$, 
which is similar to the typical value, 0.44, reported by \cite{Calzetti1997}. 
The attenuation ratio is found to be correlated with several properties of galaxies, 
e.g., $M_\star$, SFR, and metallicity, in recent works \citep[e.g.,][]{Koyama2019,Lin2020,Li2021}. 
A two-component dust model is usually adopted to explain the differences between 
$A_{V,\mathrm{\,stellar}}$ and $A_{V,\mathrm{\,narrow}}$ in star-forming galaxies \citep[e.g.,][]{Charlot2000}.
In this model, 
the massive young stars are embedded in dense birth clouds
while the long-lived stars locate 
in less-obscured diffuse ISM. 
Massive young stars can provide the bulk of the ionizing photons. 
For example, under the assumption of a constant SFR in the recent 100 Myr with the PopStar templates, 
stars younger than 6 Myr (i.e., lifetime of O-type stars, \citealt{Kennicutt2009,Calzetti2013})
contribute to 96\% of the total ionizing flux in star-forming region.
Therefore the $A_{V,\mathrm{\,narrow}}$ based on Balmer lines can trace the 
attenuation of radiation of young stars
that travels through the dense birth clouds 
and the outside diffuse ISM. 
Since the radiation from the young stars can be highly obscured,  
the observed stellar light can be dominated by the long-lived stars 
that only suffer the extinction of 
the less-obscured diffuse ISM. 
As a consequence, the continuum-based $A_{V,\mathrm{\,stellar}}$ mainly traces extinction in the diffuse ISM. 
For example, with the estimated $A_{V,\mathrm{\,narrow}}=3.0$, $A_{V,\mathrm{\,stellar}}=1.5$
and the above assumption of constant star formation, 
96\% of the observed $V$-band stellar continuum is emitted by the long-lived stars ($t>6$ Myr) in the diffuse ISM. 

\begin{figure*}
	\vspace{-4mm}
	\begin{center}
		\includegraphics[trim=0 54 0 50, clip, width=0.62\textwidth]{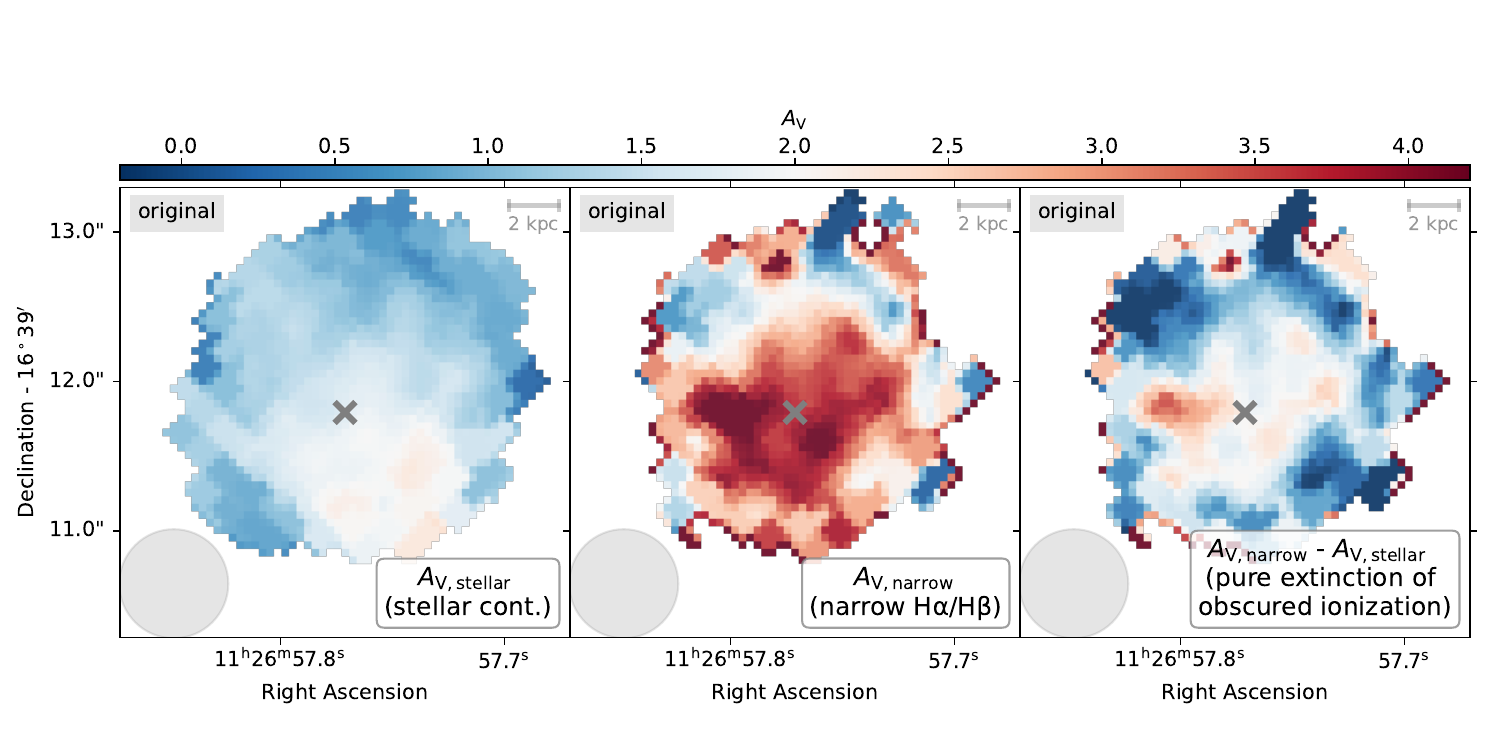}
		\includegraphics[trim=0  0 0 90, clip, width=0.62\textwidth]{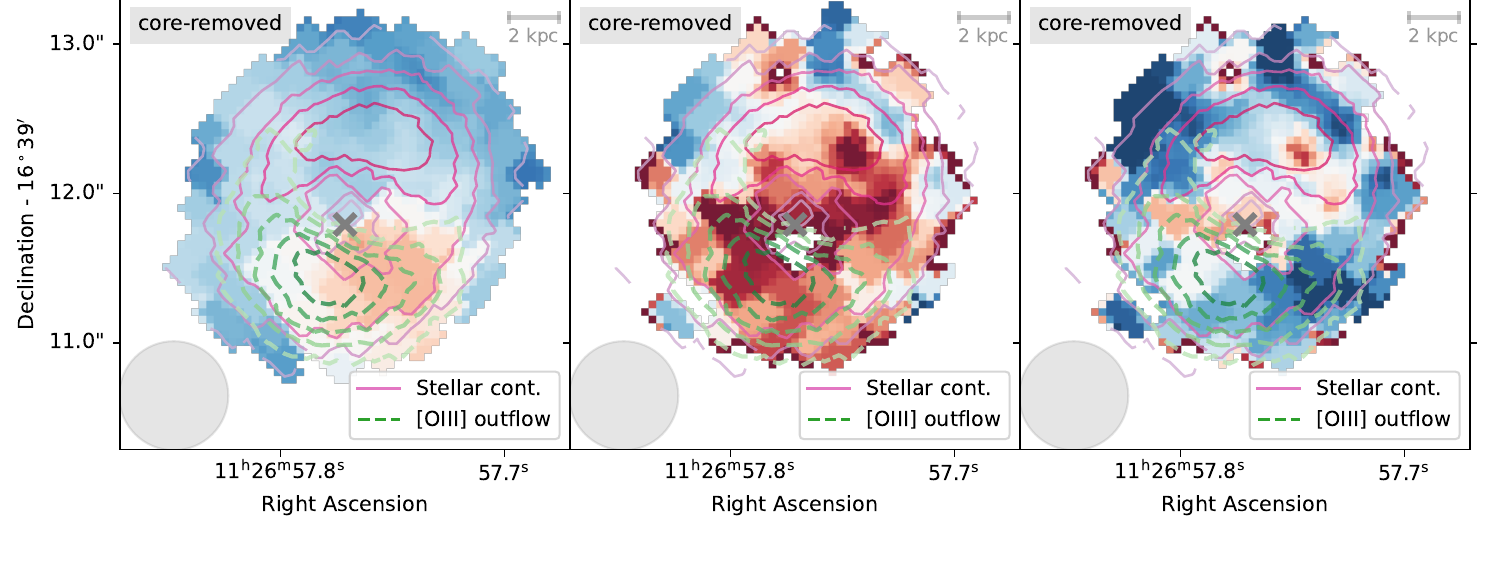}
	\end{center}
	\vspace{-9mm}
	\caption{
		Dust extinction estimated with stellar continuum fitting ($A_{V,\mathrm{\,stellar}}$, left), 
		Balmer decrement of narrow lines ($A_{V,\mathrm{\,narrow}}$, middle), 
		and the pure extinction of the obscured ionizing region ($A_{V,\mathrm{\,stellar}}-A_{V,\mathrm{\,narrow}}$, right)
		from the original (top) and core-removed (bottom) GMOS data. 
		The spatial distributions of $V$-band stellar continuum and \oiii\ outflow are shown in the bottom panels 
		to indicate the relation between the extinction and the two components.
		All of the panels share the same color bar scale. 
		Only pixels with S/N $>$ 5 for $V$-band continuum and
		S/N $>$ 3 for both of \ha\ and \hb\ fluxes are shown in the panels.
		PSF is shown in grey circles and the galaxy center in grey crosses.
	}
	\label{fig:J1126_GMOS_AV}
	\vspace{-1mm}
\end{figure*}
 
The dust extinction in J1126 can be more complicated 
than pure star-forming galaxies
due to a significant contribution of AGN/shock in ionization,
and a possible foreground extinction by the dust entrained in the extended outflow.
Following a similar assumption of the two-component dust model of star-forming galaxies,  
we consider that the measured $A_{V,\mathrm{\,stellar}}$ reflects the extinction of diffuse medium, 
which can be a mixture of the extinction of the host diffuse ISM
and that of the dust entrained in the outflow.
The pure extinction in obscured ionizing regions ($A_{V,\mathrm{\,ionizing}}$) 
either by young stars or shocks in the host disk 
can be estimated by removing the foreground extinction (e.g., by dust in diffuse ISM or outflow)
from the values based on the Balmer decrement, 
i.e., $A_{V,\mathrm{\,ionizing}}=A_{V,\mathrm{\,narrow}} - A_{V,\mathrm{\,stellar}}$. 
The maps of these estimates are shown in Figure \ref{fig:J1126_GMOS_AV}. 

Both of the original and the core-removed maps of $A_{V,\mathrm{\,stellar}}$ 
show an increasing trend towards the south. 
The enhanced extinction in the south is 
associated with the fast ionized outflow, 
suggesting that the enhancement is 
probably due to the dust carried out by the outflow
from the dense central region
that shields the underlying stellar light
in the southern outskirt. 
The difference of $A_{V,\mathrm{\,stellar}}$ between the south
(with outflow, $\sim1.0$) and north (no outflow, $\sim2.5$),
$\sim1.5$, is similar to the extinction of the outflowing gas,
$A_{V,\mathrm{\,broad}}=1.7$, estimated from the integrated spectrum, 
which supports the explanation of dust entrained in the outflow.

The original map of the pure extinction in ionizing regions shows 
$A_{V,\mathrm{\,ionizing}}=$ 2.0--3.0 in the central region of the galaxy, 
which suggests the heavy obscuration of the central starburst and/or the AGN/shock-ionizing region. 
In the core-removed map, the high $A_{V,\mathrm{\,ionizing}}$ ($\sim2.5$)
is mainly shown in the north where star formation dominates the ionization (e.g., Figure \ref{fig:J1126_GMOS_BPT_map}), 
indicating that the dusty environment is extended from the central starburst
to the star forming region in the northern outskirt.
The southern shock-dominated region exhibits
relatively low extinction $A_{V,\mathrm{\,ionizing}}\sim0.5$--2.0, 
and implies that the extended shock can occur in gas less-obscured 
than the dense star-forming clouds.

Due to the faintness of broad \hb\ (Figure \ref{fig:J1126_GMOS_Hb_intensity})
and the difficulty to decompose broad \ha\ and \nii\ (Appendix \ref{appendix:GMOS_broad_Ha_NII})
in the outskirt, 
we do not analyze the map of $A_{V,\mathrm{\,broad}}$ in this work. 


\section{Analysis method and results of the ALMA observations}
\label{sec:ALMA}

\subsection{Observations of (sub-) millimeter continua}
\label{subsec:ALMA_dust}

\begin{figure}
	\begin{center}
		\hspace{-8mm}
		\includegraphics[trim=0  0 230 40, clip, width=0.90\columnwidth]{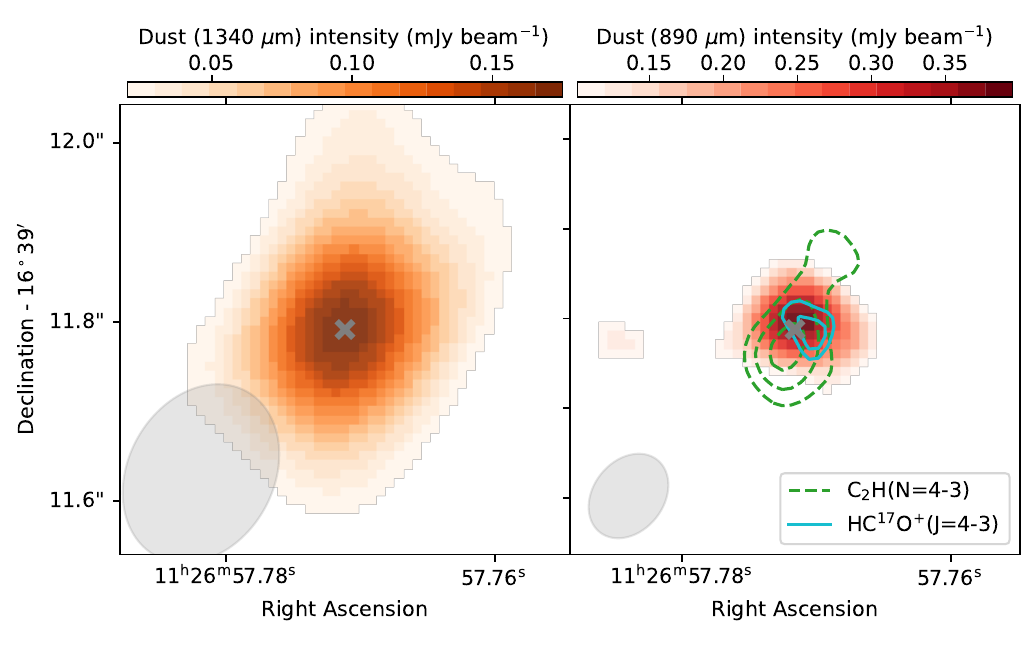}
	\end{center}
	\vspace{-9mm}
	\caption{
	Intensity maps of continua at rest 1340 \micron\ (left) and 890 \micron\ (right). 
	The contours in the right panel denote two faint emission lines, i.e., 
	C$_2$H(N=4-3) in green and HC$^{17}$O$^+$(J=4-3) in cyan.
	The continuum is mainly emitted from the central starburst region (Section \ref{subsec:ALMA_dust}), 
	while the C$_2$H(N=4-3) line mainly traces the molecular outflow (Section \ref{subsec:ALMA_C2H}). 
	Only pixels with S/N$>$3 are shown for each components, 
	except for HC$^{17}$O$^+$(J=4-3) with S/N$>$2. 
	The beam is shown in grey ellipses and the galaxy center in grey crosses.
	}
	\label{fig:J1126_ALMA_dust_continua}
\end{figure}

(Sub-)millimeter continua of J1126 are detected by ALMA
at rest 1340 and 890 \micron\
with a spatial resolution of 1.1 kpc and 0.6 kpc, respectively;
and detected at rest 660 \micron\ by an unresolved ACA observation. 
The intensity maps at rest 1340 and 890 \micron\ are shown in Figure \ref{fig:J1126_ALMA_dust_continua}. 
The entire flux at 1340 \micron\ is $0.25\pm0.02$ mJy
measured within a radius of 0.7\arcsec,
which is the total-flux-radius from the 
radial curve of growth.
The flux of the continuum at 890 \micron\ is $0.65\pm0.12$ mJy
within a total-flux-radius of 0.2\arcsec. 
Since the two total-flux-radii are much smaller than the maximum recoverable scales of the observations
i.e., 7.6\arcsec\ and 1.0\arcsec\ in 1340 \micron\ and 890 \micron, respectively, 
the derived fluxes are considered to reflect the entire fluxes of the galaxy at the two bands.  
The total flux\footnote{
	The contribution on the flux at 660 \micron\ by the companion galaxy J1126b is ignorable, 
	since it is not detected the maps of 890 and 1340\,\micron. 
	There is no other sources in the extracting region. }
at 660\,\micron, $1.85\pm0.87$\,mJy, 
is calculated in a region with a $2\times2$ beam size of ACA ($15\arcsec\times9\arcsec$).

The observed (sub-)millimeter continua can be emitted by 
the dust heated by stellar light, 
the thermal free-free emission of the ionized gas, 
and/or the non-thermal synchrotron emission.
In order to understand the contribution of these radiation mechanisms,
we perform the SED fitting with the new ALMA observations 
plus the archived data from optical to radio bands
following the method employed in \cite{Chen2020}. 
The stellar component is constrained by the spectral fitting results reported in Section \ref{subsec:GMOS_stellar}.
The SKIRTor AGN model \citep{Stalevski2016} and THEMIS ISM dust model \citep{Jones2017} are used in the fitting. 
The free-free and synchrotron components are fit with $S_\nu \propto \nu^{-\alpha}$, 
and the indexes are adopted as $\alpha_\text{ff}=0.1$ and $\alpha_\text{sync}=0.8$ \citep[e.g.,][]{Murphy2011,Tabatabaei2017}. 
The best-fit model is shown in Figure \ref{fig:J1126_image_SED} (right panel). 
It is found that the detected continua at 660 and 890 \micron\ are mainly dominated by the ISM dust radiation (97\% and 89\%).
Synchrotron and free-free emissions contribute to 26\% and 10\% of the 1340 \micron\ continuum, respectively. 

The spatial profiles of the continua at 890 and 1340 \micron\ 
give an half-light radius of 0.2 and 0.3 kpc, respectively, after correction for the beam blurring.
The compact dust continuum provides an independent evidence of the highly-obscured central environment
as indicated by the nebular attenuation (Section \ref{subsec:GMOS_extinction} and Figure \ref{fig:J1126_GMOS_AV}). 


\subsection{Spectra and moment maps of CO emission lines}
\label{subsec:ALMA_CO_main}

\begin{figure*}
	\begin{center}
		\includegraphics[trim=0 0 -30 0, clip, width=0.8\columnwidth]{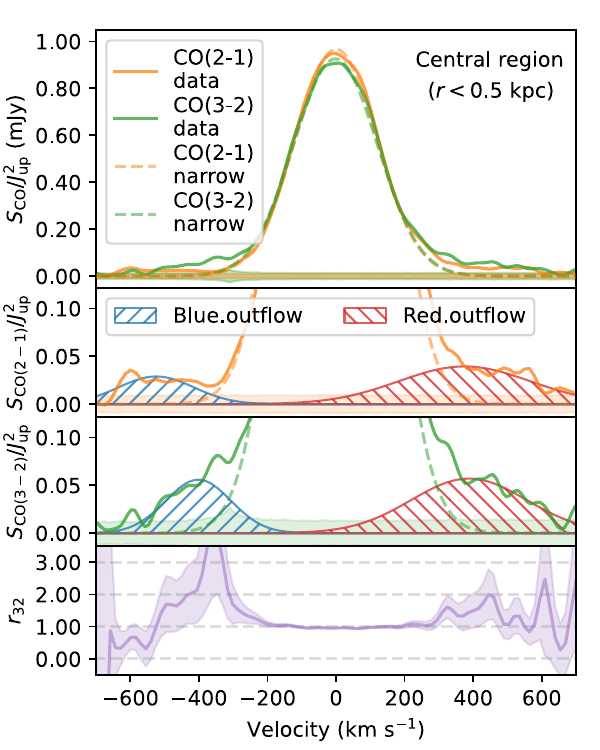}
		\includegraphics[trim=0 0 -30 0, clip, width=0.8\columnwidth]{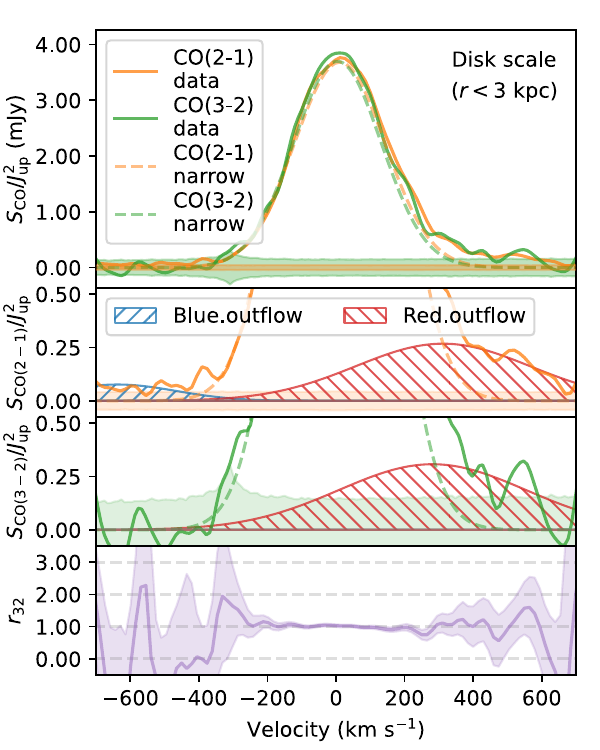}
	\end{center}
	\vspace{-7mm}
	\caption{
		\textbf{Upper:} 
		Continuum-subtracted spectra of CO(2-1) (orange) and CO(3-2) (green) 
		extracted in the central region ($r<0.5$ kpc, left) 
		and in a galaxy disk scale ($r<3$ kpc, right)
		with the beam and MRS-matched cubes. 
		The spectra are convolved by a Gaussian profile with FWHM of 50 \kms.
		The narrow components of 3-Gaussian fitting are shown in orange and green dashed curves for CO(2-1) and CO(3-2), respectively.
		The $\pm1\sigma$ error ranges are denoted in shadow regions. 
		\textbf{Middle-upper:} 
		A zoom-in at the bottom of CO(2-1) profile. 
		The blue- and redshifted outflow components are shown in blue and red hatches, respectively. 
		\textbf{Middle-lower:} 
		A zoom-in at the bottom of CO(3-2) profile. 
		The outflow components are shown as the legends used for CO(2-1). 
		\textbf{Lower:} 
		CO line ratio ($r_{32}$), The $\pm1\sigma$ range of $r_{32}$ is shown in the shadow region. 
		}
	\label{fig:J1126_ALMA_mCO_spec}
\end{figure*}

CO(2-1) and CO(3-2) lines of J1126 are observed by ALMA with different angular resolutions (AR) and 
maximum recoverable scales (MRS). 
In order to perform a direct comparison between the two CO lines, 
we mainly utilize the beam and MRS-matched data cubes (see the reduction in Section \ref{subsec:Reduction_ALMA})
of CO(2-1) and CO(3-2), 
which have a mean AR of 1.1 kpc and a MRS of 6.0 kpc, 
in Section \ref{subsec:ALMA_CO_main} and \ref{subsec:ALMA_CO_outflow}.
The original CO(2-1) cube with a larger MRS ($>40$ kpc) is used to estimate the CO flux and molecular gas mass of the entire galaxy;
while the original CO(3-2) cube with a higher AR (0.6 kpc) is used to show the detailed outflow structure in the central region
(discussed in Section \ref{subsec:ALMA_CO_outflow}).  

Figure \ref{fig:J1126_ALMA_mCO_spec} shows the spectra of CO(2-1) and CO(3-2)
extracted from the central beam-size region ($r<0.5$ kpc) and the disk scale ($r<3$ kpc),
with the beam and MRS-matched cubes. 
A multi-gaussian fitting is employed for the CO spectra
with a narrow profile for the disk component
and one or two wing components for non-circular motion
(i.e., 2-Gaussian and 3-Gaussian fit). 
The velocity of narrow component in the 3-Gaussian best-fit
of CO(2-1) in the central region ($r<0.5$ kpc)
is selected to determine the systemic redshift of the galaxy, $z_\mathrm{sys}=0.468417\pm0.000002$
(the uncertainty is estimated with Monte Carlo simulations).
The broad components that indicates the molecular outflow are discussed in Section \ref{subsec:ALMA_CO_outflow}. 

We adopt $r_{32}=(S_\text{CO(3-2)}/3^2)/(S_\text{CO(2-1)}/2^2)$
to calculate the ratios between the two CO lines \citep[e.g.,][]{Kirkpatrick2019,Molyneux2024,Montoya2024},
which is shown in the bottom panels in Figure \ref{fig:J1126_ALMA_mCO_spec}. 
The spectra of the galaxy disk scale show nearly constant $r_{32}\sim1$ 
in the velocity range of $|v_\text{s}|<250$ \kms,
which is a bit higher than the average $r_{32}$ of local (U)LIRGs \citep[e.g., 0.8][]{Montoya2023}, 
and could suggest a stronger heating of starburst in J1126 considering its high SFR of $\sim800$ \sfrunit. 
The spectra of the central region also show $r_{32}\sim1$ in the low-velocity range ($|v_\text{s}|<200$ \kms);
however, the ratio becomes higher as velocity increases, especially on the blueshifted side where $r_{32}$ even exceeds 3. 
The relation between high CO excitation and large gas dispersion (e.g., outflow) are reported in several recent works.
\cite{Cicone2018} found that the CO(2-1)/CO(1-0) ratio
increases in the spatially-resolved outflow of NGC 6240.
\cite{Montoya2024} reported the CO ladder up to $J_\text{up}=7$ for 8 local (U)LIRGs with the single-dish APEX observation
and found that the components with higher velocity tend to show higher CO excitation (thermalized or over-thermalized line ratios)
than the dynamically-quiescent components (sub-thermalized line ratios).
The relation between $r_{32}$ and velocity of J1126 suggests the existence of molecular outflow in the central region.
We discuss the association between CO excitation and outflow in details in Section \ref{subsubsec:ALMA_CO_PV_ratio}. 

\begin{figure*}[!ht]
	\begin{center}
		\includegraphics[trim=0 54 0 50, clip, width=0.62\textwidth]{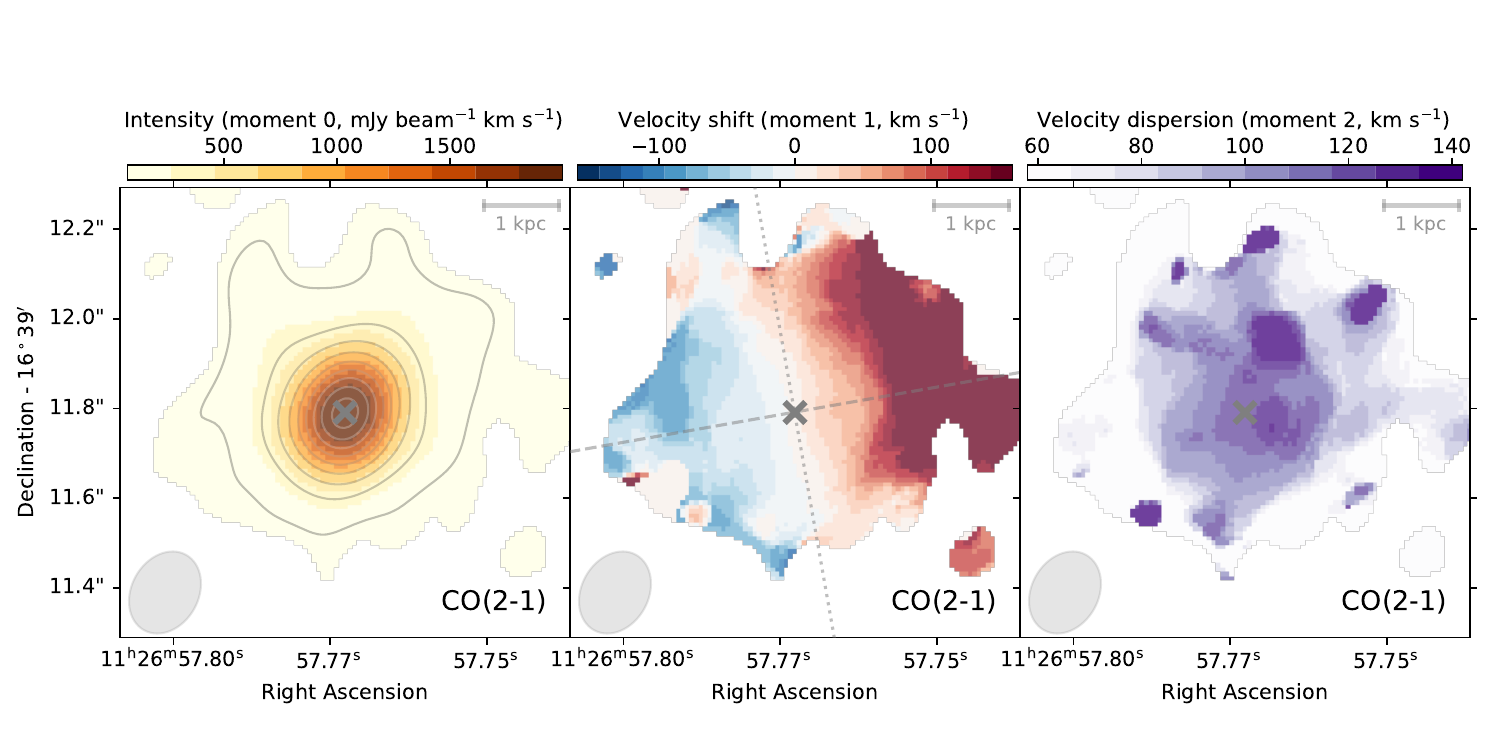}
		\includegraphics[trim=0  0 0 64, clip, width=0.62\textwidth]{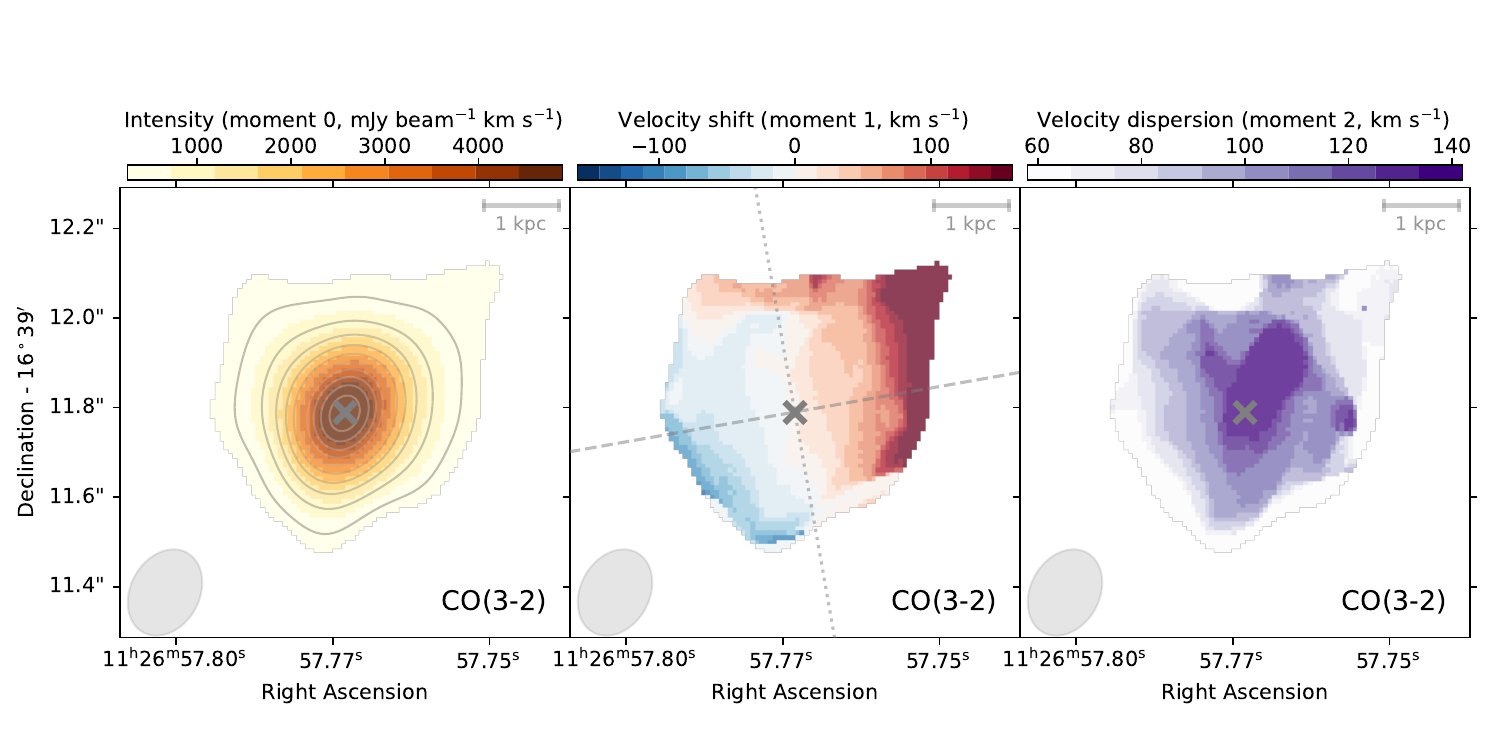}
	\end{center}
	\vspace{-9mm}
	\caption{
	Moment-0, 1, and 2 maps of 
	the beam and MRS-matched cubes of CO(2-1) (top) and CO(3-2) (bottom). 
	Only pixels with intensity S/N $>$ 3 are plotted in each panel. 
	The intensity contours in the left panel start from S/N\,=\,5, 
	with intervals of S/N of 10 for CO(2-1) and 5 for CO(3-2). 
	The dashed and dotted lines in the middle panel represent the major and minor axes of the rotating disk. 
	The grey ellipse shows the beam and the grey cross shows the galaxy center in each panel. 
	}
	\label{fig:J1126_ALMA_mCO_mom012}
\end{figure*}

The maps of intensity, velocity shift and velocity dispersion (i.e., moment-0/1/2 maps)
of the beam and MRS-matched cubes of CO(2-1) and CO(3-2)
are shown in Figure \ref{fig:J1126_ALMA_mCO_mom012}.
The intensity maps are integrated within the velocity range of $-500$ to 500 \kms,
which contributes to $>$99\% of the flux of the entire spectral profile (e.g., Figure \ref{fig:J1126_ALMA_mCO_spec}).
Both of the two CO lines have a total-flux-radius of $\sim$ 6 kpc
from the curve of growth, which results in an integrated flux
of $6.5\pm0.1$\,Jy \kms\ for CO(2-1)
and $13.1\pm0.7$\,Jy \kms\ for CO(3-2). 
The effective (half-light) radii of CO(2-1) and CO(3-2)
are 1.0 and 0.6 kpc, respectively.
CO(2-1) is more extended than CO(3-2), suggesting a lower CO excitation (e.g.,a lower temperature) in the outside region.

The maps of velocity shift of CO(2-1) and CO(3-2) (middle panels of Figure \ref{fig:J1126_ALMA_mCO_mom012})
indicate a pattern of a rotating disk, which has a major axis with a PA of $\sim280^\circ$.
The direction of the major axis is estimated to be perpendicular to the direction of pixels with a zero velocity shift.
The PA of the molecular gas disk is similar to that of the ionized gas disk
as shown in the middle panels of Figure \ref{fig:J1126_GMOS_Ha}.
The similar rotating pattern suggests that
the (non-outflow) ionized and molecular gas locates in the same disk plane. 
The velocity dispersion maps of CO(2-1) and CO(3-2) (right panels of Figure \ref{fig:J1126_ALMA_mCO_mom012})
both show an enhanced dispersion towards the northwest direction (PA\,$\sim$\,$340^\circ$), 
suggesting the existence of molecular outflow in this region. 


In the above discussions we adopt the MRS-matched CO(2-1) cube for a comparison to CO(3-2) observation,
which can loss flux in a large scale, e.g., $>1$ kpc. 
In order to estimate the entire molecular gas mass ($M_\mathrm{H2}$) of J1126, we employ the original CO(2-1) cube that has an integrated flux of $7.62\pm0.13$ Jy \kms\ within a total-flux-radius of 6 kpc
(i.e., the MRS-matched cube has a flux loss of 14\%).  
Assuming the typical value of the CO-to-H$_2$ factor
$\alpha_\mathrm{CO}=0.8$ \citep{Downes1998,Carilli2013}
and the CO(2-1)/CO(1-0) ratio $r_{21}=1.0$ 
in nearby ULIRGs \citep{Montoya2023,Molyneux2024},
the total $M_\mathrm{H2}$ of J1126 is estimated to be 
$(1.79\pm0.03)\times10^{10}$ \msun\ 
(the uncertainty reflects the measurement error).
The estimated $M_\mathrm{H2}$ corresponds to a molecular gas mass fraction of
$M_\mathrm{H2}/(M_\star+M_\mathrm{H2})=8.3\%$. 

In addition to CO(2-1) and CO(3-2), CO(4-3) line of J1126
is also detected in the unresolved ACA observation. 
The flux of CO(4-3), $18.5\pm1.2$ Jy \kms, is extracted in a $2\times2$ beam-size region
as performed for the 650 \micron\ continuum. 
The line ratio of CO(4-3) and CO(2-1) ($r_{42}$) of the entire galaxy is estimated to be 0.57
using the entire flux of the original CO(2-1) cube within 6 kpc. 
The ratio is similar to the mean value of local (U)LIRGs \citep[e.g., 0.5--0.6,][]{Montoya2024}. 
Due to a lack of resolved spatial information for CO(4-3), 
hereafter we focus on the observation of CO(2-1) and CO(3-2) for kinematic analyses. 


\subsection{Identification of molecular outflow with CO(2-1) and CO(3-2) observations}
\label{subsec:ALMA_CO_outflow}

\subsubsection{Multi-Gaussian fitting of CO lines}
\label{subsubsec:ALMA_multi_gaussian}

Since the CO emission in the sub-millimeter band is less sensitive to dust extinction, 
both of the redshifted outflow, which is shielded by the dust in the host galaxy in UV/optical observations (e.g., \oiiiblong),
and the blueshifted outflow can be observed in the CO line profiles if the outflow is bright. 
Similar to the identification of ionized outflow, 
we perform a multi-gaussian fitting for the CO lines, 
in which a narrow component is used to fit the gas motion in the galaxy disk
with one or two broad components for the outflow 
(hereafter 2-Gaussian and 3-Gaussian fit). 
The two broad components in the 3-Gaussian fit are adopted to fit the blue- and redshifted outflows separately.

The 3-Gaussian fitting results for CO(2-1) and CO(3-2) lines are shown in middle panels in 
Figure \ref{fig:J1126_ALMA_mCO_spec}. 
For the CO(2-1) line in the central region ($r<0.5$ kpc),
both of blue- and redshifted wings are detected in the 3-Gaussian fit, 
which have a FWHM of 250 \kms\ and 450 \kms,
and a flux fraction of 3\% and 6\%, respectively.
The 2-Gaussian best-fit result
has a single wing component, which is broader (FWHM = 950 \kms) and brighter (flux fraction of 14\%)
than the 3-Gaussian fitting results.
However, a half of the wing flux of the 2-Gaussian fitting
originates from the low-velocity region (e.g., $|v_\text{s}|<200$ \kms), 
which is less significant in kinematics and could be contaminated by the gas motion on the disk.
Therefore, we consider that the 3-Gaussian fitting results as a fiducial 
estimation of the molecular outflow.
In the central region ($r<0.5$ kpc), 
the outflow components of CO(3-2) have similar velocities to those of CO(2-1) and a larger flux fractions, 5\% and 8\%, on the blue- and redshifted sides, respectively. 
The detailed fitting results are listed in Table \ref{tab:mCO_gfit}. 

In the galaxy disk scale ($r<3$ kpc), 
the blueshifted outflow becomes faint and makes it difficult to be detected
(e.g., with a peak S/N $\sim$ 1 for CO(3-2)). 
On the other side, both of CO(2-1) and CO(3-2) possess 
an obvious redshifted wing with the flux fraction even larger than the central region, 
i.e., 12\% for CO(2-1) and 15\% for CO(3-2). 
The different results between the blue- and redshifted outflows  suggest
an asymmetric outflow structure:
the blueshifted outflow is faint and concentrated in the central region, 
while the redshifted outflow is bright and extended in the galaxy disk scale. 
Note that even in the center the blueshifted outflow is weaker (e.g., a lower flux fraction) than the redshifted one. 

In order to understand the details of the outflow structure, 
a spatially-resolved analysis is required. 
However, it is hard to perform the multi-Gaussian fitting for the resolved cubes due to the contamination of the disk motion. 
Unlike the fast ionized outflow ($|v|>1000$ \kms), 
the velocity of molecular outflow is moderate (300--500 \kms)
and makes it difficult to
separate the outflow from disk motion with the multi-Gaussian fitting 
in the outskirt region where disk motion shows non-zero velocity shift
(e.g., Figure \ref{fig:J1126_ALMA_mCO_mom012}, moment-1 map). 
Another method is required to address the issue. 

\subsubsection{Decomposition of disk and outflow motions with fitting of the Barolo disk model}
\label{subsubsec:ALMA_barolo}

\begin{figure*}[!ht]
	\begin{center}
		\includegraphics[trim=150 0 150 0, clip, width=0.88\textwidth]{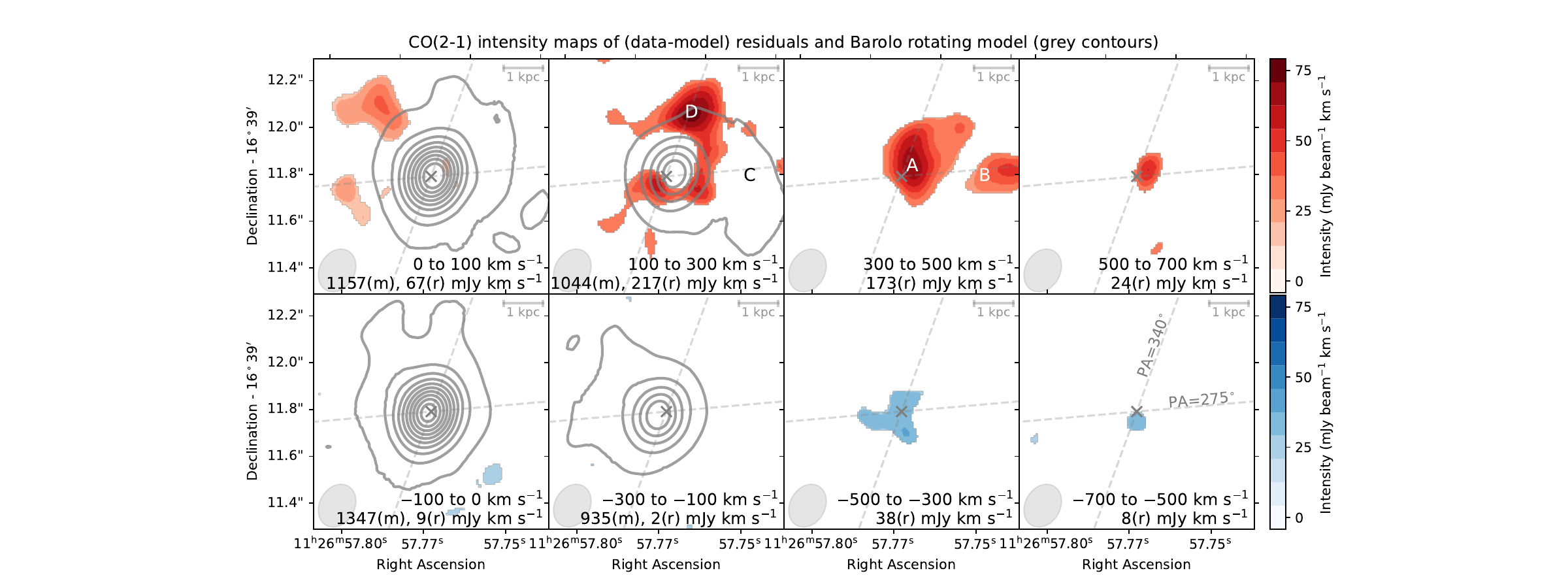}
	\end{center}
	\vspace{-8mm}
	\caption{
	Residual maps of MRS-matched CO(2-1) cube 
	after subtracting the best-fit disk model (contours) .
	The bottom and top panels show the blue- and redshifted sides, respectively. 
	Only pixels with velocity-integrated S/N $>$ 3 are shown.
	Rotation model is shown in contours starting from model S/N = 3 with an interval of S/N = 10. 
	The fluxes of model (m) and residual (r) integrated with S/N $>$ 3 pixels are shown at bottom in each panel. 
	The grey ellipses denote the beam and the grey cross indicates the galaxy center.
	The dashed lines show the PA of the PV diagrams shown in Figure \ref{fig:J1126_ALMA_mCO_PV}.
	See text in Section \ref{subsubsec:ALMA_CO_PV_ratio} for discussion of regions marked with the letter A--D. 
	}
	\label{fig:J1126_ALMA_CO21r_vel_wins}
\end{figure*}

In order to determine the spatial distribution of the molecular outflow, 
in this subsection we adopt the rotating disk model of $^\text{3D}$Barolo\citep[hereafter Barolo;][]{DiTeodoro2015}
to decompose the motion of the molecular gas disk, 
and then identify the outflow with the non-rotating residuals.
In order to reduce the degeneracy of the fitting, 
we fix the central position of the rotating disk model to the center obtained from the moment-0 map (Figure \ref{fig:J1126_ALMA_mCO_mom012}, left);
the PA of the model is set to the direction of the major axis 
estimated with the moment-1 map (Figure \ref{fig:J1126_ALMA_mCO_mom012}, middle) with a tolerance of $\pm5^\circ$; 
the systemic velocity of the disk model is fixed to the value from the narrow component of the 
3-Gaussian fitting (e.g., Table \ref{tab:mCO_gfit}). 
We set a primary S/N cut of 5 and a secondary (growing) S/N cut of 3 to determine the fitting region mask.
The maps of the best-fit disk model of CO(2-1) 
and the model-subtracted residuals 
are shown in Figure \ref{fig:J1126_ALMA_CO21r_vel_wins}.
The rotating model suggests a disk in a nearly face-on view
with an inclination angle of $\sim20^\circ$, 
which rotates in a counter-clockwise direction along the line of sight
and dominates the emission with the observed velocity (\voffabs) less than 300 \kms. 

The molecular outflow is detected with \voffabs\ up to 700 \kms\ 
on both of blue- and redshifted sides
in the non-rotating residual maps (Figure \ref{fig:J1126_ALMA_CO21r_vel_wins}). 
The redshifted outflow has a flux-weighted velocity of 350 \kms. 
It changes from an unresolved, compact distribution in \voff\ $> 500$ \kms,
to an extended structure in lower velocity ranges with a radius of 2--3 kpc, 
which could reflect the deceleration of outflowing gas at a large radius towards the north and west, 
or the changes of inclination angles of a cone-like outflow structure.
The redshifted outflow is brighter than the blueshifted one in all of the velocity bins.
The blueshifted outflow shows a flux-weighted velocity of 390 \kms, similar to that of redshifted outflow.
It has a dominant direction to the south in \voff\ $> 300$ \kms, 
which is opposite to the redshifted outflow and suggests a bi-conical structure. 
The orientation of the molecular outflows
is consistent with that of the south-tilted (blueshifted) ionized outflow 
(Section \ref{subsec:GMOS_kinematics}). 

The non-rotating residual maps suggest an asymmetric structure of the molecular outflow: 
the outflow on the redshifted side is brighter and more extended 
than that on the blueshifted side. 
However, there is still uncertainties to determine the extent of the redshifted outflow. 
For example, an outskirt region of the redshifted outflow shown with the letter
``B'' in the 300--500 \kms\ map of Figure \ref{fig:J1126_ALMA_CO21r_vel_wins}
overlaps with a spiral arm-like extension of the rotating disk in lower velocity ranges, 
i.e., the region ``C'' in the 100-300 \kms\ map. 
Such overlapping implies 
a possible non-outflow origination of 
the extended non-rotating residuals (e.g., ``B''),
e.g., tidal tails of galaxy merger interaction or an inflowing stream. 

\subsubsection{Identification of outflow by combination of non-rotating motion and enhanced CO excitation}
\label{subsubsec:ALMA_CO_PV_ratio}

\begin{figure*}
	\begin{center}
		\includegraphics[trim=0 54 0 50, clip, width=0.62\textwidth]{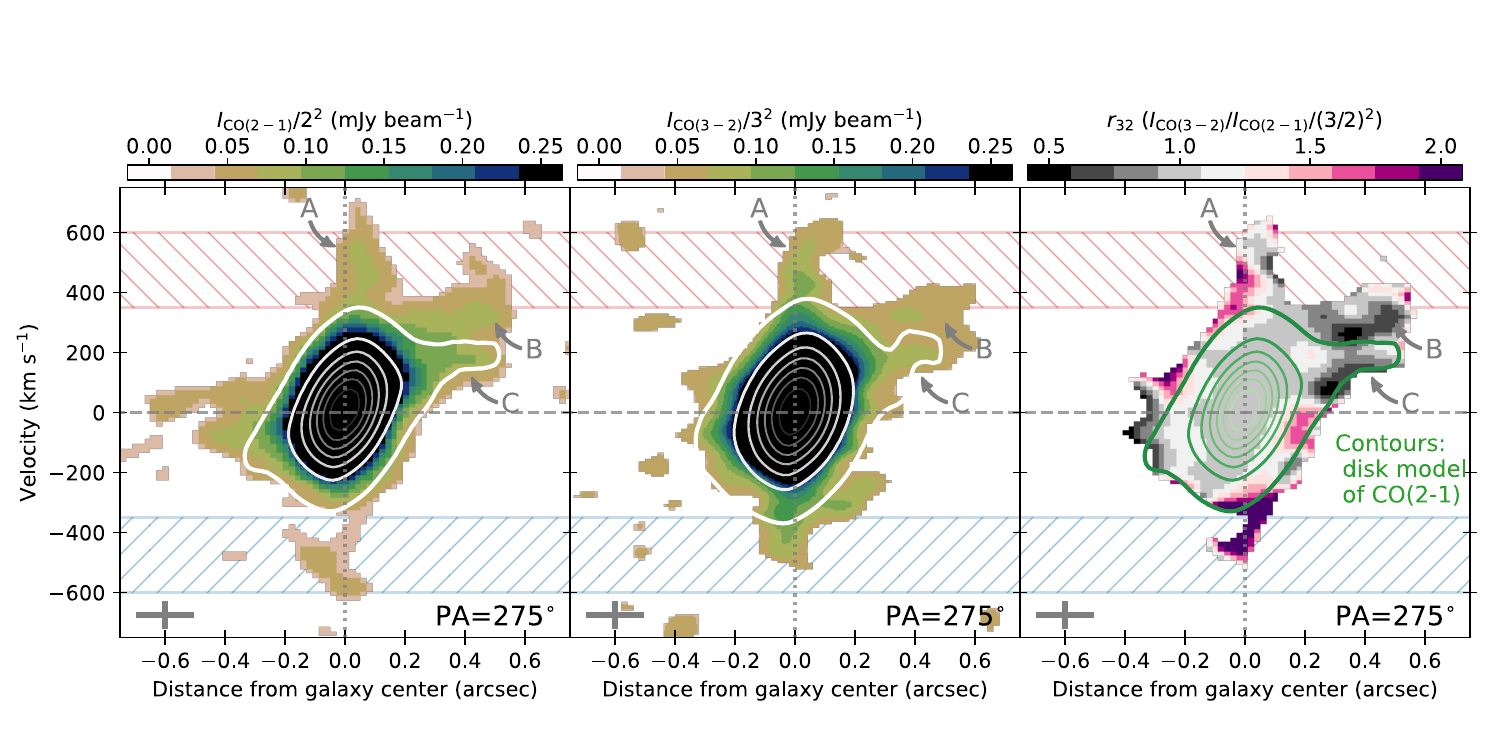}
		\includegraphics[trim=0  0 0 90, clip, width=0.62\textwidth]{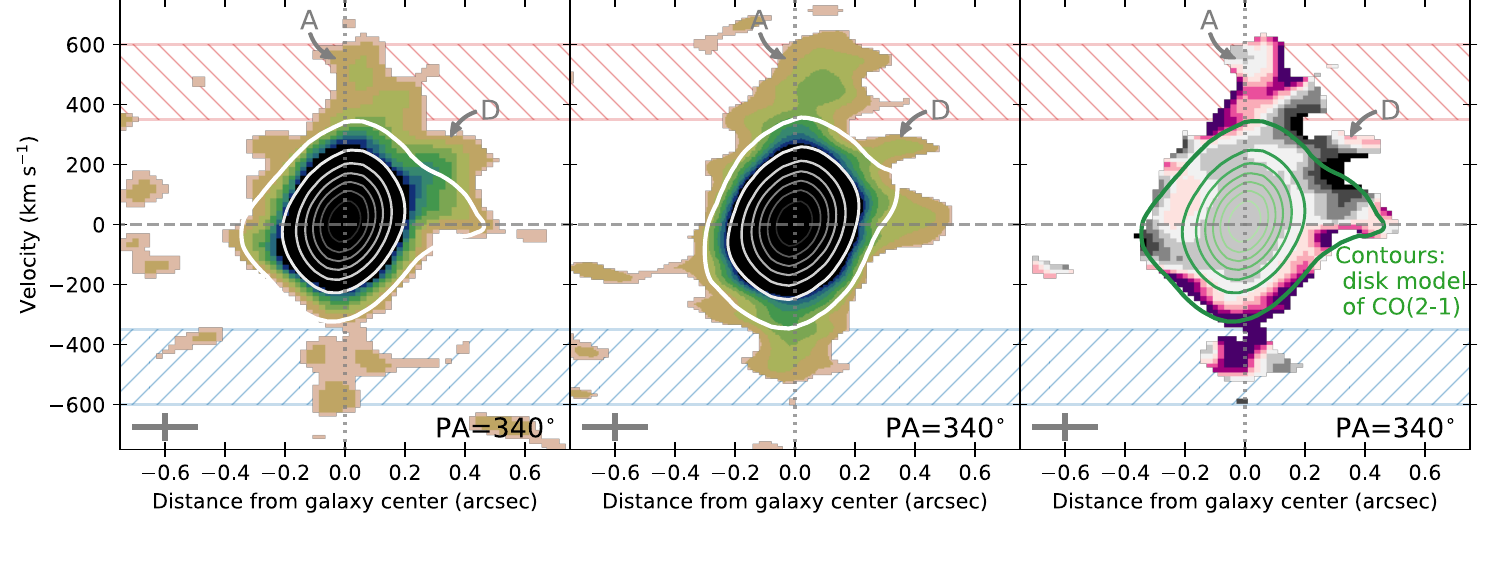}
	\end{center}
	\vspace{-9mm}
	\caption{
	\textbf{Top:} Position-velocity (PV) diagrams of CO(2-1) (left), CO(3-2) (middle), and the line ratios (right). 
	A positive $x$-axis value shows the direction towards PA = $275^\circ$ and a negative value towards PA = $95^\circ$. 
	In the left and middle panels, the map shows the spectral specific intensity of CO lines
	averaged within a width of $0.1\arcsec$. 
	Only spaxels with S/N $>$ 2 are plotted.
	The contours in the left and middle panels show the best-fit rotating disk models of CO(2-1) and CO(3-2), respectively. 
	The contours in the right panel shows the disk model of CO(2-1). 
	The grey cross in the bottom-left corner represents the spectral (50 \kms) and spatial resolution ($\sim$0.2\arcsec). 
	The red and blue hatches denote the velocity ranges 
	of outflow intensity maps in Figure \ref{fig:J1126_ALMA_mCO_outflow}. 
	\textbf{Bottom:} The same PV diagrams for PA of $340^\circ$ (positive $x$-axis) and $160^\circ$ (negative $x$-axis). 
	The legends are the same as those in top panels.
	The letter A--D correspond to the regions marked in Figure \ref{fig:J1126_ALMA_CO21r_vel_wins}. 
	See text in Section \ref{subsubsec:ALMA_CO_PV_ratio} for detailes. 
	}
	\label{fig:J1126_ALMA_mCO_PV}
\end{figure*}

\begin{figure*}
	\begin{center}
		\includegraphics[trim=0 0 0 45, clip, width=0.62\textwidth]{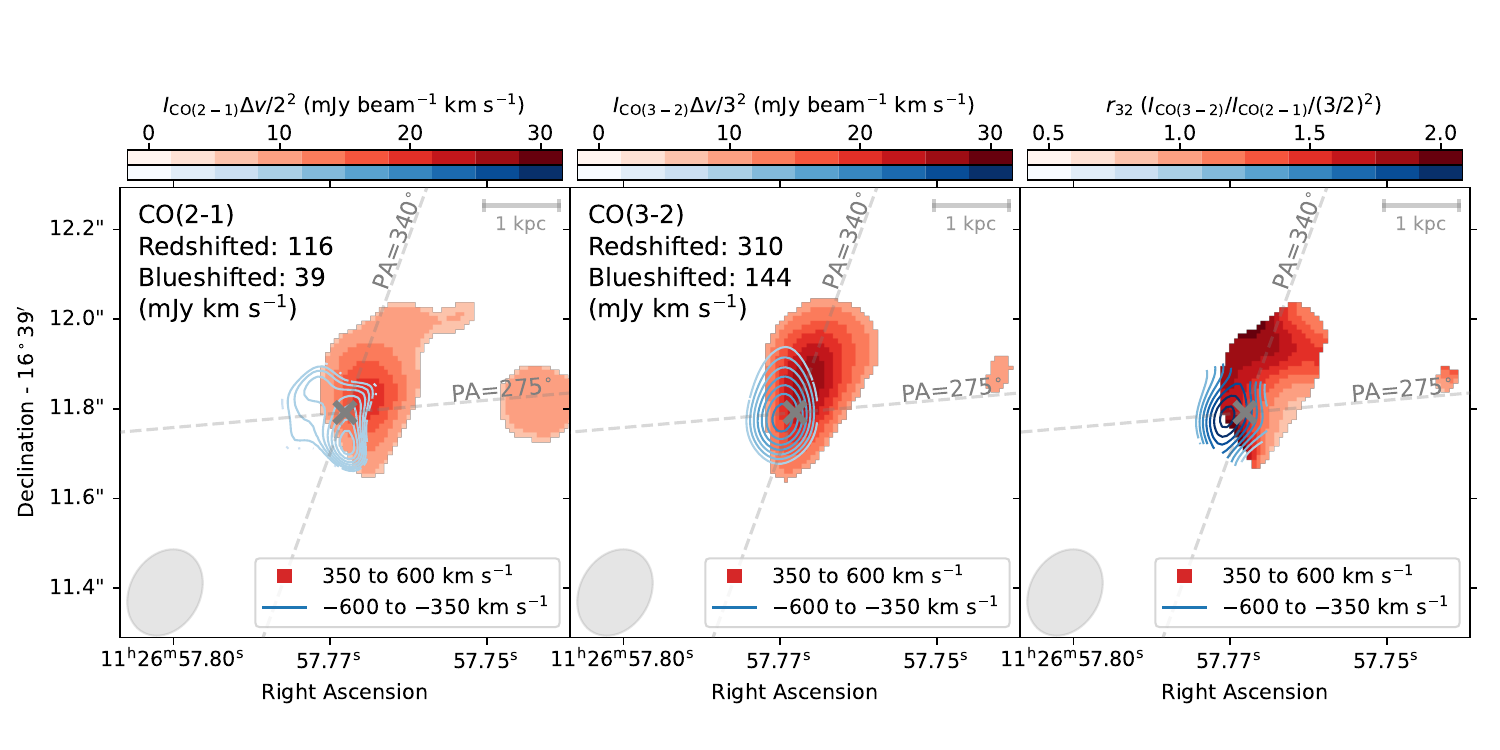}
	\end{center}
	\vspace{-9mm}
	\caption{
	Intensity maps of outflow component of CO(2-1) (left) and CO(3-2) (middle), and the line ratio maps (right). 
	The redshifted outflow is shown in red colored map while the blueshifted one shown in blue contours. 
	The fluxes extracted in pixels with S/N $>$ 3 of CO(2-1) and CO(3-2) are shown in the upper-left corner of each panel. 
	The dashed lines denote the PA of the PV diagrams shown in Figure \ref{fig:J1126_ALMA_mCO_PV}.
	The beam is shown in grey ellipses and the galaxy center in grey crosses.
	}
	\label{fig:J1126_ALMA_mCO_outflow}
\end{figure*}

The relation between molecular outflow
and an enhanced CO excitation
is shown in the integrated CO lines of J1126
(Figure \ref{fig:J1126_ALMA_mCO_spec}, left-lower panel)
and has been reported in local (U)LIRGs \citep{Montoya2024}.
In order to reduce the possible contamination by 
other irregular gas motion (e.g., tidal tails) 
in the spatial identification of the outflow, 
we combine the method of non-rotating motion with 
the analysis of spatially resolved CO line ratios
using the position-velocity (PV) diagrams.

The PV diagrams are plotted following two lines (i.e., the dashed lines in Figure \ref{fig:J1126_ALMA_CO21r_vel_wins})
that passes through the galaxy center with a width of $0.1\arcsec$. 
One line with PA of $340^\circ$ traces 
the bulk of outflow with 300--500 \kms\ (i.e., the region ``A'' in Figure \ref{fig:J1126_ALMA_CO21r_vel_wins})
and a low-velocity (100--300 \kms) non-rotating residual region extended to the north (i.e., the region ``D''). 
The line also traces the 
the main direction of the blueshifted ionized outflow
(e.g., Figure \ref{fig:J1126_GMOS_OIII}). 
The other line with PA of $275^\circ$ is used to 
understand the possible association between 
the possibly extended outflow in the east, i.e., the region ``B'' in the 300--500 \kms\ map, 
and the spiral arm-like extension of the rotating disk, i.e., the region ``C'' in the 100--300 \kms\ map. 
The second line is close to the major axis of the molecular gas disk 
(PA $\sim280^\circ$, see Figure \ref{fig:J1126_ALMA_mCO_mom012}).

The PV diagrams of the beam and MRS-matched CO(2-1) and CO(3-2) cubes 
as well as their line ratio 
are shown in Figure \ref{fig:J1126_ALMA_mCO_PV}. 
The best-fit rotating disk models for both of CO(2-1) and CO(3-2) observations 
are over-plotted as contours in Figure \ref{fig:J1126_ALMA_mCO_PV}. 
The rotating disk model correlates well with $r_{32}=1.0$ 
in the central ($r<0.2\arcsec$), low-velocity (\voffabs\ $<$ 200 \kms) regions. 
The spiral arm-like extension of the disk, e.g., the region ``C'', tends to show weaker excitation with $r_{32}\sim0.6$. 
On the contrary, the non-rotating components in the central ($r<0.1\arcsec$), high-velocity (\voffabs\ $>$ 300 \kms) regions
show stronger excitation with $r_{32}>1.5$. 
These intense excitation regions extend to a large radius (0.2\arcsec--0.3\arcsec)
as velocity decreases (e.g., due to deceleration as gas travels to the outside)
and are mixed with the edges of the rotating disk in the PV diagram. 

Although showing a non-rotating motion, 
the extended regions ``B'' and ``D'' mainly
correspond to a low $r_{32}$ of 0.5--1.0, 
which is similar to $r_{32}$ of the spiral arm-like region ``C''
and lower than that of the central outflow. 
The low $r_{32}$ indicates that 
the extended non-rotating residuals
can be dominated by non-outflow phenomena (e.g., tidal features). 

We combine the method of non-rotating motion and enhanced CO line ratio to identify a ``pure'' molecular outflow
by adopting a velocity threshold of 350 \kms, 
with which only the non-disk residuals with high $r_{32}$ values are selected. 
We note that there are several limitations of the treatment. 
For example, due to the relatively low S/N of CO(3-2) observation, 
the line ratio cannot traces all of the extended regions in CO(2-1) observation. 
Furthermore, the relation between a high CO line ratio and the molecular outflow
is not fully determined, it is also possible that a high CO excitation
only occurs in the outflow close to the galaxy center or the ionizing zone where the gas temperature is high. 

The maps of the ``pure'' outflow component (\voffabs\ $>$ 350 \kms)
of CO(2-1) and CO(3-2) and their line ratio
are shown in Figure \ref{fig:J1126_ALMA_mCO_outflow}. 
The upper bound of the velocity range of the maps, 600 \kms, is set based on 
the central outflow components in the PV diagrams to reduce the data points with a low S/N. 
On the redshifted side, 
both of CO(2-1) and CO(3-2) show an extended outflow
up to a radius of 1.5 kpc towards PA = $340^\circ$.
CO (2-1) has an extension up to $r$ = 3 kpc to the east (PA $\sim275^\circ$), 
while only a clump is shown in this region of the CO(3-2) map. 
The blueshifted outflows are fainter and more compact than the redshifted ones
for both of CO(2-1) and CO(3-2), with a size similar to the beam size. 

\begin{figure}
	\begin{center}
		\hspace{-2mm}
		\includegraphics[trim=4 0 176 40, clip, width=0.98\columnwidth]{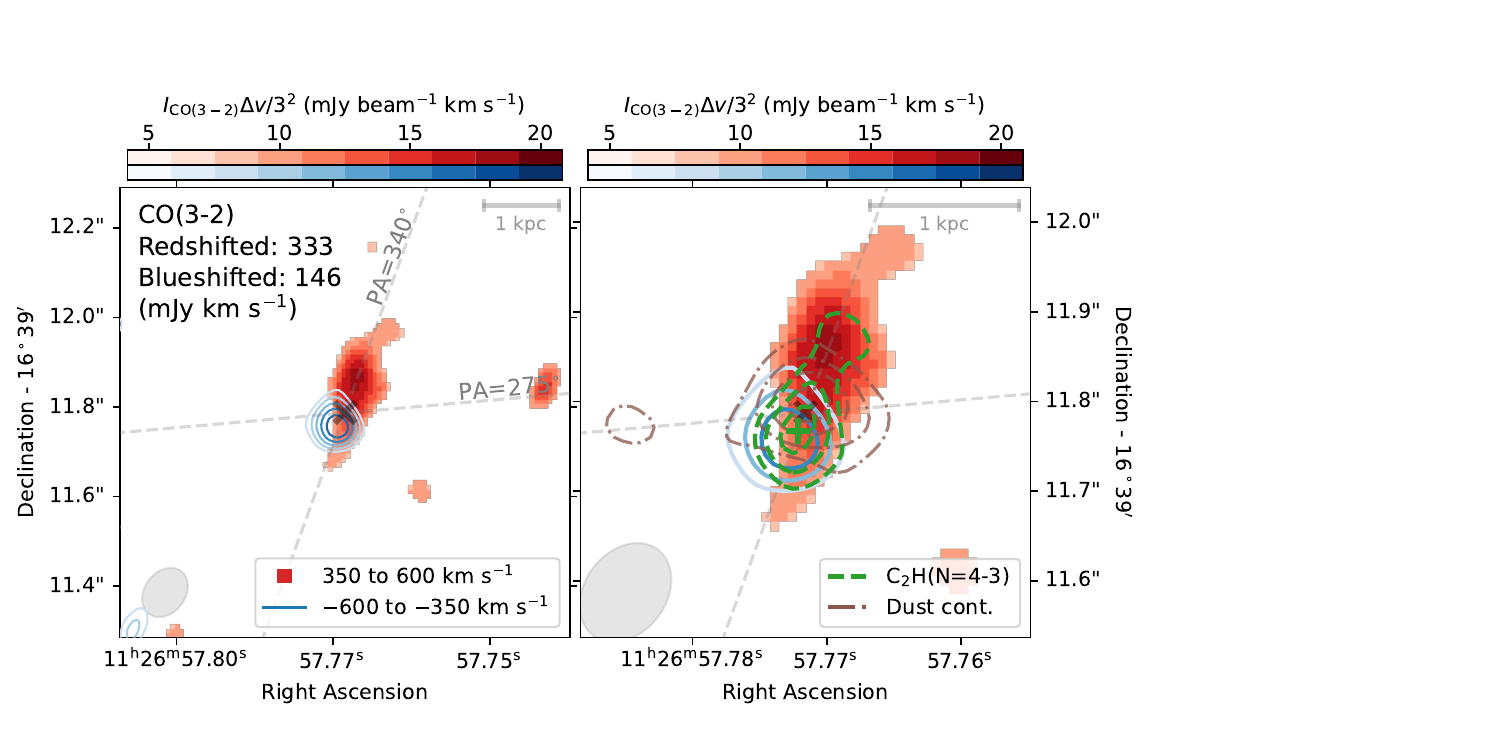}
	\end{center}
	\vspace{-9mm}
	\caption{
	\textbf{Left:}
	Intensity maps of outflow component of CO(3-2) shown in its original spatial resolution ($\sim0.6$ kpc). 
	The legends are the same as those of Figure \ref{fig:J1126_ALMA_mCO_outflow}. 
	\textbf{Right:}
	A zoom-in window of the left panel in the center with a width of 3 kpc.
	The brown dash-dotted and green dashed contours show the distribution of the dust continuum at rest 890 \micron. 
	and the C$_2$H(N=4-3) line, respectively, with S/N $>3$. 
	The green cross denotes the center of C$_2$H and 
	the grey cross shows the center of CO(3-2) moment-0 map (Figure \ref{fig:J1126_ALMA_mCO_mom012}). 
	The frequency integration range of the C$_2$H contours is shown in Figure \ref{fig:J1126_ALMA_C2H_spec}. 
	}
	\label{fig:J1126_ALMA_CO32u_outflow}
\end{figure}

Finally, we also show the outflow of CO(3-2) in its original spatial resolution, 
i.e., $\sim$0.6 kpc, in Figure \ref{fig:J1126_ALMA_CO32u_outflow}. 
The main redshifted outflow in the center is better resolved
with a cone-like structure extended to the north following PA = $340^\circ$; 
while the blueshifted outflow is still not resolved, suggesting a compact distribution. 
We discuss the possible mechanism behind the asymmetric outflow structure in Section \ref{subsec:Discuss_outflow_driven}. 

\begin{figure}
	\begin{center}
		\includegraphics[trim=0 0 -30 -10, clip, width=0.9\columnwidth]{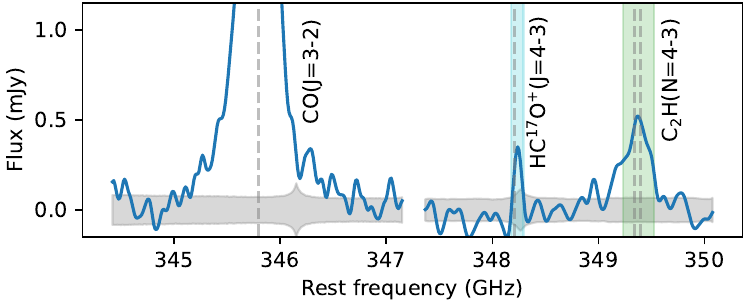}
	\end{center}
	\vspace{-6mm}
	\caption{
	Continuum-subtracted spectrum extracted in the central region ($r<0.25$ kpc). 
	The rest frequencies of CO(J=3-2), HC$^{17}$O$^+$(J=4-3), and C$_2$H(N=4-3) 
	are shown in vertical dashed lines, in which C$_2$H has two rest frequencies due to fine structure. 
	The frequency integration ranges of HC$^{17}$O$^+$(J=4-3) and C$_2$H(N=4-3) 
	are shown in cyan and green hatched regions, which are used to create the intensity plots
	shown in Figure \ref{fig:J1126_ALMA_dust_continua} and \ref{fig:J1126_ALMA_CO32u_outflow}. 
	The spectra are convolved by a Gaussian profile with FWHM of 50 \kms.
	The 1-$\sigma$ measurement error is shown in grey regions. 
	}
	\label{fig:J1126_ALMA_C2H_spec}
\end{figure}


\subsection{Association between CO-traced molecular outflow and the ethynyl radical intensity}
\label{subsec:ALMA_C2H}

The spectral setup of the ALMA Cycle 9 observations also covers 
the rotational transition of the ethynyl radical, C$_2$H(N=4-3) (hereafter C$_2$H(4-3)), 
with its spectrum shown in Figure \ref{fig:J1126_ALMA_C2H_spec}. 
C$_2$H(4-3) is split into two fine structure lines
with rest frequencies of 349.34 (J=9/2-7/2) and 349.40 GHz (J=7/2-5/2),
and each of them is split into two hyperfine structure lines\footnote{
	The hyperfine structure of C$_2$H(N=4-3) offset by 1 MHz is not resolved in the current observation. 
}, due to the spin-orbit and electron-nucleus interactions. 
Considering the complicated line structure and a moderate S/N of the detection, 
we do not perform a spatially-resolved fitting as that conducted for CO(3-2). 
A single-Gaussian fitting for the line profile extracted in a beam-size region in the center ($r<0.3$ kpc)
results in a FWHM of 200 \kms\ with a shift of $-10$ \kms\ from the mean rest frequency of the two fine structure lines. 
The intensity map of C$_2$H(4-3) integrated in the velocity range of $\pm100$ \kms\ 
is shown in Figure \ref{fig:J1126_ALMA_CO32u_outflow}, which
shows an alignment to the CO-traced molecular outflow, 
e.g., a tail extended towards the north (PA = $340^\circ$) as the redshifted outflow of CO(3-2). 
The peak of the C$_2$H(4-3) intensity is tilted to the south, which is likely correlated with the compact blueshifted outflow of CO(3-2). 

The association between the molecular outflow and the ground transition of the ethynyl radical, C$_2$H(N=1-0), 
in the nearby Seyfert galaxy NGC 1068 is discussed in \cite{Garcia2017} and \cite{Saito2022PC}. 
The abundance of C$_2$H is found to be enhanced in the outflow, 
which can be produced by the increased number of carbon atoms and ions
released by disassociated CO molecules.
The CO disassociation is possibly driven by the intense UV and X-ray radiation, 
and/or the shock-induced icy mantle spluttering as outflow interacts with ISM. 
These studies support the association between the C$_2$H(4-3) emission and molecular outflow in J1126. 

The C$_2$H(4-3) transition is reported in several nearby (U)LIRGs, e.g., IRAS 20551-4250 \citep{Imanishi2013}, 
NGC 4418 and Arp 220 \citep{Sakamoto2021}. 
\cite{Imanishi2013} argued a contribution by the rotational transition of methyl cyanide, CH$_3$CN(J=19-18) at the frequency. 
CH$_3$CN(J=19-18) has a higher dipole moment (i.e., a higher Einstein coefficient thus higher critical density) 
and a higher excitation temperature ($>200$ K vs. 42 K) than C$_2$H(4-3). 
The rotationally excited CH$_3$CN emission is known as a tracer of dense and warm gas 
in the massive star-forming regions in the Galactic Center region \citep[e.g.,][]{Churchwell1983}
and starburst galaxies \citep[e.g.,][]{Martin2011}. 
As for the line detected at the frequency 
in J1126, since its intensity map shows an alignment with
the CO(3-2)-traced outflow other than the continuum emitted from the dusty starburst (Figure \ref{fig:J1126_ALMA_CO32u_outflow}), 
we consider that C$_2$H instead of CH$_3$CN dominants the detection at the frequency. 

There is another faint emission line in the neighbor of CO(3-2) and C$_2$H(4-3), i.e., 
the rotationally excited HC$^{17}$O$^+$(J=4-3) with a rest frequency of 348.21 GHz\footnote{
	The hyperfine structure of HC$^{17}$O$^+$(J=4-3) is not resolved in the current observation. 
}. As an isotopomer of HC$^{16}$O$^+$(J=4-3), HC$^{17}$O$^+$(J=4-3) traces 
dense molecular gas ($2\times10^6$ \ccm; e.g., \citealt{Greve2009}). 
The spectrum and the intensity map integrated with \voffabs\ $<$ 40 \kms\ 
are shown in Figure \ref{fig:J1126_ALMA_C2H_spec} and \ref{fig:J1126_ALMA_dust_continua}, respectively, 
which is only detected in the galaxy center. 
Unfortunately, the marginal detection of this line prevents us from performing a detailed analysis of it. 


\section{Discussions}
\label{sec:Discussion}

\subsection{Comparison of the velocity and spacial extent of the ionized and molecular outflows}
\label{subsec:Discuss_outflow_comp}

\begin{figure*}
	\begin{center}
		\includegraphics[trim=0 0 -32 0, clip, width=0.9\columnwidth]{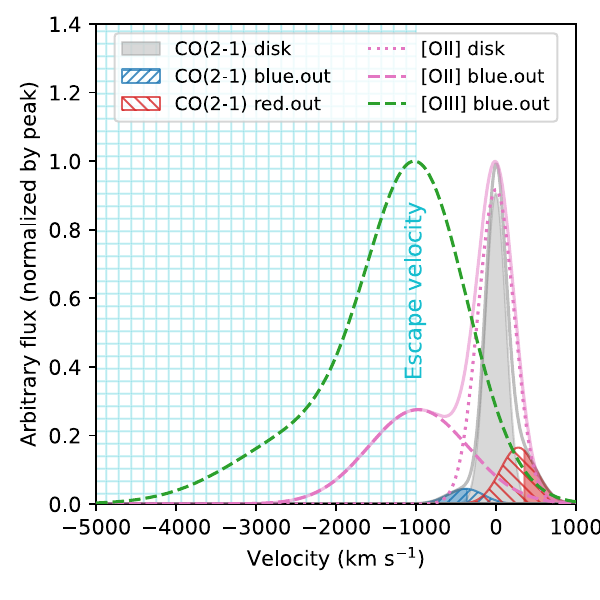}
		\includegraphics[trim=0 0 -32 0, clip, width=0.9\columnwidth]{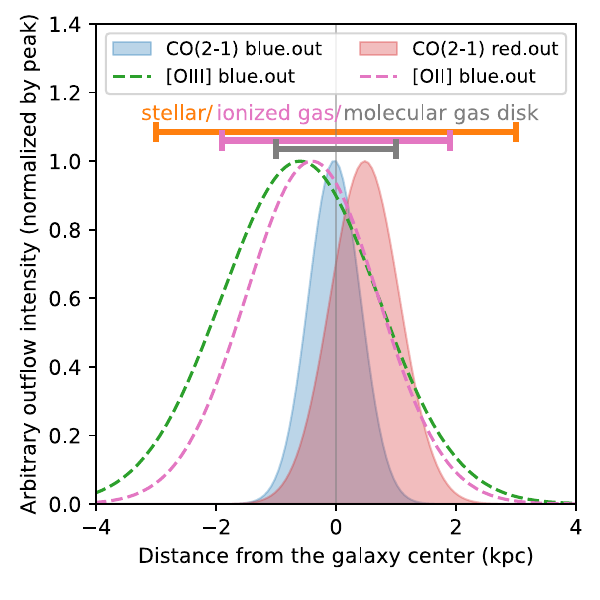}
	\end{center}
	\vspace{-9.5mm}
	\caption{
		\textbf{Left:}
		Comparison of the best-fit spectral profiles of ionized gas emission lines, \oiii\ and \oii, 
		and molecular gas emission line, CO(2-1). 
		All of the profiles are extracted from the entire galaxy
		and normalized at the peak of the total line profiles. 
		The profiles of \oiii\ (green) and \oii\ (purple) 
		have been corrected for the instrumental broadening of GMOS ($\sim300$ \kms). 
		The disk and outflow components of \oii\ are shown in dotted and dashed curves, respectively. 
		The best-fit disk component of CO(2-1) is shown in grey, 
		while the blue- and redshifted outflows are shown in blue and red hatches, respectively. 
		The CO(2-1) outflows with a velocity threshold of $\pm350$ \kms\ 
		are shown as blue and red filled regions to show a ``pure'' molecular outflow (Section \ref{subsubsec:ALMA_CO_PV_ratio}). 
		The cyan hatch denotes the velocity range which exceeds the escape velocity of the host galaxy. 
		\textbf{Right:}
		Comparison of the one-dimensional spatial profiles of the outflows traced by 
		\oiii\ (green dashed), \oii\ (purple dashed), and CO(2-1) (blue and red filled regions). 
		A positive value of the $x$-axis represents the distance from the galaxy center towards the north with PA = $340^\circ$. 
		All of the profiles have been corrected for the PSF blurring 
		and normalized at the peak of the outflow intensity profiles. 
		The PSF-corrected effective (i.e., half-light) diameters of the stellar, ionized gas, and molecular gas disks
		are shown in orange, purple, and grey horizontal lines, respectively, for a comparison with the outflows. 
	}
	\label{fig:J1126_comp_outflow}
\end{figure*}

\begin{figure}
	\begin{center}
		\includegraphics[trim=0 0 -32 0, clip, width=0.9\columnwidth]{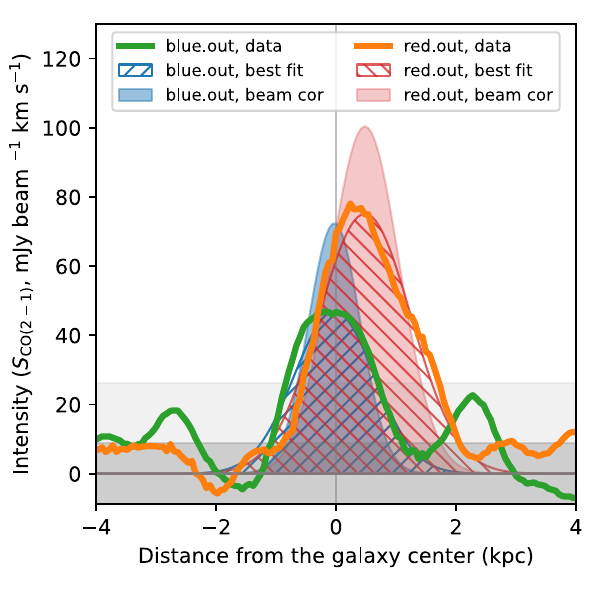}
	\end{center}
	\vspace{-9.5mm}
	\caption{
	One-dimensional spatial profiles of the CO(2-1) outflows, which are 
	integrated in [$-600$, $-350$] \kms\ and [350, 600] \kms\ on the blue- and redshifted sides, respectively.
	The observed data are shown in green and orange for the blue- and redshifted outflows, respectively.
	The best-fit profiles are marked with hatches and the beam-corrected profiles with filled regions.  
	The $\pm1\sigma$ and $\pm3\sigma$ error ranges are shown in dark and light grey, respectively. 
	}
	\label{fig:J1126_ALMA_CO21u_outflow_size}
\end{figure}

The spectra of \oiii\ and \oii\ as examples of high- and low-IP ionized emission lines,
as well as CO(2-1) that traces the neutral molecular gas, 
are shown in Figure \ref{fig:J1126_comp_outflow} (left) for a direct comparison between the multi-phase outflows. 
All of the spectra are extracted from the entire galaxy and normalized at the peak of the entire line profiles. 
The best-fit profiles instead of the observed data are shown to 
correct for the instrumental broadening and the contamination of neighboring lines.
The original CO(2-1) cube is used to obtain the entire CO(2-1) flux. 

The spectra of the ionized and molecular outflows show significant differences
in the velocity range and flux fraction of the entire line profile. 
\oiii\ shows an almost fully blueshifted line profile
with a maximum velocity ($|v_{10}|$) of 2400 \kms. 
60\% of the \oiii\ flux exceeds the escape velocity ($v_\text{esc}$) of the host galaxy, $\sim1000$ \kms, 
which is estimated following the method in \cite{Chen2020}. 
\oii\ shows a mixed contribution of host disk (i.e., narrow component)
and outflow components. 
The \oii\ outflow contributes to 45\% of the entire flux 
and 49\% of the \oii\ outflow flux exceeds the escape velocity. 
On the contrary, CO(2-1) shows a much lower outflow fraction, $<10\%$, 
and a significantly lower outflow velocity, $\sim500$ \kms, on both of blue- and redshifted sides 
(for the ``pure'' molecular outflow discussed in Section \ref{subsubsec:ALMA_CO_PV_ratio}). 
The results reveal that
a large fraction of the ionized gas outflow can escape from the gravitational potential of the host galaxy, 
while the bulk of the molecular gas has not been accelerated by the outflow
and even the swept-out molecular gas could not escape from the galaxy
\citep[e.g.,][]{Toba2017b}.

It is also worthy to compare the spatial distributions of the ionized and molecular outflows. 
Since the PSF of the GMOS observation ($\sim0.7\arcsec$) is much larger than that of 
the ALMA observation ($\sim0.2\arcsec$ for CO(2-1)), 
it is hard to directly compare the morphologies of the multi-phase outflows. 
Similar to the treatment for the spectral profiles discussed above, 
we perform a Gaussian fitting to the observed spatial profiles
and then correct for the PSF blurring in the best-fit profiles. 
For the sake of simplicity, the profile fitting and PSF correction are employed 
in one-dimension following the primary direction of the ionized and molecular outflows, i.e., PA of $160^\circ$ and $340^\circ$.
The spatial profile fitting of the CO(2-1) outflow
is shown as an example in Figure \ref{fig:J1126_ALMA_CO21u_outflow_size}.
The observed one-dimensional profiles are obtained from original CO(2-1) cube, 
which are integrated in [$-600$, $-350$] \kms\ and [350, 600] \kms\ on the blue- and redshifted sides, respectively, 
and averaged in a width of $0.1\arcsec$ following PA of $160^\circ$ and $340^\circ$. 
The best-fit profiles after correction for the beam blurring ($\sim1$ kpc)
(filled regions in Figure \ref{fig:J1126_ALMA_CO21u_outflow_size})
are used to represent the intrinsic extent 
of the molecular outflows on the blue- and redshifted sides. 
The PSF-corrected profiles of the ionized outflows traced with broad \oiii\ and \oii\ lines
are obtained with the same method and the velocity ranges shown in the left panel of Figure \ref{fig:J1126_comp_outflow}. 
The corrected profiles of the ionized and molecular outflows 
are shown in the right panel of Figure \ref{fig:J1126_comp_outflow} (normalized at each profile peak).
The effective (i.e., half-light) radii of the stellar, 
ionized gas, and molecular gas disks
are also shown for a comparison between the sizes of outflow and disk components. 
The \oiii\ outflow has a shifted center with a distance 
($d_\text{cen}$; a negative $d_\text{cen}$ represents the distance towards the south) 
of $-0.7$ kpc 
and an extent ($\Delta d_\text{\tiny FWHM}$) of 3.2 kpc.
The \oii\ outflow shows a similar distribution with $d_\text{cen}=-0.5$ kpc and $\Delta d_\text{\tiny FWHM}$ = 2.8 kpc. 
Both of the high- and low-IP ionized outflows 
show an extended distribution in a galaxy-scale, i.e., comparable to the effective radii of the stellar or ionized gas disks. 
The blueshifted CO(2-1) outflow shows a compact size 
($\Delta d_\text{\tiny FWHM}$ = 1.1 kpc) 
with its center close to the center of the molecular gas disk ($d_\text{cen}$ = $-0.1$ kpc). 
Note that although the blueshifted ionized and molecular outflows have different extents, 
they all show small ratios of 
$|d_\text{cen}|/\Delta d_\text{\tiny FWHM}\sim$ 0.1--0.2, 
which suggest a nearly face-on view of the outflow cone on the blueshifted side. 
The different $d_\text{cen}$ and $\Delta d_\text{\tiny FWHM}$ of the ionized and molecular outflows
indicate different positions on the outflow cone, i.e., 
the molecular outflow locates at the bottom and close to the disk center,
while the ionized outflow can travels to a larger distance with a wider distribution. 

On the redshifted side only the molecular outflow is detected
due to the shielding from the galaxy disk of the UV/optical emissions. 
The PSF-corrected profile of the redshifted outflow has 
$d_\text{cen}$ = 0.5 kpc and $\Delta d_\text{\tiny FWHM}$ = 1.5 kpc. 
A larger ratio, $d_\text{cen}/\Delta d_\text{\tiny FWHM}$ = 0.3, 
of the redshifted outflow is consistent with its elongated morphology 
towards the north (e.g., Figure \ref{fig:J1126_ALMA_mCO_outflow}). 

\begin{figure}
	\vspace{2mm}
	\begin{center}
		\includegraphics[trim=0 0 0 0, clip, width=0.9\columnwidth]{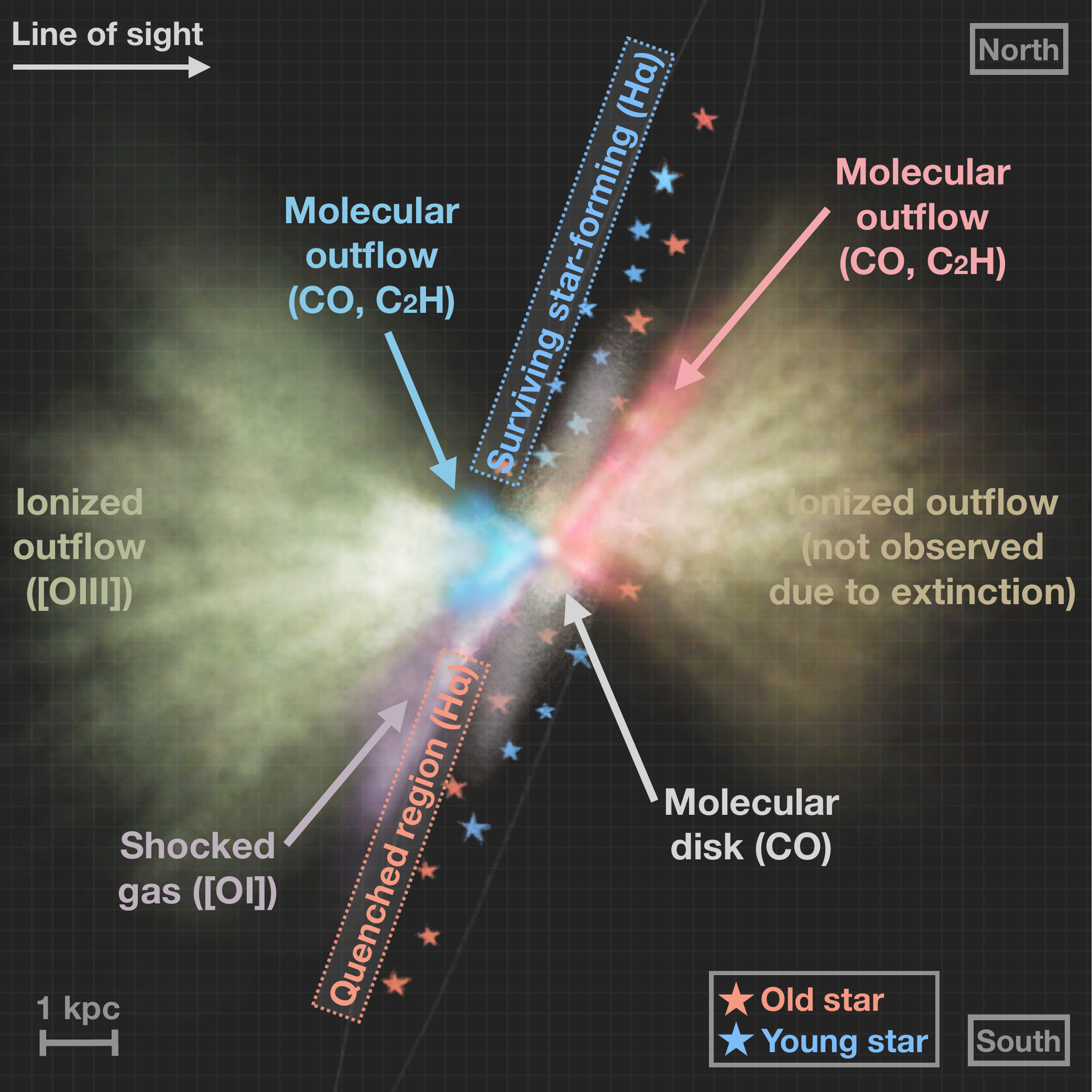}
	\end{center}
	\vspace{-4mm}
	\caption{
	A schematic view of the multi-phase outflows of J1126 in the section along PA of $160^\circ$ (south) and $340^\circ$ (north). 
	The inclination of the molecular disk plane (grey) is estimated from the Barolo disk model (Figure \ref{fig:J1126_ALMA_CO21r_vel_wins}). 
	The ionized outflow (green fluid) and the shock ionization (violet fluid) regions on the blueshifted side 
	(i.e., near side to the observer) are estimated
	with the \oiii\ intensity map (Figure \ref{fig:J1126_GMOS_OIII}) 
	and the \oi-traced ionization map (Figure \ref{fig:J1126_GMOS_BPT_map}), respectively.
	The molecular outflows on the two sides (blue and red fluids) are estimated with the CO outflow intensity maps
	(Figure \ref{fig:J1126_ALMA_mCO_outflow}). 
	The multi-shell biconical outflow structure is discussed in Section \ref{subsec:Discuss_outflow_driven}. 
	A star-forming region survives in the north (near side), 
	which is not significantly affected by the outflow; 
	while the star formation in the southern outflow region (near side) is suppressed
	(Figure \ref{fig:J1126_GMOS_BPT_map} and \ref{fig:J1126_SFH_sigmaSFR}). 
	}
	\label{fig:J1126_scheme}
\end{figure}

A schematic plot of the spatial structure of the multi-phase outflows are shown in Figure \ref{fig:J1126_scheme}
in a section along the primary outflow direction. 
In the scheme the inclination of the galaxy disk is estimated 
with the Barolo fitting of CO(2-1) cube
(Figure \ref{fig:J1126_ALMA_CO21r_vel_wins}). 
An extended ionized outflow and a compact molecular outflow 
are plotted on the blueshifted side (i.e., near side to the observer). 
The redshifted molecular outflow could have an orientation 
close to the disk plane to form a cone-like outflow cavity 
(e.g., Figure \ref{fig:J1126_ALMA_mCO_outflow}).
We estimate the dynamical properties of the outflows in Section \ref{subsec:Discuss_outflow_dynamics} 
and discuss the possible mechanisms to launch and shape the multi-phase outflows in Section \ref{subsec:Discuss_outflow_driven}. 


\subsection{Estimation and comparison of dynamical quantities of the multi-phase outflows}
\label{subsec:Discuss_outflow_dynamics}

We start the analyses of the dynamical quantities from the estimation of gas mass
in the ionized and molecular outflows. 
The mass of the ionized outflow is estimated using the broad components of the \ha, \hb, 
and the combination of \oiiblong\ and \oiiiblong\ lines following the method 
adopeted in \cite{Chen2019} (see also \citealt{Carniani2015}). 
The ionized gas mass from the \ha\ emission can be estimated as:
\begin{equation}\label{equ:Mout_ha}
    \frac{ M_{\rm ion} }{10^8\ \rm M_{\odot} } = 
    \frac{3.33 L_{\rm H\alpha}}{10^{43}\ \rm {erg\ s^{-1}} } \left( \frac{n_{\rm e}}{100\ \rm{cm^{-3}} } \right)^{-1}.
\end{equation} 
A 10\% contribution of helium in the ion number density is assumed in the equation. 
The equation based on \hb\ can be obtained replacing the factor 3.33 in Equation \ref{equ:Mout_ha} with 9.52
following the same assumptions. 
Similarly, we can obtain the estimated ionized gas from the oxygen lines as:
\begin{equation}\label{equ:Mout_oiii}
    \frac{ M_{\rm ion} }{10^8\ \rm M_{\odot} } = 
    \frac{0.88 L_{\rm [O II]} + 0.46 L_{\rm [O III]} }{10^{43}\ \rm {erg\ s^{-1}} } \left( \frac{n_{\rm e}}{100\ \rm{cm^{-3}} } \right)^{-1}.
\end{equation}
In this equation we assume the oxygen ions are mainly in the singly and doubly ionized phases
with a solar oxygen abundance in the ionized gas. 
Solar metallicity is observed as
a typical value in local (U)LIRGs \citep[e.g.,][]{Perez2021}.
The dust extinction corrected luminosity of the broad \ha, \hb, \oii\ and \oiii\ lines 
are used to estimate the ionized gas mass, 
where the extinction, $A_{V}=1.7$, 
is estimated using the broad \ha/\hb\ ratio (Section \ref{subsec:GMOS_extinction}). 
The electron density of the outflow\footnote{
	The electron density of the host disk is estimated to be $370\pm100$ \ccm\ with the ratio of the narrow \sii\ doublets, 
	$f_{6716}/f_{6731}=1.05\pm0.07$. 
}, $n_\text{e}$ = $200\pm150$ \ccm, 
is estimated with the ratio of the broad \sii\ doublets of the integrated spectrum ($f_{6716}/f_{6731}=1.18\pm0.15$)
and the calibration equation of \cite{Proxauf2014}.
The derived ionized gas mass in the outflow is 
$1.8\times10^8$ \msun, $1.8\times10^8$ \msun, and $1.6\times10^8$ \msun, 
based on \ha, \hb, and oxygen lines, respectively. 
We adopt $M_\text{ion}=1.8\times10^8$ \msun\ in later analysis. 

The molecular gas mass in the outflow can be estimated with the equation:
\begin{equation}\label{equ:Mout_CO21}
    M_{\rm mol} = \alpha_\text{CO} \frac{L^\prime_\text{CO(2-1)}}{r_{21}},
\end{equation}
where $L^\prime_\text{CO(2-1)}$ is the luminosity of the outflow component of CO(2-1) 
in the unit of K \kms\ pc$^2$;
$r_{21}$ is the brightness ratio of CO(2-1) over CO(1-0);
$\alpha_\text{CO}$ is the mass conversion factor of CO(1-0) in the unit of \msun\ (K \kms\ pc$^2$)$^{-1}$. 
In this work we assume the CO(2-1) is thermalized with CO(1-0), i.e., $r_{21}=1$, 
which is found as an average value in the nearby (U)LIRGs and AGN host galaxies \citep[e.g.,][]{Montoya2023,Molyneux2024}.
$\alpha_\text{CO}=0.8$ is adopted in the calculation, which is a common assumption 
in the analyses of (U)LIRGs \citep[e.g.,][]{Downes1998,Carilli2013,Cicone2014,Fiore2017,Veilleux2020},
$M_{\rm mol}$ in the outflow of J1126 is estimated to be 
$7.4\times10^8$ \msun\ and $1.7\times10^9$ \msun\ on the blue- and redshifted sides, respectively, 
for the ``pure'' molecular outflow\footnote{
	The estimated gas mass in the redshifted molecular outflow can be twice larger 
	if we also consider the low-velocity non-rotating residuals with \voff\ $<$ 350 \kms.}
with \voffabs\ $>$ 350 \kms\ (see discussion in Section \ref{subsubsec:ALMA_CO_PV_ratio}), 
which contributes to 4\% and 9\% of the total molecular gas mass.  

The time-averaged mass-loss rate of the outflow can be derived as $\dot{M}_\text{out}=M_\text{out}/\Delta t$, 
where $M_\text{out}$ is the gas mass in the outflow as estimated above 
and $\Delta t$ is the timescale of the outflow,
which can be estimated as the ratio of the radius of the outflow over its velocity.
For the ionized outflow traced by \oiii\, 
the simplest way to estimate $\Delta t$ is to use the ratio of the radius and velocity
of the bulk of the profile, i.e., $\Delta d_{\sigma}/v_{50}=1.1$ Myr, where $\Delta d_{\sigma}=\Delta d_\text{\tiny FWHM}/2.355=1.4$ kpc 
is the PSF-corrected dispersion of the outflow spatial profile (Figure \ref{fig:J1126_comp_outflow}, right)
and $|v_{50}|=1210$ \kms\ is the 50th percentile velocity of the line spectral profile (Figure \ref{fig:J1126_comp_outflow}, left).
If we replace $\Delta d_{\sigma}$ with 
the distance of the center of \oiii\ outflow, 
i.e.,  $|d_\text{cen}|=0.7$ kpc, 
the corresponding $\Delta t$ is 0.6 Myr. 
Note that in the above estimation, the inclination of the outflow is not considered, e.g., 
the velocity is measured along the line of sight while the radii are the values projected on the sky plane. 
Considering the nearly face-on morphology of the \oiii\ outflow, 
we assume the expansion of the outflow cone in the direction of the sky plane 
is dominated by the velocity dispersion 
and the dispersion is isotropic and then, 
$\Delta t$ can be estimated as $\Delta d_\text{\tiny FWHM}/\Delta v_{80}=1.3$ Myr, 
where $\Delta v_{80}=2370$ \kms\ is the 80\% width of \oiii\ spectral profile. 
The maximum distance and velocity can also be used to derive the timescales of outflows \citep[e.g.,][]{Fiore2017}. 
If we adopt $v_\text{max}$ as $|v_{10}|=|v_{50}|+\Delta v_{80}/2$ 
following \cite{Chen2020}
and a similar definition of $d_\text{max}$ as $|d_\text{cen}|+\Delta d_\text{\tiny FWHM}/2$, 
the corresponding $\Delta t$ is 0.9 Myr. 

For the molecular outflow traced by CO(2-1), 
$\Delta t$ from the $\Delta d_{\sigma}/|\bar{v}_\text{s}|$ are 1.0 Myr and 1.4 Myr on the blue- and redshifted sides, respectively, 
where $|\bar{v}_\text{s}|\sim450$ \kms\ is the mean velocity of the ``pure'' molecular outflow. 
$\Delta t$ is estimated to be 0.5 Myr using the center radius of the redshifted outflow
(which is not applicable for the blueshifted outflow with a center radius close to zero). 
Simulation studies indicate the bulk of gas mass of the molecular outflow 
locates close to the disk plane \citep[e.g.,][]{Menci2019}, 
therefore the expansion with a spherically symmetric dispersion is likely unreasonable for the molecular outflow
and we do not estimate $\Delta t$ with this scenario. 
Adopting a similar definition of $d_\text{max}$ and $v_\text{max}$ as the ionized outflow, 
$\Delta t$ is estimated to be 1.0 and 2.0 Myr for the blue- and redshifted outflows, respectively, 
with the maximum velocities derived from 
the 3-Gaussian fitting results in the central region ($r<0.5$ kpc, see Table \ref{tab:mCO_gfit}), 
i.e., $v_{10}=-660$ \kms\ on the blueshifted side and $v_{90}=620$ \kms\ on the redshifted side\footnote{
	Note that the low-velocity range ($|v_\text{s}|<350$ \kms) is included in the Gaussian fitting to derive the maximum velocities, 
	while it is not used in the estimation of the gas mass in the ``pure'' molecular outflow. 
}. 

The timescales from the different estimations of the ionized and molecular outflows 
are approximately consistent with 1 Myr, 
implying that the multi-phase outflows could be launched in the same period. 
We adopt $\Delta t = 1$ Myr in the later analysis. 
The corresponding time-averaged mass-loss rate is 
180 \sfrunit\ for the blueshifted ionized outflow; 
740 \sfrunit\ and 1700 \sfrunit\ for the molecular outflow on the blue- and redshifted sides, respectively. 
The results can increase by a factor of 3 if we adopt the definition of the instantaneous outflow rate
assuming a spherical sector with a uniform density \citep[e.g.,][]{Cicone2014,Fiore2017}. 

The momentum rates and the kinetic energy ejection rates (i.e., the kinetic powers) can be derived as
$\dot{P}_\text{out}=\dot{M}_\text{out} v_\text{out}$ and
$\dot{E}_\text{k,out}=\dot{M}_\text{out} v_\text{out}^2 /2$. 
We adopt $v_\text{out}$ as the maximum velocity (i.e., $|v_{10}|$ or $|v_{90}|$)
which is 2400 \kms\ for the ionized outflow traced with \oiii, 
660 \kms\ and 620 \kms\ for the blue- and redshifted molecular outflows traced with CO(2-1), respectively.
With the time-averaged $\dot{M}_\text{out}$, 
the derived $\dot{P}_\text{out}$ and $\dot{E}_\text{k,out}$ of the blueshifted ionized outflow 
are $2.7\times10^{36}$ dyne and $3.3\times10^{44}$ \lumcgs, respectively. 
The $\dot{P}_\text{out}$ and $\dot{E}_\text{k,out}$ of the molecular outflow are
$3.1\times10^{36}$ dyne and $1.0\times10^{44}$ \lumcgs\ on the blueshifted side,
$6.6\times10^{36}$ dyne and $2.1\times10^{44}$ \lumcgs\ on the redshifted side, respectively. 
These results depend on the choice of $\dot{M}_\text{out}$ 
(e.g., an triple increasing using the instantaneous $\dot{M}_\text{out}$), 
and the definition of $v_\text{out}$. 
Taking the ionized outflow as an example, 
the derived $\dot{P}_\text{out}$  
changes by a factor of 1.5 with $v_\text{out}$ defined as $|v_{50}|+2\Delta v_{\sigma}$
(i.e., $|v_{02}|$ or $|v_{98}|$ for a single Gaussian profile; e.g., \citealt{Veilleux2013,Fiore2017});
or by a factor of 0.8 with $v_\text{out}=\Delta v_{80}/1.3$ \citep[e.g.,][]{Liu2013}. 
The derived $\dot{E}_\text{k,out}$
changes by a factor of 0.7 if we adopt 
$\dot{E}_\text{k,out}=\dot{M}_\text{out} (v_{50}^2 /2 + 3\Delta v_{\sigma}^2 /2)$
assuming a spherically symmetric turbulent motion. 

Assuming symmetric ionized outflows on the blue- and redshifted sides,  
the total $\dot{M}_\text{out}$, $\dot{P}_\text{out}$, and $\dot{E}_\text{k,out}$
of the ionized + molecular outflows are
2800 \sfrunit, $1.5\times10^{37}$ dyne, and $9.6\times10^{44}$ \lumcgs, respectively;
or 8400 \sfrunit, $4.5\times10^{37}$ dyne, and $2.9\times10^{45}$ \lumcgs\
if the instantaneous $\dot{M}_\text{out}$ is adopted, 
e.g., to be compared with the results in the literature \citep[e.g.,][]{Cicone2014,Fiore2017}. 
The contribution of the neutral atomic gas is not considered in the estimations. 
Although the absorption lines of neutral sodium at 5896 and 5890\AA\ are covered by the GMOS and SDSS spectra,
they are too faint to be detected 
(e.g., Figure \ref{fig:J1126_GMOS_spec_int}). 
The mass ratio of the outflowing gas in the neutral atomic phase over the ionized phase
varies from unity \citep[e.g., Arp 220;][]{Perna2020} to several tens \citep{Avery2022} in galaxies in the local universe, 
therefore the above estimates of J1126 is considered as a lower estimation 
of the total outflow dynamics. 

\begin{figure}
	\begin{center}
		\includegraphics[trim=0 0 -32 14, clip, width=0.9\columnwidth]{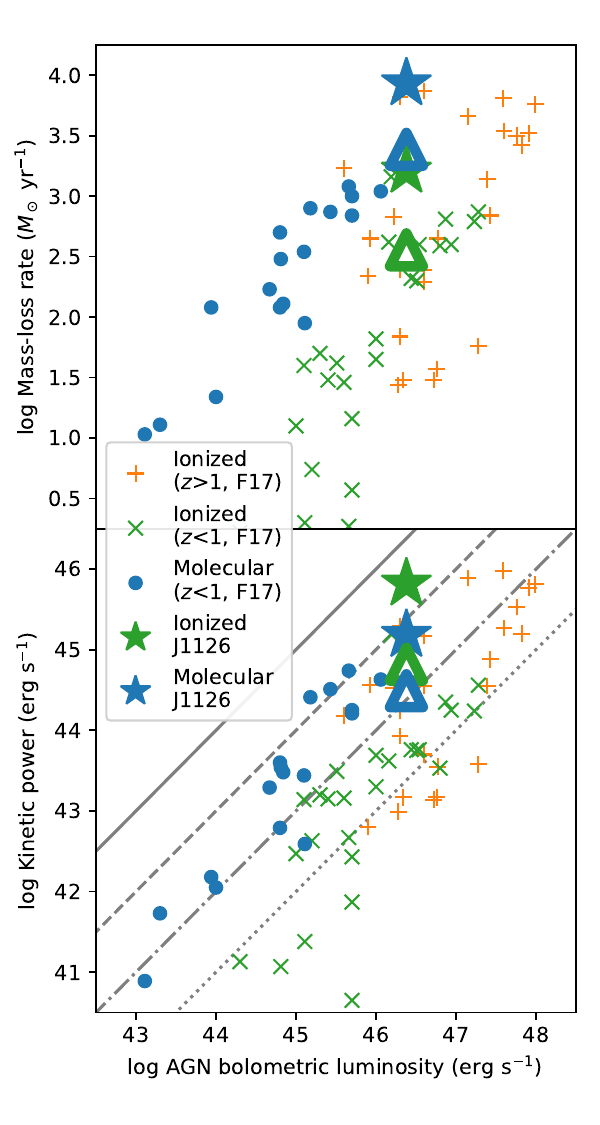}
	\end{center}
	\vspace{-11mm}
	\caption{
	Mass-loss rate (top) and kinetic power (bottom) vs. AGN bolometric luminosity of AGNs
	in the literature, which are summarized in \cite{Fiore2017} with a uniform set of assumptions.
	The ionized outflows at $z>1$ and $z<1$ are shown in orange and green crosses;
	the molecular outflows at $z<1$ are shown in blue circles. 
	In order to make a direct comparison between J1126 and the literature studies, 
	the kinetic values of J1126 (stars) are re-calculated following the same method and assumptions used in \cite{Fiore2017}, 
	e.g., the instantaneous $\dot{M}_\text{out}$ with outflow velocity defined as $v_{02}$. 
	The blue stars denote the kinetic values of the sum of blue- and redshifted molecular outflows of J1126.
	The green stars denote twice of the kinetic values of the blueshifted ionized outflow under the assumption
	of symmetric ionized outflows on the blue- and redshifted sides. 
	The blue and green open triangles denote the corresponding values with the method adopted in the paper. 
	The grey lines in the bottom panel mark positions with a loading factor of the kinetic power 
	of 0.1\% (dotted), 1\% (dashdotted), 10\% (dashed), and 100\% (solid). 
	}
	\label{fig:dMout_Ek_comp}
\end{figure}

Figure \ref{fig:dMout_Ek_comp} exhibits the comparison of $\dot{M}_\text{out}$ and $\dot{E}_\text{k,out}$
of multi-phase outflows between J1126 and the sample summarized by \cite{Fiore2017}.
Under the same assumptions adopted by \cite{Fiore2017}, 
e.g., the instantaneous $\dot{M}_\text{out}$ with outflow velocity defined as $v_{02}$,
J1126 tends to show the highest kinetic values at $z<1$ for both of ionized and molecular outflows;
the values of the ionized outflow of J1126 is comparable to those of the most luminous AGNs at high-$z$ universe. 


\subsection{Mechanisms behind the launching of the multi-phase outflows}
\label{subsec:Discuss_outflow_driven}

The high velocity and the Seyfert-type ionization (Section \ref{subsec:GMOS_ionization})
of the ionized outflow suggest that it is driven by a luminous AGN. 
Since the molecular outflow is aligned with the ionized outflow with a similar primary direction and has a similar outflow timescale, 
we consider the molecular outflow is also related to AGN activity. 
Considering the intense star formation activity in J1126, the contribution of starburst to launch
the molecular outflow cannot be fully ignored.
However, the molecular outflow has different morphologies from that of the central dusty starburst region, 
i.e., the outflow is elongated while the starburst shows a compact disk-like morphology 
(Figure \ref{fig:J1126_ALMA_CO32u_outflow})
and thus, at least the starburst is not the predominant energy source of the outflow. 

\subsubsection{Galactic outflow driven by an AGN nuclear wind}
\label{subsubsec:Discuss_driven_blast}

The so-called ``blast-wave'' scenario (also called thermal pressure-driven model or wind energy-driven model) 
is favored to explain powerful galactic outflows in AGN-host galaxies \citep[e.g.,][]{Richings2018b,Menci2019,Costa2020}.
In the scenario, a highly ionized, ultrafast (e.g., $>10^4$ \kms) wind is launched in the vicinity of the SMBH, 
which is thermalized by a reverse shock as it interacts with the ambient medium and expands adiabatically as a hot bubble. 
The hot expanding wind sweeps out gas in the host galaxy by its thermal pressure and produce a large-scale outflow. 
The pre-existing cold molecular gas in the swept-out medium can be disassociated into 
warm neutral and ionized atomic phases by hydrodynamical instability
of a forward shock \citep[e.g., section 2.3.1 in ][and references therein]{Veilleux2020};
while an effective cooling of the swept-out medium 
can protect the molecular clouds and/or
enable the in-situ formation of cold gas in the outflow 
\citep[e.g.,][]{Richings2018a,Veilleux2020}. 
The blast-wave scenario 
predicts large momentum boosts, e.g., $\dot{P}_\text{out}/(L_\text{AGN,bol}/c) \sim$ 10--20
(where $L_\text{AGN,bol}$ is the AGN bolometric luminosity and $c$ is the speed of light),
and high loading factors of kinetic powers, e.g., $\dot{E}_\text{k,out}/L_\text{AGN,bol} \sim$ 1\%--5\%,
with fiducial parameters \citep[e.g.][]{King2015,Richings2018b,Costa2020}. 

In order to examine whether the energetics of the observed outflow in J1126 matches the prediction of the blast-wave scenario, 
it is necessary to obtain the estimation of $L_\text{AGN,bol}$.
$L_\text{AGN,bol}$ can be estimated with 
the \oiii\ luminosity of the AGN-ionized outflow, 
or the IR luminosity from the dusty torus of AGN. 
The IR-based method depends on the torus covering factor 
\citep[e.g.,][]{Stalevski2016,Ichikawa2019,Toba2021}.
Due to the heavy dust obscuration in the galaxy center
(Section \ref{subsec:GMOS_extinction} and \ref{subsec:ALMA_dust}), 
the torus covering factor could be higher than typical AGNs, 
which leads to a large uncertainty in the estimated $L_\text{AGN,bol}$. 
In addition, the torus emission traces the recent AGN activity, 
with a light traveling and reprocessed
time of $\sim$ 10 years in a 10 pc scale \citep[e.g.,][]{Ichikawa2017}, 
which is too short compared to the outflow timescale of $\sim$ 1 Myr 
and makes the IR-based $L_\text{AGN,bol}$ 
not directly reflecting the launching mechanism of the outflow. 
Therefore, the extinction corrected\footnote{
	The shown values in the main text is based on the extinction from the broad component of \ha\ and \hb\ lines.
	If we adopt the extinction from the narrow Balmer lines, the estimated $L_\text{AGN,bol}$
	increases with a factor of 2.6. 
} $L_{\rm [O III]}$ is adopted to estimate $L_\text{AGN,bol}$. 
Adopting the empirical relation between the X-ray and \oiii\ luminosities \citep{Berney2015}
and the X-ray bolometric corrector \citep{Ricci2017,Ichikawa2019}, 
we obtain $L_\text{AGN,bol}=3.2\times10^{46}$ \lumcgs;
a lower estimation, $1.6\times10^{46}$ \lumcgs\ is obtained with the 
X-ray and \oiii\ luminosity relation of \cite{Ueda2015}. 
We adopt the mean of the two \oiii-based $L_\text{AGN,bol}$, i.e., 
$2.4\times10^{46}$ \lumcgs, in the later analysis. 

Adopting the time-averaged, total outflow dynamical properties in Section \ref{subsec:Discuss_outflow_dynamics}, 
the momentum boost and the loading factor of the kinetic power are
$\dot{P}_\text{out}/(L_\text{AGN,bol}/c)=19$
and 
$\dot{E}_\text{k,out}/L_\text{AGN,bol}=4\%$, 
which are consistent with the predictions of the blast-wave scenario. 
If we adopt the same assumptions used by \cite{Fiore2017}, 
J1126 will have an extreme loading factor with $\dot{E}_\text{k,out}/L_\text{AGN,bol}\sim20\%$
(Figure \ref{fig:dMout_Ek_comp}), 
which is among the highest values in the literature and 
is close to the upperlimit value adopted in simulations on galaxy evolution \citep[e.g.,][]{Schaye2015}.
With the SMBH mass ($M_\text{SMBH}$) of $6\times10^7$ \msun, which is estimated from the empirical
$M_\text{SMBH}$-$M_\star$ relation in the local universe \citep[e.g.,][]{Reines2015}, 
the \oiii-based $L_\text{AGN,bol}$ corresponds to a super-Eddington accretion, i.e., $\lambda_\text{Edd}=3.4$, 
which is efficient to launch a nuclear ultrafast wind \citep{King2015}. 
In addition to the powerful energetics, 
the blast-wave scenario is also supported by 
the outflow-induced shock ionization in the galaxy scale (Section \ref{subsec:GMOS_ionization}), 
and the C$_2$H-traced molecule dissociation in the outflow (Section \ref{subsec:ALMA_C2H}). 

The multi-phase outflows in J1126
can be explained with the blast-wave scenario accompanying with the galaxy disk \citep[e.g.,][]{Menci2019,Costa2020}, 
i.e., a massive, low-speed outflow dominated by cold molecular phase locates close to the disk plane, 
and a less-massive, fast outflow dominated by hot/warm ionized phase travels in the polar direction. 
The low speed of the molecular outflow is due to the resistance of the galaxy disk with a high ambient density\footnote{
	Since J1126 is observed with a nearly face-on view of the galaxy disk (e.g., Figure \ref{fig:J1126_scheme}), 
	the measured velocity of the molecular outflow 
	can be even lower than its intrinsic value due to the projection in the line of sight. 
}, while the dominated molecular phase is the result of the efficient cooling (i.e., due to the low velocity and high density)
and the consequent enhancement of the in-situ molecular gas formation \citep[e.g.,][]{Richings2018a}. 
The fast outflow can persist as it travels through the less dense ambient gas in the polar direction. 
The strong radiation and fast shock can enhance the ionization and dissociation of molecules; 
while the high velocity (i.e., high post-shock temperature) and low density 
reduce the cooling efficiency and suppress the in-situ molecular gas formation. 
In consequence, a fast, ionized phase dominated outflow is elongated in the polar direction of the disk. 

Different from the disk-axisymmetric structure induced by an isotropic nuclear wind in the standard blast-wave model,
it is suggested that J1126 possesses inclined outflows, 
which are tilted to the south and north 
on the blue- and redshifted sides, respectively, 
roughly following the opposite orientations 
(PA $\sim160^\circ$ and $340^\circ$; 
e.g., Figures \ref{fig:J1126_GMOS_OIII} and \ref{fig:J1126_ALMA_mCO_outflow}). 
Interestingly, the outflows on the two sides could have different inclination angles to the disk plane, 
which are indicated by the asymmetric line profiles (e.g., Figure \ref{fig:J1126_ALMA_mCO_spec}) 
and different morphologies (e.g., Figure \ref{fig:J1126_ALMA_mCO_outflow})
between the molecular outflows on the two sides. 
The redshifted outflow could have an inclination closer to the disk plane (as shown in Figure \ref{fig:J1126_scheme}), 
which carves a larger volume of gas out of the disk 
and create a brighter, more elongated morphology than the blueshifted outflow. 
The asymmetric inclinations can also explain the different CO excitation between outflows on the two side
(e.g., Figure \ref{fig:J1126_ALMA_mCO_spec} and \ref{fig:J1126_ALMA_mCO_PV}), i.e., 
the gas clouds in the blueshifted outflow are more directly exposed in the strong radiation/shock in the galaxy center
and are more intensely excited with $r_{32}$ up to 3;
while disk shielding on the inclined redshifted outflow could weaken the excitation with a lower $r_{32}$. 

\subsubsection{Other possible mechanisms to drive the outflow}
\label{subsubsec:Discuss_driven_other}

The AGN in J1126 could be highly obscured.
An almost fully buried AGN with a dust covering factor over 90\% 
is implied by the SKIRTor torus model \citep{Stalevski2016}
based on the IR luminosity of the AGN torus 
($2.2\times10^{46}$ \lumcgs, estimated from the SED fitting, e.g., Figure \ref{fig:J1126_image_SED})
and the \oiii-based $L_\text{AGN,bol}$.
The highly obscured AGN can also be indicated by the observational evidence that
although the ionized outflow exhibits a nearly face-on view, 
there is no features of Type-1 AGNs (e.g., extremely broad Balmer lines, Balmer continuum, and iron pseudo continua)
in the optical spectra. 
The long spanning between the GMOS and SDSS observations ($>10$ years) minimizes the possibility 
that a compact dusty spot locate along the line of sight incidentally 
to explain the absence of Type-1 AGN features. 

The highly obscured AGN indicates the scenario of the radiation pressure-driven outflow \citep[e.g.,][]{Costa2018}, 
in which the gas is accelerated via 
the momentum transfer as the optical/UV radiation being absorbed by the dusty medium. 
Although it is usually considered that the radiation pressure-driving mechanism cannot yield large momentum boosts 
as that output by the thermal pressure-driven scenario, 
recent works reported that
the model can also provide large momentum boosts when 
including the trapping of reprocessed IR photons \citep[e.g.,][]{Costa2018,Ishibashi2021}. 
It is likely that the observed galactic outflow of J1126
is produced with a combination of 
the thermal pressure-driving by the nuclear hot wind
(e.g., high kinetic power and shock-induced ionization/heating)
and the radiation pressure-driving by the IR photons
(e.g., highly obscured AGN with luminous IR radiation). 

The radio jet-driven scenario is considered to be dominated in radio-loud AGNs
\citep[e.g.,][and references therein]{Harrison2024}. 
The radio loudness ($f_{\nu,\text{5GHz}}/f_{\nu,\text{4400\AA}}$) of J1126 is 
1 or 6 with the extinction-corrected or uncorrected 4400\AA\ flux
and an assumed radio power-law index of 0.8 \citep[e.g.,][]{Murphy2011,Tabatabaei2017}, 
which categorizes J1126 as a radio-quiet galaxy 
($f_{\nu,\text{5GHz}}/f_{\nu,\text{4400\AA}}<10$, e.g., \citealt{Macfarlane2021}). 
Therefore, the jet driving is not a dominant mechanism for J1126. 
After subtracting the star formation-contributed radio luminosity 
that is estimated with the IR luminosity of ISM dust 
and the empirical IR-radio relation in star-forming galaxies \citep[e.g.,][]{Murphy2011}, 
the AGN-related radio luminosity is estimated to be $\nu L_\text{1.4GHz} = 1.3 \times 10^{40}$ \lumcgs, 
which is 40\% of the total radio radiation. 
The AGN radio luminosity is consistent with the 
empirical relation between $\nu L_\text{1.4GHz}$ and \oiii\ velocity \citep{Zakamska2014,Hwang2018},  
implying that it could originate from the 
synchrotron emission of electrons accelerated in shocks of the outflow.

Finally, we note that although the star formation driving may not be the dominant mechanism 
of the outflow (e.g., because of a different morphologies), 
the stellar wind may enhance the AGN-driven outflow via carrying dust into the ISM, 
which can either increase the efficiency of the in-situ molecular gas formation in the outflow \citep[e.g.,][]{Richings2018a}
or boost the radiation pressure-driving mechanism \citep[e.g.,][]{Ishibashi2021}. 


\subsection{Feedback effect of the powerful outflow on the host star formation activity} 
\label{subsec:Discuss_outflow_feedback}

The GMOS and ALMA observations of J1126 reveal a series of effects of the powerful AGN-driven outflow, e.g., 
the strong ionization by the direct AGN radiation and outflow-induced shocks (Section \ref{subsec:GMOS_ionization}), 
the enhanced excitation of molecular gas (Section \ref{subsubsec:ALMA_CO_PV_ratio}), 
and the large momentum and kinetic power coupled into the multi-phase ambient medium 
in the host galaxy (Section \ref{subsec:Discuss_outflow_dynamics}). 
In this subsection we discuss the possible instantaneous and long-term effects of the powerful galactic outflow
on the stellar mass build-up in the host galaxy of J1126. 

\subsubsection{A possible instantaneous negative feedback effect} 
\label{subsubsec:Discuss_feedback_short}

\begin{figure*}
	\vspace{-8mm}
	\begin{center}
		\includegraphics[trim=20 -8 0 0, clip, width=0.85\columnwidth]{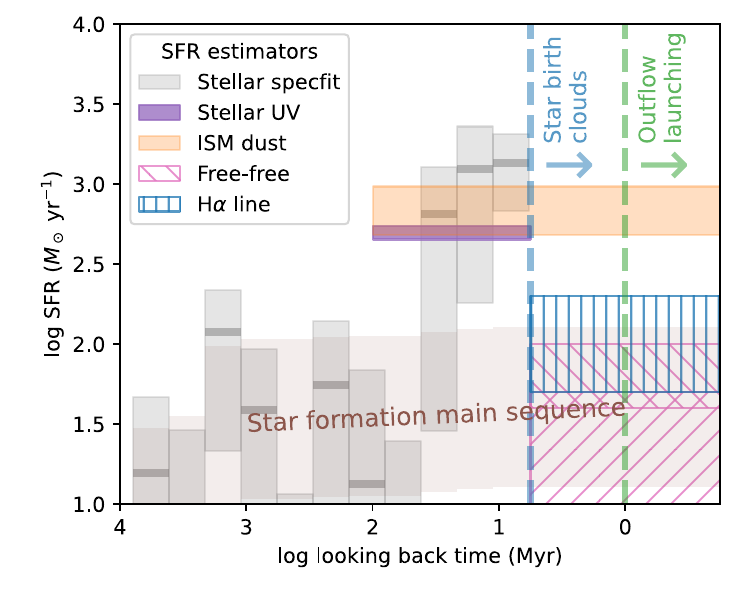}
		\includegraphics[trim=0 0 220 0, clip, width=1.25\columnwidth]{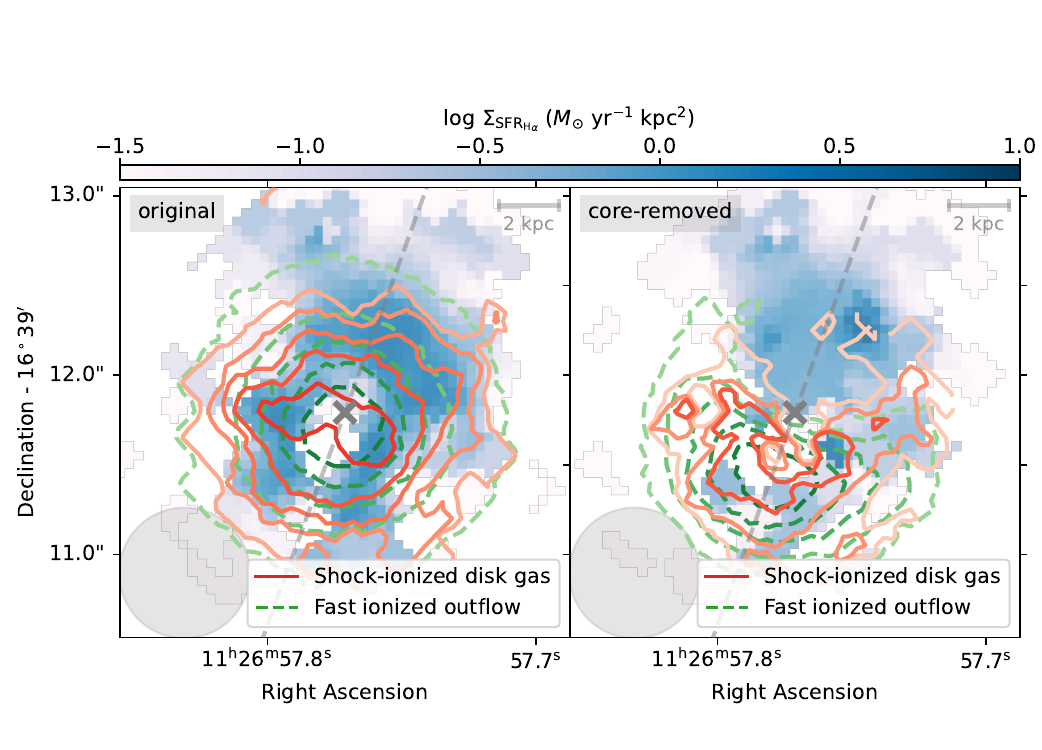}
	\end{center}
	\vspace{-10.5mm}
	\caption{
		\textbf{Left:} 
		Star formation history (SFH) of J1126 with the SFR estimated from different estimators in different timescales. 
		The grey thick lines represent the best-fit non-parametric SFH from the stellar continuum fitting
		with the grey filled bars showing the $\pm1\sigma$ uncertainty in each age bin. 
		The violet bar shows the SFR from the UV flux of the stellar continuum (rest 3500 \AA)
		with the $\pm1\sigma$ uncertainty from the extinction correction. 
		The orange bar denotes the SFR estimated from the IR luminosity contributed by star formation;
		its height reflects the uncertainty from the SED fitting and the different calibrations \citep{Murphy2011,Calzetti2013}.
		The blue and purple hatches show
		the SFR estimated from \ha\ line 
		and free-free emission (``//'' for $\alpha_\text{syn}=0.8$ and 
		``\textbackslash\textbackslash'' for $\alpha_\text{syn}=1.0$), respectively. 
		The upper bounds of these hatches contain the contamination by AGN activity, 
		while the lower bounds show the contamination-corrected SFR. 
		The blue and green vertical dashed lines show the timescales of the stellar birth clouds and the outflow launching, respectively.
		The brown shadow region represents the star formation main sequence \citep{Peng2014}
		calculated with the accumulated stellar mass in each age bin. 
		See discussion in Appendix \ref{appendix:SFH} for details of the estimations. 
		\textbf{Middle:} SFR surface density ($\Sigma_{\rm SFR_{H\alpha}}$) 
		estimated from the intensity map of narrow \ha\ line after correction for the shock ionization 
		(see Section \ref{subsubsec:Discuss_feedback_short} for details of the correction). 
		The intensity of shock-ionized \ha\ in the disk (i.e., narrow line) is shown in red contours. 
		The fast ionized outflow traced by \oiii\ are shown in green dashed contours (see also Figure \ref{fig:J1126_GMOS_Ha}).
		The grey dashed line denotes the primary outflow direction with PA of $160^\circ$ and $340^\circ$ (e.g., Figure \ref{fig:J1126_ALMA_CO32u_outflow}). 
		PSF are shown in grey circles.
		\textbf{Right:} The same map as the middle panel for the core-removed GMOS data. 
	}
	\label{fig:J1126_SFH_sigmaSFR}
	\vspace{1mm}
\end{figure*}

The archived and newly obtained multi-band observations of J1126 provide
different SFR estimators, which trace the star formation activity on different timescales
and enable us to reconstruct the star formation history (SFH) of J1126. 
The method for these SFR estimators are described in details in Appendix \ref{appendix:SFH}. 
The reproduced SFH is shown in the left panel of Figure \ref{fig:J1126_SFH_sigmaSFR}.
The spectral fitting and UV luminosity of the integrated stellar continuum 
as well as the IR radiation emitted by the ISM dust suggest 
an averaged SFR of 500--900 \sfrunit\ in the recent 100 Myr. 
However, the recombination lines of ions (e.g., \ha)
that traces the ionizing photons from the newly formed massive stars, 
reveal a lower current SFR (timescale of several Myr) of 200 and 50 \sfrunit\ 
before and after correcting for the contamination of AGN/shock ionization, respectively.
The millimeter free-free emission also suggests a low current SFR with several tens \sfrunit. 
The SFR decay in the recent several Myr implies that J1126 is experiencing a transition 
from a starburst phase back to the regime of the star formation main sequence \citep[e.g.,][]{Peng2014}.
Since the transition timescale is close to the outflow traveling time, $\sim$ 1 Myr, 
a straightforward explanation is that the current star formation activity
is suppressed by the outflow as the fueling gas swept-out and/or heated by the powerful outflow, 
i.e., an instantaneous negative feedback effect. 

Another evidence for the negative feedback of outflow is shown by 
the core-removed intensity map of the narrow \ha\ line (Figure \ref{fig:J1126_GMOS_Ha}), i.e., 
the \ha\ emission is fainter in the southern outflow region
even though the contribution of shock ionized \ha\ emission is larger
in the south (Figure \ref{fig:J1126_GMOS_BPT_map}). 
In order to reduce the contamination of shock ionization to obtain the 
the pure star formation ionized \ha\ emission and the corresponding SFR surface density ($\Sigma_{\rm SFR_{H\alpha}}$), 
we perform a rough approximation with the following equation:
\begin{equation}\label{equ:SF_ha}
	\Sigma_{L_{\rm H\alpha,SF}} = \frac{\beta_{\rm \,shock} \Sigma_{L_{\rm H\alpha,tot}} - \Sigma_{L_{\rm [NII],tot}}}
	                          {\beta_{\rm \,shock} - \beta_{\rm \,SF}},
\end{equation}  
where $\Sigma_{L_{\rm H\alpha,tot}}$ and $\Sigma_{L_{\rm [NII],tot}}$ 
are the extinction corrected, total luminosity surface densities of \ha\ and \nii\ lines, 
while $\Sigma_{L_{\rm H\alpha,SF}}$ shows the \ha\ emission only from star formation ionizing region; 
$\beta_{\rm \,SF}$ and $\beta_{\rm \,shock}$ are the ratios, $f_{\rm [NII]}/f_{\rm H\alpha}$, 
for a pure star formation or shock ionization. 
In the approximation, we arbitrarily select log$\beta_{\rm \,SF}$ = $-0.3$
and log$\beta_{\rm \,shock}$ = 0.1, 
which are the boundaries of the composite regime in the BPT diagram \citep{Kewley2001,Kauffmann2003}
with a mean log($f_{\rm [OIII]}/f_{\rm H\beta}$) = $-0.35$
(see discussion in Section \ref{subsec:GMOS_ionization} and Figure \ref{fig:J1126_GMOS_BPT_int}).
The derived $\Sigma_{L_{\rm H\alpha,SF}}$ is then converted to $\Sigma_{\rm SFR_{H\alpha}}$
with the calibration of \cite{Murphy2011},  
which are shown in the middle and right panels of Figure \ref{fig:J1126_SFH_sigmaSFR}
for the original and core-removed GMOS data. 
In the $\Sigma_{\rm SFR_{H\alpha}}$ map of the original GMOS data, 
a cavity of low $\Sigma_{\rm SFR_{H\alpha}}$ is shown along the primary direction of the outflow towards the south (PA = $160^\circ$) . 
In the cavity, the \ha\ emission is dominated by the shock ionization that is aligned with fast ionized outflow. 
The $\Sigma_{\rm SFR_{H\alpha}}$ map of the core-removed GMOS data 
suggests the suppression of $\Sigma_{\rm SFR_{H\alpha}}$ in the south outflow region more apparently\footnote{
	Due to the shielding of the galaxy disk of UV/optical emissions, 
	the observed \ha\ map could be dominated by that emitted from the side near to the observer 
	(e.g., Figure \ref{fig:J1126_scheme}) and thus,
	Figure \ref{fig:J1126_SFH_sigmaSFR} (right panel) indicates a strong feedback towards the south. 
	On the side far from the observer, a profound feedback on star formation in the north
	is expected, which is indicated by the redshifted CO outflow (e.g., Figure \ref{fig:J1126_ALMA_mCO_outflow}). 
}. 
We note that the derived values of $\Sigma_{\rm SFR_{H\alpha}}$
rely on the arbitrarily selected $\beta_{\rm \,SF}$ and $\beta_{\rm \,shock}$, 
however, a relative suppression of star formation by the outflow can be implied by these results. 

The declined current SFR (e.g., from \ha) compared to the SFR in a longer timescale (e.g., from ISM dust radiation)
as well as the suppressed $\Sigma_{\rm SFR_{H\alpha}}$ in the outflow region
both suggest an instantaneous negative feedback effect of the powerful outflow. 
However, the negative feedback may not be the only answer 
to explain the low SFR (and its surface density) estimated from the faint \ha\ emission, 
which is also possibly due to the absorption of Lyman continuum photons (Lyc) 
by dust embedding the newly born massive stars \citep[e.g.,][]{Calzetti2013}. 
The Lyc absorption can directly remove the ionizing photons emitted by young stars
and lead to the faintness of recombination lines (e.g., \ha)
and the thermal free-free emission and thus, an underestimation of SFR from these tracers. 
The Lyc absorption have been studied for the Galactic massive star-forming regions (MSFRs) \citep[e.g.,][]{Inoue2001,Binder2018}. 
Adopting the averaged Lyc absorbed fraction ($f_\text{C,abs}$) of 50\% 
from the most luminous MSFRs in the sample of \cite{Binder2018}, 
the \ha-based SFR of J1126 could be corrected to 100 \sfrunit\ 
after removal of the contamination of AGN/shock ionization. 
Although the Lyc absorption-corrected current SFR is still less than the average SFR in a long timescale, 500--900 \sfrunit, 
an instantaneous negative feedback is still difficult to be determined because of the 
large uncertainties to correct for the Lyc absorption and the AGN/shock ionization. 
An enhanced Lyc absorption (i.e., a larger $f_\text{C,abs}$) in the central and southern regions of J1126
could also provide a possible non-outflow-quenching explanation for the low measured $\Sigma_{\rm SFR_{H\alpha}}$ 
in these locations (Figure \ref{fig:J1126_SFH_sigmaSFR}), 
where the dust extinction tends to be intenser than that in the northern region (e.g., Figure \ref{fig:J1126_GMOS_AV}). 

\subsubsection{Implication of long-term feedback effects of the powerful outflow} 
\label{subsubsec:Discuss_feedback_long}

The intense kinematic effect as well as the strong ionizing/exciting power of the outflow of J1126
indicate the potential to regulate star formation in the host galaxy in a long term. 
A fraction of 60\% of the ionized outflow travels faster than the escape velocity of the host galaxy 
(Figure \ref{fig:J1126_comp_outflow}, left) with a mass-loss rate of 200 \sfrunit, 
assuming the redshifted ionized outflow has the same properties as the observed one on the blueshifted side.
Such an amount of gas can be directly removed from the host gas reservoir. 
Moreover, the ejected-out warm 
ionized outflow could heat the circumgalactic medium (CGM) with the interaction of them, 
which can affects cold accretion of CGM onto the galaxy that fuels star formation. 
A more massive outflow carries molecular gas out of the nuclear star-forming region
with a mass-loss rate of 2500 \sfrunit, which is 3--5 times of the average SFR 
in the recent starburst duration.
Although the molecular outflow cannot escape from the gravitational potential of the host galaxy, 
the enhanced excitation in the outflow could suggest the heating of gas 
as it is swept out of the nuclear star-forming region. 
Consequently, the kinematic removal and/or heating of the powerful outflow 
can induce a quenching of star formation in the host galaxy due to starvation of cold fueling gas, 
i.e., a so-called preventive or delayed feedback \citep[e.g.,][and references therein]{Veilleux2020}. 

However, there are also several factors that can limit the feedback effects of the outflow on the host.
For instance, the star formation surface density and the ionization maps (e.g., Figure \ref{fig:J1126_GMOS_BPT_map} and \ref{fig:J1126_SFH_sigmaSFR})
suggest that only the star formation along the orientation of outflow
is profoundly affected, 
while an active star-forming region can survive in the region away from the main direction of the outflow. 
Moreover, the kinematic analysis of the molecular gas in the host galaxy 
reveals an extended spiral arm-like structure with a lower excitation than that of the central disk region
(Figure \ref{fig:J1126_ALMA_CO21r_vel_wins} and \ref{fig:J1126_ALMA_mCO_PV}), 
which probably suggests the inflow of cold fueling gas that locates close to the disk plane 
and is shielded from the destruction of the outflow. 

On the other hand, the powerful galactic outflow 
also implies the existence of a strong nuclear wind, 
which is predicted by the blast-wave scenario (Section \ref{subsubsec:Discuss_driven_blast})
and has been found in several ULIRGs/AGNs \citep[e.g.,][and references therein]{Fiore2017}.
The nuclear wind may induce the self-feedback on the growth of the central SMBH
by clearing out the fueling gas in the vicinity of SMBH, 
and as a result, reduce the AGN luminosity or even halt the current active accreting period
(so-called ``fading/dead AGN''; e.g., \citealt{Ichikawa2016,Ichikawa2019c,Ichikawa2019b,Chen2020L,Pflugradt2022}). 
A typical timescale of the high-accretion rate period is 0.01--1 Myr 
from simulation studies \citep[e.g.,][]{Hickox2014,Ishibashi2022}, 
which is much shorter than the estimated starburst duration of J1126 ($\sim$ 30 Myr, Figure \ref{fig:J1126_SFH_sigmaSFR}) 
and the depletion time by star formation (20--30 Myr, assumming a constant SFR of the past starburst period). 
Although it is probably for the central SMBH to experience multiple active accreting periods, 
these short periods could result in the discontinuous energy output to the outflow and as a consequence, {}
limit the cumulative feedback effect on the host galaxy.


\section{Summary}
\label{sec:Summary}

In this paper we report the recent Gemini/GMOS IFU and ALMA observations
of a HyLIRG, J1126 at $z=0.46842$, which is selected from the SDSS-WISE-AKARI cross-matched catalog
with an apparent co-existence of a fast ionized outflow and an intense starburst activity.
The new observations reveal a powerful, multi-phase outflows 
with a series of feedback effects on the host galaxy. 

\begin{itemize}
	\item The fast ($v_{10}=2400$ \kms), \oiii-traced ionized outflow is found to be extended to several kpc
	with a $\dot{M}_\text{out}$ of 180 \sfrunit\
	on the blueshifted side. 
	Molecular outflows are detected on both of the blue- and redshifted sides, 
	which have a moderate velocity ($v_{10}=600$ \kms) with asymmetric morphologies on the two sides; 
	a high $\dot{M}_\text{out}$ of 2500 \sfrunit\ is entrained
	in the molecular outflow 
	(Section \ref{subsec:Discuss_outflow_comp}, and \ref{subsec:Discuss_outflow_dynamics}).

	\item A large momentum boost and kinetic power suggest the powerful outflow is driven by the blast-wave scenario, i.e., the adiabatic expansion of a hot nuclear wind of AGN.
	The radiation pressure-driving mechanism also probably contributes to the launching of the outflow with a highly obscured AGN
	(Section \ref{subsec:Discuss_outflow_driven}). 

	\item In addition to the high kinetic energy coupled into the ambient ISM, the powerful outflow also affect the host galaxy 
	via inducing a shock ionization (e.g., traced by \oi; Section \ref{subsec:GMOS_ionization})
	and enhancing the excitation of the molecular gas (e.g., traced by CO line ratio; Section \ref{subsubsec:ALMA_CO_PV_ratio}).
	Dissociation of molecular gas in the outflow is also suggested with the C$_2$H observation (Section \ref{subsec:ALMA_C2H}). 

	\item A possible instantaneous negative feedback effect is indicated by the declined current SFR 
	compared to the average SFR in a long timescale 
	and the suppressed SFR surface density in the outflow region. 
	However, the low SFR estimated from \ha\ is also likely 
	due to the absorption of Lyman continuum photons in the dusty birth clouds of young massive stars
	(Section \ref{subsubsec:Discuss_feedback_short}).

	\item The high kinetic power and
	the intense gas ionization/excitation of the outflow 
	suggest the potential to regulate the growth of the host galaxy in a long term
	via expelling-out and/or heating the cold fueling gas of star formation. 
	However, the cumulative feedback effect could be restricted both
	in the spatial scale (e.g., a limited volume of the outflow)
	and the time scale (e.g., a short AGN duty cycle) 
	(Section \ref{subsubsec:Discuss_feedback_long}). 
\end{itemize}

The current GMOS and ALMA observations of J1126 reveal one of the most powerful outflows at the intermediate redshift
(Figure \ref{fig:dMout_Ek_comp}), 
multiple feedback mechanisms (e.g., enhanced excitation and dissociation of molecular gas), 
and the evidence of an in-action negative impact on star formation in the host, 
which make J1126 a unique laborotary to discover the localized, instantaneous feedback effect of powerful outflows 
on growth of host galaxies.

There are several remaining puzzles on the nature of the powerful outflow in J1126 and 
the addressing of them requires future follow-up observations. 
For example, the kinematics of the ionized gas is unresolved
in a stellar bulge scale ($r<2$ kpc) due to the moderate resolution of ground-based optical observation by Gemini, 
which prevents a direct comparison of morphologies 
of the ionized outflow to the molecular outflow observed by ALMA.
JWST NIR observation can open a window to resolve the ionized outflow
with a higher resolution (e.g., $r<0.5$ kpc). 
Moreover, currently the enhanced excitation of molecular gas
in the outflow is only confirmed with one line ratio (e.g., $r_{32}$)
that leads to degeneracies to determine the gas properties, e.g., 
temperature and density. 
ALMA observations of higher-$J$ CO lines can break such degeneracies
to reveal more details of the feedback on molecular gas by the outflow.
Furthermore, as mentioned in Section \ref{subsubsec:Discuss_feedback_long} the powerful, galactic outflow
could suggest a strong nuclear wind, which may induce 
a self-feedback on the growth of SMBH
and halt the outflow ejection into the host.
The feedback in the nuclear region can be investigated 
with X-ray observations (e.g., NuSTAR) to estimate the current AGN activity, 
and ALMA observations with super-high resolution (e.g., 100 pc)
to determine the properties of the nuclear gas reservoir. 
We are conducting these multi-wavelength follow-ups 
and will report the results in future papers.  

\section*{Acknowledgments}
We appreciate the anornymous referee for the constructive suggestions on this paper. 
X.C. is supported by the ALMA Japan Research Grant of NAOJ ALMA Project, NAOJ-ALMA-349.
This work is supported by NAOJ ALMA Scientific Research Grant Code 2023-24A. 
M.A. is supported by JSPS KAKENHI grant No. JP24K00670.
Y.T. is supported by JSPS KAKENHI grant No. JP23K22537. 
T.K. is supported by Yamada Science Foundation and the Sumitomo Foundation (2200605). 
D.I. is supported by JSPS KAKENHI grant No. JP23K20870.
M.I. is supported by JSPS KAKENHI grant No. JP21K03632.
H.N. aknowledges the support by JSPS KAKENHI No. 19K21884, 20KK0071, 23K20239, 24K00672. 
The observations were carried out within the framework of Subaru-Gemini time exchange program which is operated by the National Astronomical Observatory of Japan. We are honored and grateful for the opportunity of observing the Universe from Maunakea, which has the cultural, historical and natural significance in Hawaii.
This paper makes use of the following ALMA data: ADS/JAO.ALMA \#2019.2.00085.S, ADS/JAO.ALMA \#2021.1.01496.S, ADS/JAO.ALMA \#2022.1.01376.S. ALMA is a partnership of ESO (representing its member states), NSF (USA) and NINS (Japan), together with NRC (Canada), NSTC and ASIAA (Taiwan), and KASI (Republic of Korea), in cooperation with the Republic of Chile. The Joint ALMA Observatory is operated by ESO, AUI/NRAO and NAOJ.
Data analysis was carried out on the Multi-wavelength Data Analysis System operated by the Astronomy Data Center (ADC), National Astronomical Observatory of Japan.

\facilities{
	Gemini-N/GMOS, ALMA, 
	SDSS, Subaru/HSC, 2MASS, WISE, AKARI, VLA
} 

\software{
	Astropy \citep{Astropy},
	$^\text{3D}$Barolo \citep{DiTeodoro2015},
	CASA \citep{McMullin2007},
	Gemini IRAF \citep{GeminiIRAF},
	L.A.Cosmic \citep{vanDokkum2001},
	Matplotlib \citep{Matplotlib}
} 


\appendix
\restartappendixnumbering

\section{Method of spectral fitting for the GMOS data}
\label{appendix:GMOS_fitting}

We perform the spectral fitting with stellar continuum and emission line components. 
The stellar continuum is fit with an non-parametric SFH \citep[e.g.,][]{Hernandez2000,Iyer2019,Ciesla2023}
using the single stellar population (SSP) library of PopStar \citep{Molla2009,Millan-Irigoyen2021}.
The initial mass function of \cite{Kroupa2001}
with stellar masses of 0.1--100 \msun\ is adopted. 
Since there is no nebular continua features (e.g., hydrogen free-bound continuum) of very young stars, 
we only employ the population with stellar age over 5.5 Myr. 
A solar metallicity is fixed to reduce the degeneracy in the fitting, 
which is found as a typical value in local (U)LIRGs \citep[e.g.,][]{Perez2021}.

The GMOS spectra covers a wide emission line range 
from \nevlong\ to \siilong. 
Two Gaussian profiles are used to fit each line, 
with a narrow profile (FWHM $<$ 750 \kms) for the emission from the disk gas, 
and a broad profile (750 $<$ FWHM $<$ 2500 \kms) for the emission from outflowing gas. 
Since J1126 is a Type-2 AGN, the Broad-Line-Region of AGN 
does not contributes to the observed broad line component of permitted line (e.g., \ha). 
In order to reduce the degeneracy of fitting, for either narrow or broad components, 
we tie the kinematic parameters, i.e., velocity shift ($v_\mathrm{s}$) and FWHM, of each emission line (e.g., \oii\ and \ha).
An additional broad component (the purple dashed line in bottom panels of Figure\,\ref{fig:J1126_GMOS_spec_int}) 
with $v_\mathrm{off} > 2000$ \kms\ is required for \oiii\ and \ha-\nii\ complex, 
for their blueshifted wing if the peak of the additional component shows S/N $>$ 3. 
The flux ratios of neighboring lines of the same ion, i.e., 
\oiilong, \oiiilong, and \niilong, 
are fixed to the values calculated with PyNeb \citep{Luridiana2015}
under an electron temperature of $10^4$ K and an electron density of 100 \ccm\ to reduce the degeneracy of the fitting 
(Table \ref{tab:lines}). 

The model spectra (stellar continuum plus emission lines) are convolved by the spectral PSF of 
the GMOS observation ($R\sim1000$) before fitting to the observed spectra and thus,
the output velocity width of both stellar continuum and emission lines reflect the PSF-corrected value. 
Monte-Carlo simulations are used to estimate the uncertainty of each fitting parameter. 
For each observed spectrum, we generate 100 mock spectra via adding random noise normalized by the measurement errors, 
and perform spectral fitting for the mock spectra. 
The standard deviations of best-fit parameters of the simulated spectra are used as the 
uncertainties of best-fit parameters of observed spectra. 
An example of the spectral fitting for the integrated spectra of J1126 is shown in Figure\,\ref{fig:J1126_GMOS_spec_int}.

The same spectral fitting is performed for both of the original and the core-removed cubes. 
For either the original or the core-removed cube, we bin adjacent pixels in the faint regions to achieve
S/N $>$ 5 for the $V$-band stellar continuum and S/N $>$ 10 at the peak of the \oiii\ line. 

\subsection{Comparison between the best-fit results of the SDSS and GMOS integrated spectra} 
\label{appendix:GMOS_SDSS_comp}

We also perform the same spectral fitting for the archived SDSS spectrum of J1126, 
which is shown as cyan in Figure \ref{fig:J1126_GMOS_spec_int} for a comparison with the results 
from the GMOS integrated spectrum. 

The best-fit fitting of the integrated GMOS spectrum shows the stellar continuum has a 
velocity shift of \voff\ = $18\pm21$ \kms\ and FWHM of $777\pm49$ \kms. 
The stellar \voff\ is in relative to the systemic redshift from CO(2-1) ($r$$<$0.5 kpc; Section \ref{subsec:ALMA_CO_main})
and is consistent with zero considering the uncertainty. 
The estimated FWHM has been corrected for spectral PSF, and it is consistent with the estimation from SDSS spectrum $688\pm96$ \kms\
considering the uncertainty. 
However, there is a relatively large offset between \voff\ from GMOS and SDSS results, which may be due to 
uncorrected systematic errors between the two instruments 
since a similar \voff\ offset is also shown in the fitting of narrow emission lines (see Table \ref{tab:kinematics}). 

Table\,\ref{tab:kinematics} lists \voff\ and FWHM of the narrow and broad line components 
in the fitting of the GMOS and SDSS integrated spectra. 
The narrow components of GMOS and SDSS spectra both show 
blueshifted velocity in relative to the \voff\ of their stellar continua. 
As for the estimated FWHM of the narrow components, as well as both of \voff\ and FWHM of the two broad components, 
the fitting for GMOS and SDSS spectra show consistent results. 
All of the line widths in Table\,\ref{tab:kinematics} have been
corrected for the instrumental blurring. 

\begin{table}[!ht]
	\caption{Kinetic measurements of different components.}
	\vspace{-4mm}
	\centering
	\begin{tabular}{r|DD}
		\hline
		\hline
		Component & \multicolumn2c{$v_\mathrm{s}$\footnote{The velocity is in relative to the systemic redshift from CO(2-1) ($r$$<$0.5 kpc; Tabel \ref{tab:mCO_gfit}).} (\kms)} 
		& \multicolumn2c{FWHM\footnote{The width has been corrected for the instrumental broadening.} (\kms)}\\
		\hline
		\decimals
		Stellar   		& 17.8$\pm$21.2\footnote{The upper value from GMOS integrated spectrum and 
		the bottom value from SDSS spectrum. The same for emission lines.}\footnote{Note that all of the uncertainties only reflect those from the measurement noise (via Monte-Carlo simulation), and do not contain systematic errors between different instruments.} 
											& 776.5$\pm$48.8 \\
		\  		  		& 124.6$\pm$33.5	& 687.7$\pm$95.9 \\
		Narrow line   	& -33.4$\pm$2.8   	& 532.0$\pm$13.2  \\
		\ 			  	& 33.1$\pm$13.7    	& 570.6$\pm$47.7   \\
		Broad line, 1 	& -981.3$\pm$23.3 	& 1459.8$\pm$39.6 \\
		\  				& -920.0$\pm$45.5  	& 1622.4$\pm$130.2  \\
		Broad line, 2\footnote{Only used for \oiii\ and \ha-\nii\ complex.}   
						& -2414.4$\pm$155.3 & 2082.5$\pm$215.6  \\
		\			  	& -2643.8$\pm$266.4 & 1716.0$\pm$634.3  \\
		\hline
	\end{tabular}
	\label{tab:kinematics}
\end{table}


\section{Kinematic properties and intensity maps of the ionized emission lines}
\label{appendix:GMOS_ionized_lines}

\begin{figure}
	\vspace{-1.5mm}
	\begin{center}
		\hspace{-8mm}
		\includegraphics[trim=0 50 230 30, clip, width=0.90\columnwidth]{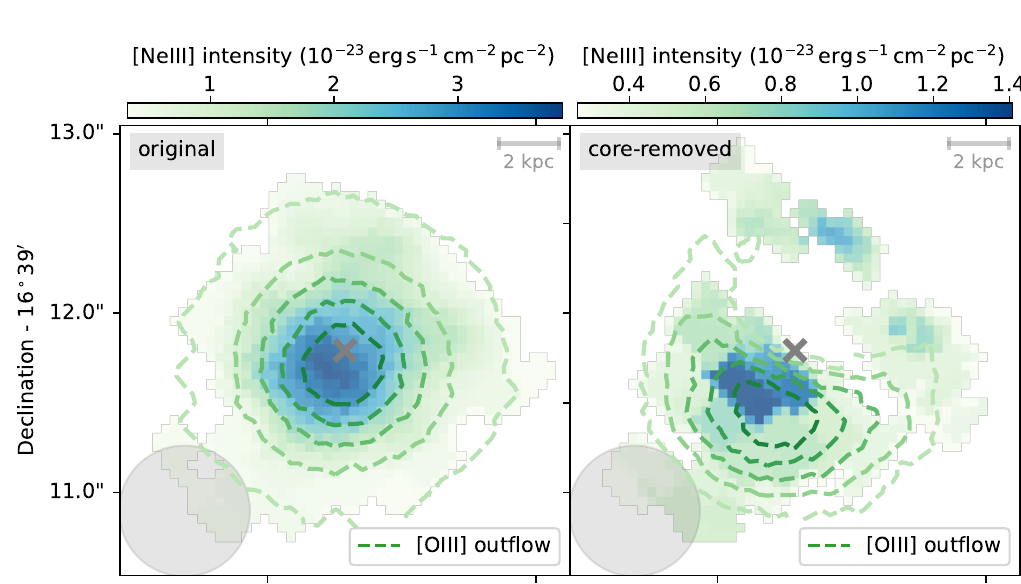}
	\end{center}
	\vspace{-5mm}
	\begin{center}
		\hspace{-8mm}
		\includegraphics[trim=0 0 230 50, clip, width=0.90\columnwidth]{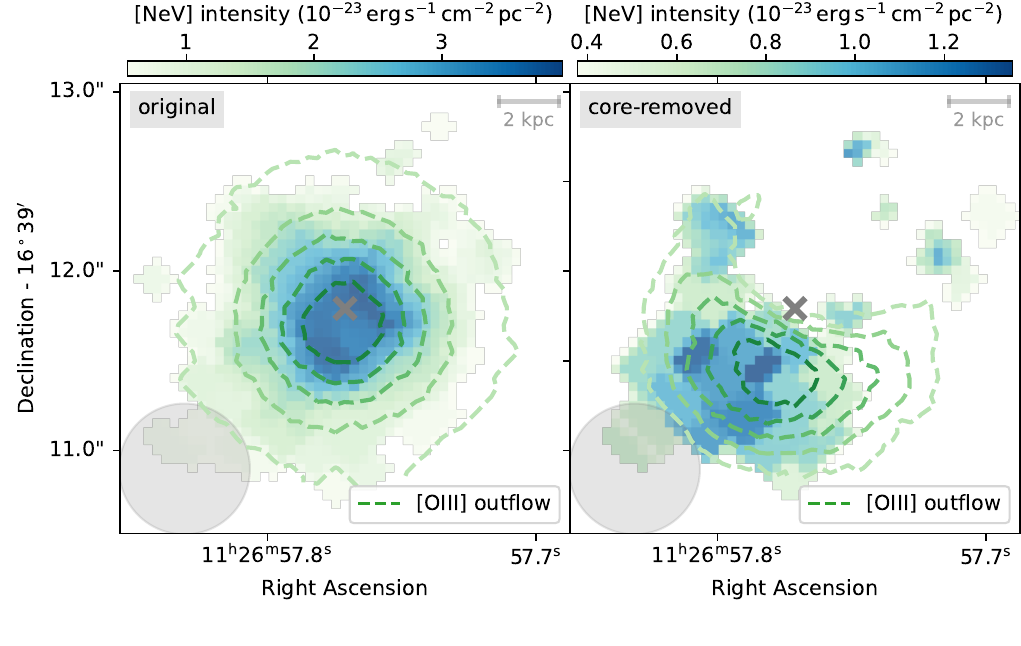}
	\end{center}
	\vspace{-9mm}
	\caption{
		Intensity maps of the outflow components of \neiiialong\ 
		and \nevlong\ 
		from the original (left) and core-removed (right) GMOS data. 
		Both of the two lines show a dominant broad component
		in its line profile, i.e., as \oiii. 
		The \oiii\ outflow is shown in green dashed contours for a reference. 
		Only pixels with S/N $>$ 3 are shown in the panels.
	}
	\label{fig:J1126_GMOS_mHiIP_intensity}
\end{figure}

\begin{figure}
	\vspace{-1.5mm}
	\begin{center}
		\hspace{-8mm}
		\includegraphics[trim=0 0 230 30, clip, width=0.90\columnwidth]{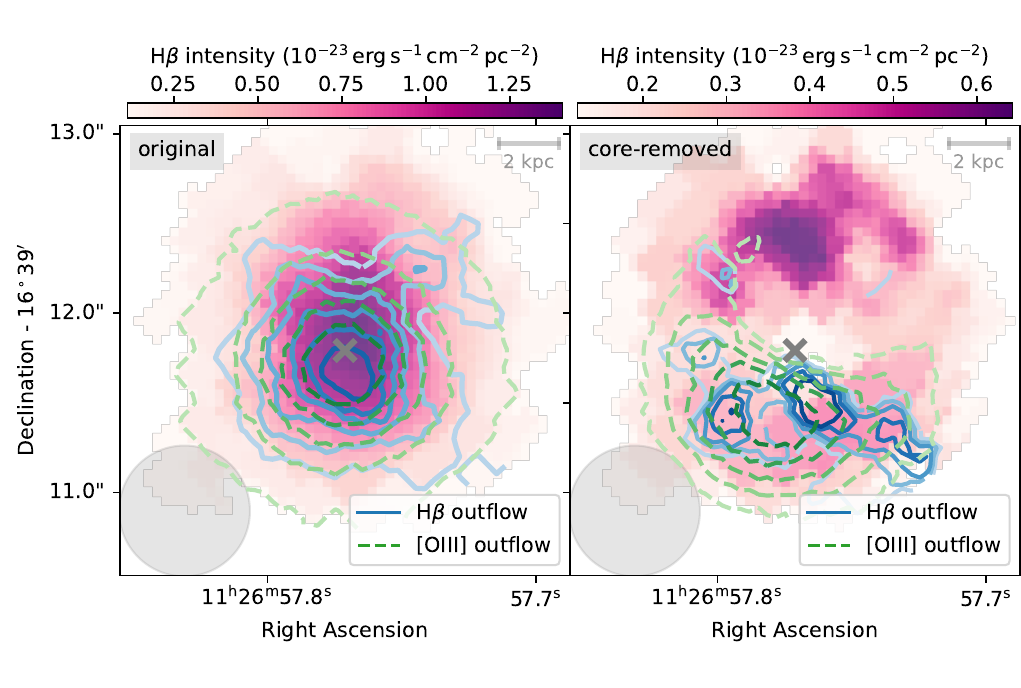}
	\end{center}
	\vspace{-9mm}
	\caption{
		Intensity maps of the narrow components of \hb\
		from the original (left) and core-removed (right) GMOS data. 
		The \hb\ outflow is shown in blue contours.
		The \oiii\ outflow is shown in green dashed contours for a reference. 
		Only pixels with S/N $>$ 3 are shown in the panels.
	}
	\label{fig:J1126_GMOS_Hb_intensity}
\end{figure}

\begin{figure}
	\vspace{-1.5mm}
	\begin{center}
		\hspace{-8mm}
		\includegraphics[trim=0 50 230 30, clip, width=0.90\columnwidth]{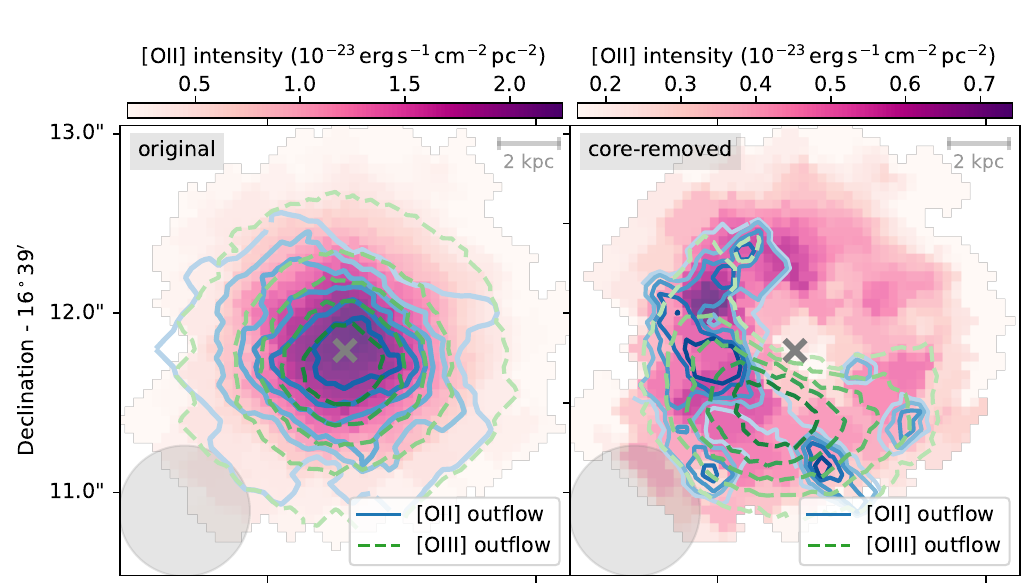}
	\end{center}
	\vspace{-5mm}
	\begin{center}
		\hspace{-8mm}
		\includegraphics[trim=0 50 230 50, clip, width=0.90\columnwidth]{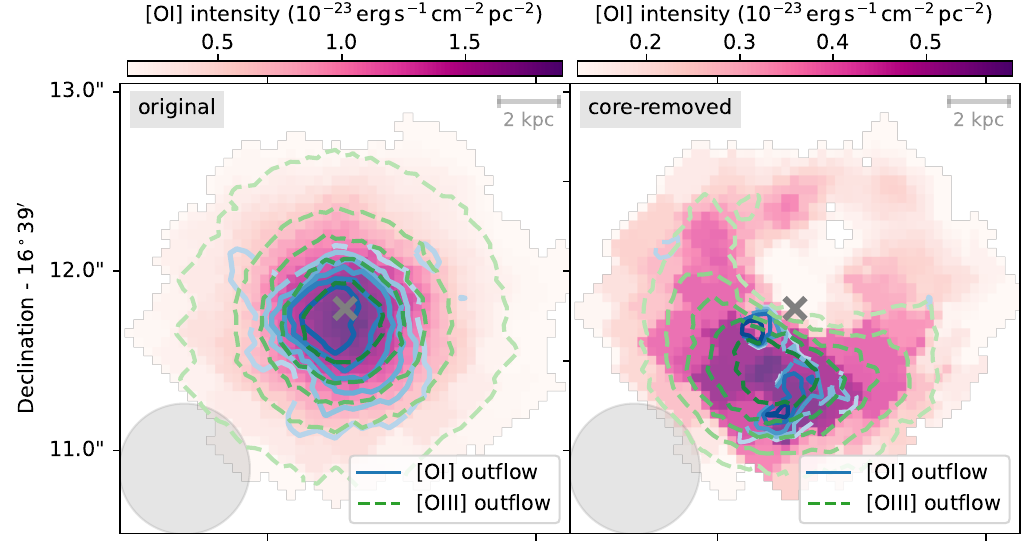}
	\end{center}
	\vspace{-5mm}
	\begin{center}
		\hspace{-8mm}
		\includegraphics[trim=0 50 230 50, clip, width=0.90\columnwidth]{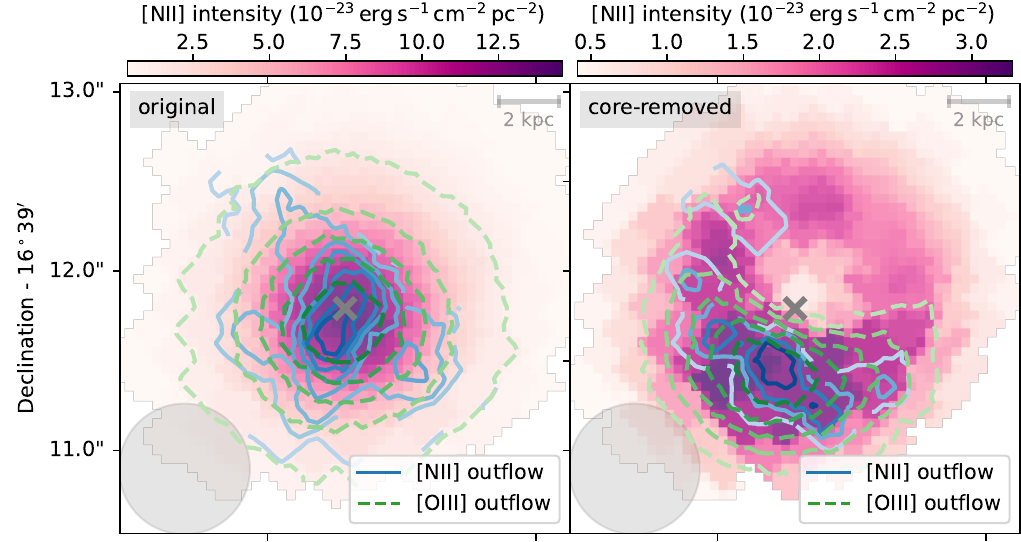}
	\end{center}
	\vspace{-5mm}
	\begin{center}
		\hspace{-8mm}
		\includegraphics[trim=0 0 230 50, clip, width=0.90\columnwidth]{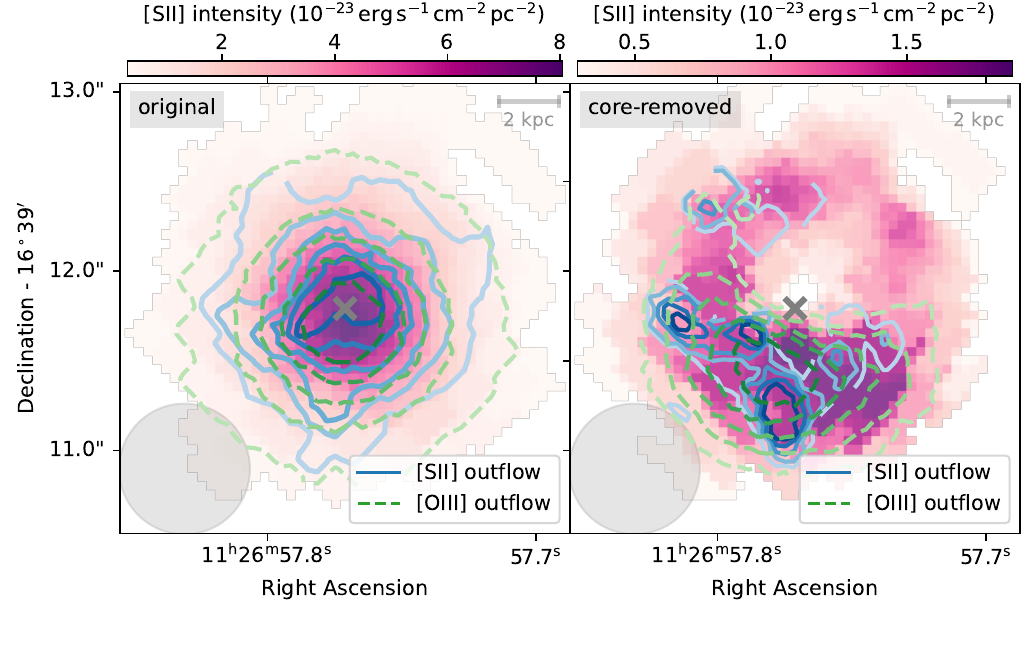}
	\end{center}
	\vspace{-9mm}
	\caption{
		The same plots as Figure \ref{fig:J1126_GMOS_Hb_intensity} for
		\oiiblong, \oialong, \niiblong, and the sum of \siilong\ doublets. 
	}
	\label{fig:J1126_GMOS_mLoIP_intensity}
\end{figure}

The velocities, fluxes and luminosities of each emission line are listed in Table \ref{tab:lines}. 
Figure \ref{fig:J1126_GMOS_mHiIP_intensity} show the intensity maps
of the lines with high ionization potentials (IP). 
Figure \ref{fig:J1126_GMOS_Hb_intensity}, and \ref{fig:J1126_GMOS_mLoIP_intensity} 
show the intensity maps of the low-IP lines (and \oi). 
The maps of \oiii\ and \ha\ are shown in Figure 
\ref{fig:J1126_GMOS_OIII} and \ref{fig:J1126_GMOS_Ha}
in the main text. 

\begin{table*}
	\caption{Velocities, fluxes, and luminosities of emission lines from the GMOS integrated spectrum.}
	\hspace{-12mm}
	\begin{tabular}{r|DDDDDD}
		\hline
		\hline
		Line \tablenotemark{\footnotesize a}
		& \multicolumn2c{$v_{50}$} 
		& \multicolumn2c{$\Delta v_{80}$} 
		& \multicolumn2c{$f_\mathrm{broad}$}
		& \multicolumn2c{Observed flux}
		& \multicolumn2c{Luminosity\tablenotemark{\footnotesize b}} 
		& \multicolumn2c{IP\tablenotemark{\footnotesize c}} \\
		\ & \multicolumn2c{(\kms)} & \multicolumn2c{(\kms)} 
		& \multicolumn2c{(\%)} & \multicolumn2c{($10^{-15}$\,erg s$^{-1}$ cm$^{-2}$)} & \multicolumn2c{($10^{43}$\,erg s$^{-1}$)} & \multicolumn2c{(eV)} \\
		\hline
		\decimals
		\nevlong\ 	 & -1096.4$\pm$23.4 & 1654.7$\pm$41.3 & 99.8$\pm$0.2 & 1.61$\pm$0.13 & 1.93$\pm$1.76 & 97.1 \\
		\oiiblong 	 &  -353.1$\pm$13.7 & 1709.3$\pm$25.8 & 44.7$\pm$1.9 & 1.94$\pm$0.05 & 6.08$\pm$1.72 & 13.6 \\
		\neiiialong\ & -1082.7$\pm$23.4 & 1641.6$\pm$42.1 & 93.7$\pm$3.4 & 1.27$\pm$0.08 & 1.41$\pm$0.91 & 41.0 \\
		\hb\ 		 &  -370.3$\pm$22.0 & 1725.1$\pm$24.2 & 48.6$\pm$2.1 & 1.42$\pm$0.04 & 1.86$\pm$0.41 & 13.6 \\
		\oiiiblong\  & -1209.6$\pm$9.6  & 2371.2$\pm$41.3 & 96.7$\pm$1.0 & 10.35$\pm$0.06 & 5.89$\pm$3.19 & 35.1 \\
		\oialong\ 	 &   -92.4$\pm$2.8  &  668.4$\pm$10.8 & 28.5$\pm$3.1 & 0.74$\pm$0.04 & 0.58$\pm$0.10 & - \\
		\ha\ 		 &  -258.2$\pm$17.8 & 1636.6$\pm$65.0 & 39.0$\pm$2.9 & 8.08$\pm$0.50 & 5.20$\pm$1.12 & 13.6 \\
		\niiblong\ 	 &  -315.9$\pm$37.7 & 2232.1$\pm$224.7 & 43.1$\pm$3.7 & 9.48$\pm$0.49 & 5.73$\pm$0.88 & 14.5 \\
		\siialong\ 	 &  -256.0$\pm$19.0 & 1627.1$\pm$37.8 & 38.0$\pm$2.7 & 2.06$\pm$0.12 &  1.24$\pm$0.21 & 10.4 \\
		\siiblong\ 	 &  -238.1$\pm$24.9 & 1597.4$\pm$50.8 & 35.4$\pm$3.6 & 1.89$\pm$0.07 &  1.16$\pm$0.19 & 10.4 \\
		\hline
	\end{tabular}
	\tablenotetext{a}{The following line doublets are tied with a fixed flux ratio, i.e., 
	$f_\mathrm{[OII]3726}/f_\mathrm{[OII]3729}=0.74$, $f_\mathrm{[OIII]4959}/f_\mathrm{[OIII]5007}=0.34$, 
	and $f_\mathrm{[NII]6548}/f_\mathrm{[NII]6583}=0.34$, for both of their narrow and broad components.
	These values are calculated with PyNeb under an electron temperature of $10^4$ K and an electron density of 100 \ccm.}
	\tablenotetext{b}{Extinction corrected luminosity. The narrow and broad components are corrected with $A_{V,\mathrm{\,narrow}}=3.0$ and $A_{V,\mathrm{\,broad}}=1.7$ (Section \ref{subsec:GMOS_extinction}), respectively. The shown errors reflect both uncertainties from the measurement error and the extinction correction. }
	\tablenotetext{c}{Ionization potential. The values are adopted from http://astronomy.nmsu.edu/drewski/tableofemissionlines.html. The binding energy of a hydrogen atom, 13.6 eV, is adopted for Balmer lines.}
	\label{tab:lines}
\end{table*}

\subsection{Decomposition of broad \ha\ and \nii\ lines in the faint outskirt region}
\label{appendix:GMOS_broad_Ha_NII}

\begin{figure}[!ht]
	\begin{center}
		\hspace{-9mm}
		\includegraphics[trim=0 0 0 0, clip, width=\columnwidth]{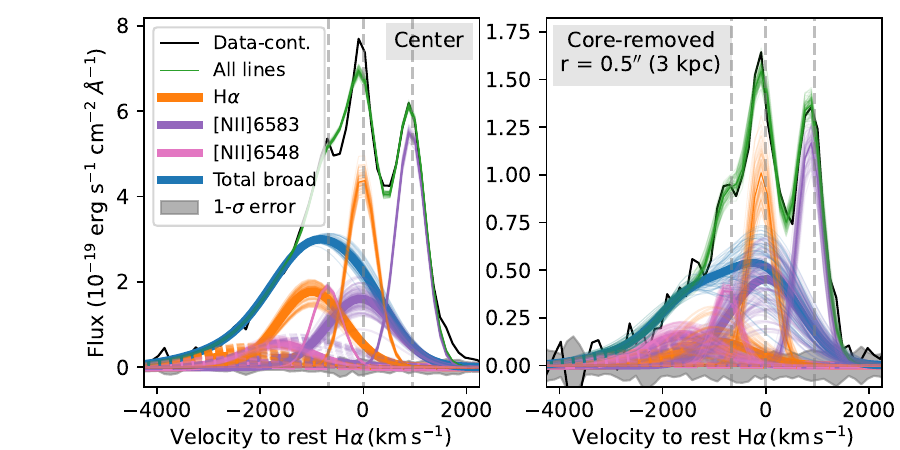}
	\end{center}
	\vspace{-6mm}
	\caption{
		Fitting of the \ha-\nii\ line complex in the central pixel (left) from the original cube,
		and an outskirt pixel with a distance of 6.5\arcsec\ from the core-removed cube 
		(shown in an red cross in Figure \ref{fig:J1126_GMOS_NIIHaB_intensity}). 
		The continuum-subtracted data is shown in black, 
		while the total line model shown in green. 
		\ha, \niiblong, and \niialong\ are shown in orange, violet, and purple, respectively.
		The narrow components are marked with thin curves that are close to 
		the rest wavelengths (vertical dashed lines) at the systemic redshift.
		The two broad components of each line are shown in thick solid and dashed curves, respectively.
		The sum of broad \ha\ and \nii\ lines is shown in blue. 
		For all of the fitting components, the thick curve marks the best-fit model of the data
		while the thin curves denote the results of the mocked spectra to shown the fitting uncertainty.
		The $\pm1\sigma$ error range is marked in grey regions. 
	}
	\label{fig:J1126_GMOS_spec_NIIHa}
	\vspace{-1mm}
\end{figure}

The fitting uncertainty of the broad \ha\ and \nii\ lines could be large in the outskirt region where S/N is moderate. 
For an example. the spectra of the central pixel and an outskirt pixel with a distance of 6.5\arcsec\ (4 kpc)
are shown in Figure \ref{fig:J1126_GMOS_spec_NIIHa} 
with the fitting uncertainty shown using the mocked spectra. 
For the bright central spectrum, the broad \ha\ and \nii\ lines can be clearly decomposed
with S/N of 9 and 7 for the fluxes of the two lines, respectively.
However, for the fainter spectrum in the outskirt pixel, the decomposition of broad \ha\ and \nii\ lines
becomes unstable with the S/N decreasing down to 2 and 3 for the two lines, respectively.
The sum of broad \ha\ and \nii\ lines is also shown in Figure \ref{fig:J1126_GMOS_spec_NIIHa}, 
which can be detected a bit more robustly with the extended blueshifted wing feature of the \ha-\nii\ complex
and have a higher S/N in the fitting (5 in this case). 

\begin{figure}
	\vspace{-1.5mm}
	\begin{center}
		\hspace{-8mm}
		\includegraphics[trim=0 0 230 30, clip, width=0.90\columnwidth]{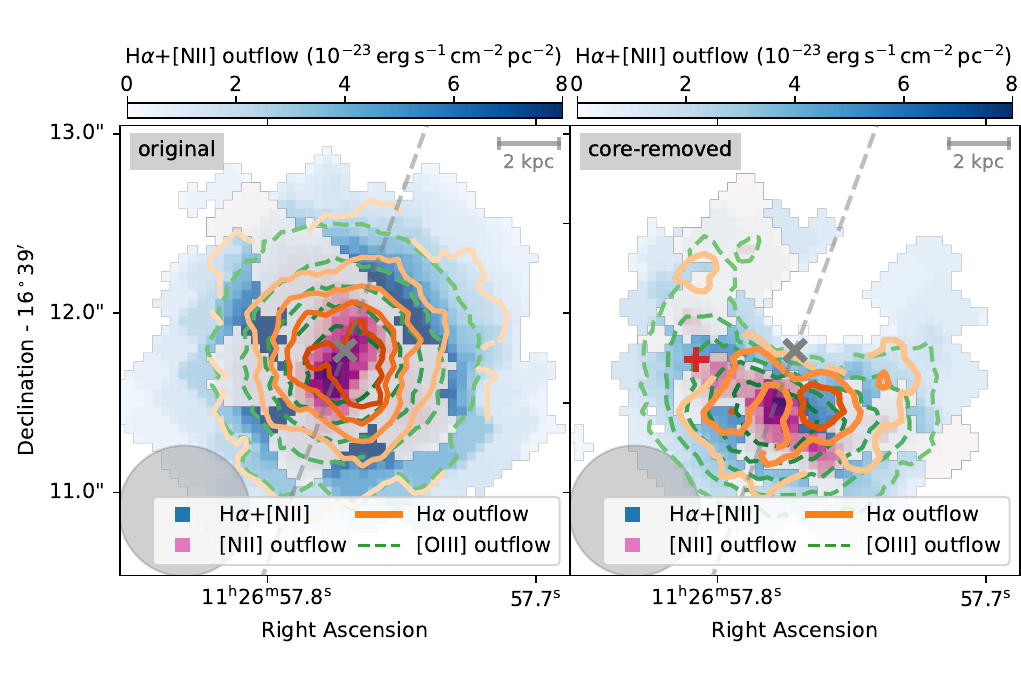}
	\end{center}
	\vspace{-9mm}
	\caption{
		Intensity maps of the sum of broad \ha\ and \nii\ lines (blue)
		from the original (left) and core-removed (right) GMOS data. 
		The the \nii\-only outflow is over-plotted in purple. 
		The \ha-only outflow is shown in orange thick contours
		while the \oiii-traced outflow shown in green dashed contours. 
		Only pixels with S/N $>$ 3 are shown for each components.
		The grey dashed line denotes the primary outflow direction with PA of $160^\circ$ and $340^\circ$. 
		The spectra of pixels locating at the grey (galaxy center) and red crosses 
		are shown in Figure \ref{fig:J1126_GMOS_spec_NIIHa}.
	}
	\label{fig:J1126_GMOS_NIIHaB_intensity}
\end{figure}

We plot the intensity maps of \ha\ and \nii\ broad lines 
and the sum of them
in Figure \ref{fig:J1126_GMOS_NIIHaB_intensity}.
Either of \ha\ or \nii\ broad components follows the distribution of \oiii-traced outflow, 
e.g., enhanced towards the south in the core-removed map.
The \ha+\nii\ sum intensity shows a closer association with \oiii, 
which could be a more robust tracer of ionized outflow than the \ha- or \nii-only broad line
since the sum has a higher S/N in the fitting. 

The \ha- and \nii-only broad lines seems to have different morphologies, 
e.g., an \ha-cavity with a \nii\ bright spot 
following the primary direction of the outflow (PA = $160^\circ$) 
in the south of the core-removed map.
The \nii-bright, \ha-faint cavity suggests a strong shock in the outflowing gas, 
where the broad \oi\ and \sii\ lines are also enhanced (Figure \ref{fig:J1126_GMOS_mLoIP_intensity}), 
and could be related to the shock in the host disk gas (i.e., narrow lines) at the same direction
(Figure \ref{fig:J1126_GMOS_BPT_map} and \ref{fig:J1126_SFH_sigmaSFR}). 
However, the hardness of the decomposition of broad \ha\ and \nii\ lines in outskirt region
prevents us from a detailed ionization analysis 
since all of the line ratios in the BPT diagnostics rely on \ha. 
Diagnostics with lines that are not affected by blurring of adjacent lines, 
e.g., the ratio of \hbox{[Fe \sc{ii}]} (1.25 \micron) and Pa$\beta$ \citep{Bianchin2024}, 
is required to address this question. 

\section{Kinematic properties of multi-Gaussian fitting of CO lines} 
\label{appendix:ALMA_CO_mGaussianFit}

The velocities and the flux fractions of the disk and outflow components of the 3-Gaussian fitting are listed in Table \ref{tab:mCO_gfit}. 
The line profiles of CO(2-1) and CO(3-2) are extracted from the beam and MRS-matched cubes. The line profile of CO(4-3) is extracted from the unresolved ACA data.

\begin{table}
	\caption{Results of 3-Gaussian fitting of CO lines}
	\vspace{-4mm}
	\centering
	\begin{tabular}{r|DDD}
		\hline
		\hline
		Component & \multicolumn2c{$v_\mathrm{s}$\footnote{Relative to systemic redshift from CO(2-1) ($r$$<$0.5 kpc).}} 
		& \multicolumn2c{FWHM\
		}
		& \multicolumn2c{Fraction} \\
		$\ $ & \multicolumn2c{(\kms)} & \multicolumn2c{(\kms)} & \multicolumn2c{\%} \\
		\hline
		\decimals
		\multicolumn7c{CO(2-1) ($r$$<$0.5 kpc)}   \\
		disk				&  0.0$\pm$0.7 		& 274.4$\pm$1.5 	& 91.5$\pm$0.7 \\
		blue-wing			& -524.1$\pm$11.9 	& 253.6$\pm$23.9 	&  2.5$\pm$0.2 \\
		red-wing			&  375.0$\pm$31.1 	& 449.6$\pm$45.1 	&  6.0$\pm$0.7 \\
		\hline
		\multicolumn7c{CO(2-1) ($r$$<$3.0 kpc)}   \\
		disk				&  11.5$\pm$1.0 	& 305.7$\pm$2.2 	& 85.4$\pm$0.9 \\
		blue-wing			& -637.5$\pm$39.8 	& 401.0$\pm$54.6 	&  2.3$\pm$0.4 \\
		red-wing			&  312.1$\pm$23.1 	& 596.3$\pm$31.3 	& 12.2$\pm$0.8 \\
		\hline
		\multicolumn7c{CO(3-2) ($r$$<$0.5 kpc)}   \\
		disk				& -0.7$\pm$1.1 		& 282.8$\pm$2.9 	& 87.7$\pm$1.7 \\
		blue-wing			& -400.7$\pm$19.8 	& 218.3$\pm$83.3 	&  4.5$\pm$1.0 \\
		red-wing			&  388.6$\pm$25.9 	& 394.1$\pm$53.2 	&  7.8$\pm$1.1 \\
		\hline
		\multicolumn7c{CO(3-2) ($r$$<$3.0 kpc)\footnote{The blueshifted wing component is too faint to be detected
		with a peak S/N $\sim$ 1.}}  \\
		disk				&  4.9$\pm$2.6 		& 296.4$\pm$7.0 	& 83.6$\pm$3.7 \\
		red-wing			&  282.1$\pm$79.6 	& 641.0$\pm$93.4 	& 14.9$\pm$3.6 \\
		\hline
		\multicolumn7c{CO(4-3) (total)\footnote{Only the disk component is used in the fitting due to a moderate S/N.}}  \\
		disk 				&  4.3$\pm$5.7  	& 360.5$\pm$9.2 	& $\,$ $\,$ $\,$ $-$ \\
		\hline
	\end{tabular}
	\label{tab:mCO_gfit}
\end{table}




\section{Estimation of star formation history} 
\label{appendix:SFH}

The archived and newly obtained multi-band observations of J1126 provide
different SFR estimators, which trace the star formation activity on different timescales,
and have different model dependencies and limitations, e.g., the contamination of AGN. 
We try to reconstruct the SFH of J1126 with these SFR estimators.
All of them are based on the stellar initial mass function of \cite{Kroupa2001}
with stellar masses of 0.1--100 \msun\ and a solar metallicity. 

We perform the stellar continuum fitting (Appendix \ref{appendix:GMOS_fitting})
assuming a non-parametric SFH \citep[e.g.,][]{Hernandez2000,Iyer2019,Ciesla2023}
utilizing the SSP library of PopStar \citep{Millan-Irigoyen2021}
with stellar ages ranging from the age of the Universe at the redshift
to the timescale of stellar birth clouds ($t_\mathrm{BC}$).  
Stars younger than $t_\mathrm{BC}$ are highly obscured 
with a small contribution on the observed stellar light (see discussion in Section \ref{subsec:GMOS_extinction})
and thus, they are not considered in the spectral and SFH fitting. 
Young stars older than $t_\mathrm{BC}$ and the evolved old stars 
are assumed to embedded in the diffuse dust with the same extinction amount. 
We assume $t_\mathrm{BC}\sim$ 5.5 Myr in the fitting. 
The best-fit SFH is shown as the grey bars in left panel of Figure \ref{fig:J1126_SFH_sigmaSFR}, 
which consists of an old population with $\sim1$ Gyr
and a starburst in the recent 5.5--30 Myr with an average SFR of $1700\pm500$ \sfrunit. 
The average SFR in the recent 5.5--100 Myr is estimated to be $450\pm140$ \sfrunit. 
We also estimate SFR using the calibration with 
the stellar continuum flux in the UV band (e.g., rest 3500\AA) 
under the assumption of a constant SFR in the recent 100 Myr \citep{Calzetti2013}, 
which gives $500\pm70$ \sfrunit\ with the uncertainty from the extinction correction, 
i.e, $A_{V,\mathrm{\,stellar}}=1.5\pm0.1$ of the diffuse dust (the violet bar in Figure \ref{fig:J1126_SFH_sigmaSFR}, left panel). 
Since the UV continuum of the newly born stars in the birth clouds is highly obscured, 
the UV-based SFR is also considered to reflect the formation of stars older than $t_\mathrm{BC}$,
which value is consistent with the averaged SFR in the 5.5--100 Myr by the non-parametric SFH fitting. 

The IR radiation reprocessed by the dust heated by the UV and optical light of stars 
is commonly used to estimate the SFR in recent several tens to 100 Myr \citep[e.g.,][]{Calzetti2013,Murphy2011}.
The total IR luminosity (1--1000 \micron) contributed by the star formation heated dust 
is estimated to be $L_\text{SF,IR}=(5.5\pm1.0)\times10^{12}$ \lsun\ from the multi-band SED fitting (Section \ref{subsec:ALMA_dust}),
which corresponds to an averaged SFR of 820 and 590 \sfrunit\
with the calibration of \cite{Murphy2011} and \cite{Calzetti2013} for a timescale of 100 Myr, respectively
(the orange bar in Figure \ref{fig:J1126_SFH_sigmaSFR}, left panel). 
The difference between the two estimates is mainly due to the extinction wavelength range of the two calibrations, i.e.,
\cite{Murphy2011} assumes the Balmer continuum (912--3646 \AA) is absorbed by the dust
while \cite{Calzetti2013} assumes the entire stellar bolometric radiation is absorbed. 
These estimated SFR contains the contribution of both of the stars embedded in the dusty birth clouds
and those older than $t_\mathrm{BC}$, i.e., in the diffuse dust.
With the best-fit stellar population and the extinction from the stellar continuum fitting, 
we can derive the absorbed stellar luminosity, $L_\text{SF,abs}=4.4\times10^{12}$ \lsun. 
Since the derived $L_\text{SF,abs}$ only considers the absorbed stellar radiation by the diffuse light, 
we can the estimate the IR luminosity purely from the dusty birth clouds as,
$L_\text{SF,IR} - L_\text{SF,abs}$.
The corresponding SFR for stars younger than $t_\mathrm{BC}$
is then estimated to be 150 \sfrunit\ 
with the calibration of \cite{Calzetti2013} for a timescale of 10 Myr.


Hydrogen recombination lines (e.g., \ha) and the forbidden metal lines (e.g., \oiialong) are correlated 
to the ionizing photons of stars and thus provide estimators of SFR. 
Since the ionization by stellar light is dominated by massive stars that have a short lifetime, 
e.g., $\sim$ 5 Myr for O-type stars \citep{Calzetti2013,Kennicutt2009}, 
and are highly obscured (see discussion in \ref{subsec:GMOS_ionization}),
these SFR estimators are considered to sample the current star formation in the dense birth clouds, i.e., $t<t_\mathrm{BC}$. 
Utilizing the extinction corrected narrow line components from the GMOS integrated spectrum
and the calibrations of \cite{Kennicutt2009}, 
the SFR are estimated to be 220 and 210 \sfrunit\ with \ha\ and \oiialong\ lines, respectively. 
As discussed in Section \ref{subsec:GMOS_ionization}, 
there is a large contribution of shock ionization for the narrow lines
especially in the southern region in the direction the ionized outflow,
therefore the SFR estimated above should be considered as upper limits.
The contamination by the shock ionization can be approximately corrected 
with the method described in Section \ref{subsubsec:Discuss_feedback_short} and Equation \ref{equ:SF_ha}. 
The pure \ha-based SFR is then estimated to be 50 \sfrunit. 
The two SFR from the total and shock-corrected \ha\ luminosities
are shown as upper- and lower bounds of the blue hatch in the left panel of Figure \ref{fig:J1126_SFH_sigmaSFR}. 

Finally we estimate the SFR from the free-free emission. 
The free-free emission in star-forming galaxies originates from the 
Coulomb interaction between free electrons and ions in thermal equilibrium, 
which is also correlated to the ionizing photons as the hydrogen recombination lines
and provides another current SFR tracer \citep{Calzetti2013,Murphy2011}.
The free-free luminosity of J1126 can be estimated from the multi-band SED fitting 
(Section \ref{subsec:ALMA_dust} and Figure \ref{fig:J1126_image_SED}), 
which is faint and depends on the assumed index of the synchrotron radio emission ($\alpha_\text{syn}$). 
The corresponds SFR is 50 \sfrunit\ for $\alpha_\text{syn}=0.8$
and 100 \sfrunit\ for $\alpha_\text{syn}=1.0$. 
Note that there is also a probable contamination by AGN such as 
the free-free cooling in the AGN-driven galactic outflow. 
We estimate the contamination by AGN following the 
theoretical relation between free-free and \oiii\ luminosities of \cite{Baskin2021}
and obtain a contamination-corrected SFR of 
0 (i.e., the free-free is fully from the cooling of outflow)
and 40 \sfrunit\ for $\alpha_\text{syn}$ of 0.8 and 1.0, respectively.
These results are shown as purple hatches in the left panel of Figure \ref{fig:J1126_SFH_sigmaSFR}. 


\bibliography{J1126_multi_outflow_v1}{}
\bibliographystyle{aasjournal}

\end{document}